\documentclass[aps,prd,10pt,nofootinbib,superscriptaddress,altaffilletter,
twocolumn,floatfix]{revtex4-2}

\usepackage{amsmath,amssymb}
\usepackage{graphicx}
\usepackage{multirow,booktabs}
\usepackage{capt-of}
\usepackage{hyperref}
\usepackage{orcidlink}
\usepackage{xcolor}

\hypersetup{
  colorlinks=true,
  linkcolor=blue,
  citecolor=blue,
  urlcolor=blue,
  filecolor=blue,
  linktoc=all,
  pdftitle={Reconstructing sign-switching dark energy histories: Scalar-field regularity, conditional potential comparison, and representative dynamics},
  pdfauthor={Shahnawaz A. Adil, Ozgur Akarsu, Mariam Bouhmadi-Lopez, Benat Ibarra-Uriondo, Nihan Katirci, J. Alberto Vazquez},
  pdfsubject={Scalar-field regularity and representative dynamics for sign-switching dark energy histories},
  pdfkeywords={sign-switching dark energy, scalar-field reconstruction, phantom field, ECDM, SSCDM, Lambda ladder}
}

\newcommand{\dd}{\mathrm{d}}

\begin{document}

\title{Reconstructing sign-switching dark energy histories:
Scalar-field regularity, conditional potential comparison, and
representative dynamics}

\author{Shahnawaz A. Adil\,\orcidlink{0000-0003-4999-7801}}
\email{shahnawaz@icf.unam.mx}
\affiliation{Instituto de Ciencias F\'{\i}sicas, Universidad Nacional Aut\'{o}noma de M\'{e}xico, Cuernavaca,
Morelos, 62210, M\'{e}xico}

\author{\"{O}zg\"{u}r Akarsu \orcidlink{0000-0001-6917-6176}}
\email{akarsuo@itu.edu.tr}
\affiliation{Department of Physics, Istanbul Technical University, Maslak 34469 Istanbul, T\"{u}rkiye}

\author{Mariam Bouhmadi-L\'{o}pez \orcidlink{0000-0002-1529-1889}}
\email{mariam.bouhmadi@ehu.eus}
\affiliation{IKERBASQUE, Basque Foundation for Science, 48011, Bilbao, Spain}
\affiliation{Department of Physics, University of the Basque Country UPV/EHU, 48080 Bilbao, Spain}
\affiliation{EHU Quantum Center, University of the Basque Country UPV/EHU, 48080 Bilbao, Spain}

\author{Be\~{n}at Ibarra-Uriondo \orcidlink{0009-0008-8064-2340}}
\email{benat.ibarra@ehu.eus}
\affiliation{Department of Physics, University of the Basque Country UPV/EHU, 48080 Bilbao, Spain}

\author{Nihan Kat{\i}rc{\i} \orcidlink{0000-0002-9492-3791}}
\email{nkatirci@dogus.edu.tr}
\affiliation{Department of Electrical-Electronics Engineering, Do\u{g}u\c{s} University, \"{U}mraniye 34775 Istanbul, T\"{u}rkiye}

\author{J. Alberto V\'{a}zquez\,\orcidlink{0000-0002-7401-0864}}
\email{javazquez@icf.unam.mx}
\affiliation{Instituto de Ciencias F\'{\i}sicas, Universidad Nacional Aut\'{o}noma de M\'{e}xico, Cuernavaca,
Morelos, 62210, M\'{e}xico}

\begin{abstract}
Phenomenologically similar sign-switching dark energy histories need not have comparable scalar-field realizations. We reconstruct minimally coupled scalars with fixed kinetic sign for three prescribed histories: the error-function model (ECDM), the smooth-step model (SSCDM), and the ladder-like model (L$\Lambda$CDM). Continuous negative-to-positive density crossings select the phantom branch. ECDM yields a smooth on-shell potential at every finite redshift; its equation-of-state pole at the density zero is only a ratio singularity. Exact SSCDM has a regular trajectory but a $C^1$, non-$C^2$ endpoint potential, $V-V_e\propto|\phi-\phi_e|^{4/3}$. Its non-Lipschitz force permits delayed departures from frozen plateaus, so the prescribed history is not uniquely generated by the reconstructed potential and plateau data. The exact Ladder requires distributional kinetic stress and has no ordinary classical realization in the adopted one-field action. In the two conditional synthetic comparisons, the sigmoid--Gaussian family ranks highest for ECDM and the generalized axion-like family for SSCDM, although the fitted axion exponents $n<1/2$ imply a divergent force at an included endpoint. Representative regular forward solutions exhibit sign changes in the scalar energy density, with the potential zero preceding the density zero for the displayed time orientation. In a direct closure test, the unretuned forward evolution of the top-ranked ECDM template tracks the target only approximately, $\max|\Delta\widetilde\Omega_\phi|\simeq0.16$, whereas closure fails structurally for compact SSCDM because no potential with a locally Lipschitz force can reproduce its finite-duration frozen plateaus from the corresponding exactly frozen initial data. Field-map existence, invertibility, and endpoint regularity must therefore be assessed before a phenomenological history is interpreted as scalar dynamics. The phantom action is used only as a homogeneous effective proxy.
\end{abstract}

\maketitle

%%-----------------------------------------------------------------------------%%
\section{Introduction}
\label{intro}
%%-----------------------------------------------------------------------------%%
The discovery of late-time cosmic acceleration \cite{SupernovaSearchTeam:1998fmf,SupernovaCosmologyProject:1998vns} and the subsequent precision measurements of the cosmic microwave background (CMB), baryon acoustic oscillations (BAO), Type Ia supernovae (SN~Ia), and large-scale structure have established the spatially flat $\Lambda$CDM model as a highly successful description of the observable Universe \cite{Planck:2018vyg,eBOSS:2020yzd,Scolnic:2021amr,Brout:2022vxf,DESI:2024mwx,AtacamaCosmologyTelescope:2025blo,SPT-3G:2025bzu}. Nevertheless, the physical origin and radiative stability of the cosmological constant remain unresolved \cite{Weinberg:1988cp,Sahni:2002kh}, and the parameters inferred from different datasets exhibit several persistent discrepancies \cite{DiValentino:2021izs,Perivolaropoulos:2021jda,Abdalla:2022yfr,CosmoVerseNetwork:2025alb}. The DESI Data Release 2 BAO analyses have sharpened interest in time-dependent dark energy (DE) when BAO are combined with CMB and supernova data \cite{DESI:2025zgx,DESI:2025fii}. The complementary DESI DR2 Ly$\alpha$-forest full-shape analysis measures the Alcock--Paczy\'nski effect with approximately one-percent precision at $z_{\rm eff}=2.33$ \cite{DESI:2026lnd}, providing a particularly precise expansion-history anchor above redshift unity. These measurements make the regime relevant to sign-switching scenarios increasingly testable \cite{Akarsu:2025dmj,Bouhmadi-Lopez:2026vyc}, but they do not, by themselves, establish that a separately defined DE density becomes negative or changes sign. Likewise, a preference for time-dependent $w(z)$, or for a phantom-divide crossing, is not evidence for a zero of $\rho_{\rm de}$: a density zero and an equation-of-state (EoS) crossing are mathematically distinct events. The complementary theory question is, therefore, whether a proposed phenomenological history belongs to the configuration space of a stated field theory. Given $\rho_{\rm de}(z)$, what homogeneous scalar trajectory and on-shell potential, if any, reproduce the same background evolution?

Among these cosmological discrepancies, the best-known is the $H_0$ tension between local distance-ladder measurements and the value inferred from early-Universe observations within base $\Lambda$CDM \cite{Planck:2018vyg,Riess:2021jrx,Breuval:2024lsv,H0DN:2025lyy}. A second, less significant discrepancy concerns the clustering combination $S_8\equiv\sigma_8\sqrt{\Omega_{\rm m0}/0.3}$, as inferred from CMB and weak-lensing or large-scale structure probes \cite{DiValentino:2020vvd,KiDS:2020suj,Wright:2025xka,DES:2026fyc}. Both the interpretation of these discrepancies and their susceptibility to systematics remain under active discussion \cite{Abdalla:2022yfr,CosmoVerseNetwork:2025alb}. They nevertheless illustrate that late-time inference can depend nontrivially on the assumed dark sector and on whether early-Universe calibration quantities, in particular the baryon-drag sound horizon \cite{Eisenstein:1997ik}, are modified, as in early-dark energy proposals \cite{Karwal:2016vyq,Poulin:2018cxd,Niedermann:2019olb,Kamionkowski:2022pkx}, or remain effectively fixed while the subsequent expansion changes, as in some late-time interacting-dark-sector scenarios \cite{DiValentino:2019jae,Gomez-Valent:2020mqn,Nunes:2022bhn,Escamilla:2023shf}. A late-time modification cannot be assessed from $H_0$ alone: it must also preserve the relevant distance integrals and remain compatible with BAO, supernova, CMB anisotropy, and growth information. This is especially important for sign-changing densities because a negative DE contribution at intermediate redshift can alter both the expansion rate and the accumulated distance while being subdominant at recombination. The present work does not perform that multi-probe test. Its purpose is instead to study the background-level field descriptions associated with selected late-time DE histories.

Negative or sign-changing DE densities have a long phenomenological history. Model-independent diagnostics based on the BOSS Ly$\alpha$ BAO measurement at $z\simeq2.34$ \cite{Delubac:2014aqe,BOSS:2014hhw} suggested that the DE density at that epoch lay below its present value and might even be negative \cite{Sahni:2014ooa}, while subsequent reconstructions and parametric analyses have repeatedly found that a negative cosmological constant, or a DE density that changes sign, is compatible with---and in some analyses mildly favored by---combinations of low- and high-redshift data \cite{Wang:2018fng,Dutta:2018vmq,Visinelli:2019qqu,Calderon:2020hoc,Sen:2021wld,Adil:2023ara,Malekjani:2024xjr,Escamilla:2023oce,Sabogal:2024qxs,Akarsu:2026anp}. A particularly economical route to such dynamics is provided by the graduated-DE (gDE) framework \cite{Akarsu:2019hmw}, which introduces a minimal dynamical departure from the vacuum-energy condition of vanishing inertial mass density, $\mathcal I_{\rm de}\equiv\rho_{\rm de}+p_{\rm de}=0$. Its simplest member, simple graduated dark energy (simple-gDE) \cite{Acquaviva:2021jov,Escamilla:2026eks}, promotes this condition to $\mathcal I_{\rm de}=\mathrm{const.}$: instead of taking the DE density itself to be constant, as for a cosmological constant, it takes the inertial mass density to be constant, with that constant allowed to be nonzero. This seemingly modest step already has nontrivial consequences. Energy conservation gives $\rho_{\rm de}(a)=\rho_{{\rm de},0}-3\mathcal I_{\rm de}\ln a$, so that $\mathcal I_{\rm de}<0$ produces a gradual evolution from negative density in the past to positive density at late times and leads asymptotically to the little sibling of the big rip (LSBR) \cite{Bouhmadi-Lopez:2014cca,Albarran:2015cda,Bouali:2019whr}. This example also illustrates the value of describing sign-changing DE through the ratio-free variables $\rho_{\rm de}$, $p_{\rm de}$, and $\mathcal I_{\rm de}$, rather than through $w_{\rm de}=p_{\rm de}/\rho_{\rm de}$ alone, which is generally ill-defined when $\rho_{\rm de}$ crosses zero. More general choices within gDE can make the sign transition progressively sharper; its rapid-transition regime motivated the $\Lambda_{\rm s}$CDM proposal, in which a cosmological constant switches from negative to positive at $z_\dagger\sim2$. The model was subsequently developed as a candidate for alleviating the $H_0$ tension together with other cosmological tensions \cite{Akarsu:2021fol,Akarsu:2022typ,Akarsu:2023mfb,Yadav:2024duq,Toda:2024ncp,Khandelwal:2026btl}, with its consequences for bound structures investigated in Ref.~\cite{Paraskevas:2024ytz}. The wider phenomenological landscape includes the omnipotent-DE parameterization \cite{Adil:2023exv}, oscillatory late-time features \cite{Akarsu:2022lhx}, and the smooth sign-switching families of Refs.~\cite{Bouhmadi-Lopez:2025ggl,Bouhmadi-Lopez:2025spo,Ibarra-Uriondo:2026zbp}. On the theory side, AdS-to-dS transitions have been realized through Casimir forces in the dark-dimension scenario \cite{Anchordoqui:2022svl,Anchordoqui:2023woo}, within type-II minimally modified gravity \cite{Akarsu:2024qsi}, and through homogeneous phantom-field dynamics \cite{Akarsu:2025gwi,Akarsu:2026lva}.

AdS vacua and negative vacuum energies are familiar in string and supergravity constructions \cite{Maldacena:1997re,Kachru:2003aw}. Controlled dS constructions remain debated, with no-go results, conjectural obstructions, and explicit proposals all represented in the literature \cite{Maldacena:2000mw,Obied:2018sgi,Kachru:2003aw}. This observation motivates studying a negative vacuum-energy-like DE regime, but it does not, by itself, derive or select a late-time sign-switching cosmology. Cosmologically, such a component is negligible deep in the radiation era and becomes relevant only when its magnitude is no longer small compared with the matter density. A transition near that epoch can then leave measurable signatures in the late expansion and growth histories. The aim here is not to propose a microscopic vacuum transition but to determine what an effective homogeneous scalar representation of prescribed sign-switching histories does and does not require.

Scalar fields provide a standard language for dynamical DE \cite{Copeland:2006wr,Tsujikawa:2013fta}, including phantom fields with $w<-1$ \cite{Caldwell:1999ew}. The inverse program---inferring a field trajectory and on-shell potential from a prescribed expansion or density history---has a long history \cite{Starobinsky:1998fr,Huterer:1998qv,Saini:1999ba,Sahni:2006pa,Boiza:2024azh,Adil:2026kfn}. For a separately conserved differentiable DE sector,
\begin{equation*}
\mathcal I_{\rm de}\equiv\rho_{\rm de}+p_{\rm de}=\frac{1+z}{3}\frac{{\rm d}\rho_{\rm de}}{{\rm d}z}
\end{equation*}
remains meaningful when $\rho_{\rm de}=0$, whereas $w_{\rm de}=p_{\rm de}/\rho_{\rm de}$ develops a ratio pole
\cite{Bouhmadi-Lopez:2007xco,Ozulker:2022slu,Akarsu:2025gwi,Akarsu:2026anp,Gokcen:2026pkq,Akarsu:2026pia}.
For a real minimally coupled scalar with a fixed kinetic sign $\xi=\pm1$, $\mathcal I_\phi=\xi\dot\phi^2$; hence, a regular single-field branch requires $\xi\,{\rm d}\rho_{\rm de}/{\rm d}z\geq0$. A zero of the density need not be a physical singularity, but a reversal of this derivative cannot be realized by one real scalar without changing the kinetic sector or adding degrees of freedom. This is the density-language form of the well-known result that a single minimally coupled scalar with a fixed kinetic sign cannot cross the phantom divide \cite{Vikman:2004dc,Hu:2004kh,Caldwell:2005ai,Kunz:2006wc,Nesseris:2006er}, an obstruction that is circumvented in two-field quintom constructions \cite{Feng:2004ad,Cai:2009zp} or in scalar--tensor and other modified-gravity settings \cite{Perivolaropoulos:2005yv}. Conversely, an effective phantom reconstruction does not establish a fundamental phantom field \cite{Mishra:2026tzn}. The two-derivative $\xi=-1$ action used below has a negative kinetic eigenvalue and is a genuine quantum ghost if interpreted fundamentally \cite{Carroll:2003st,Cline:2003gs}. Stable violation of the null energy condition requires kinetic structures beyond this action, as in ghost condensation and related effective theories \cite{Arkani-Hamed:2003pdi,Creminelli:2006xe,Rubakov:2014jja}, which we do not invoke. We therefore use the $\xi=-1$ action only as a formal homogeneous proxy for the prescribed background; no perturbative stability or ghost-free ultraviolet completion is claimed.

This distinction also limits what inverse reconstruction can determine. Eliminating redshift fixes $V(\phi)$ only along the field interval traversed by the chosen homogeneous solution; it does not determine a unique global or off-shell action. The integration leaves an arbitrary field translation and a physically equivalent reflection about an arbitrary field-space center $\phi_0$, $\phi\mapsto2\phi_0-\phi$, accompanied by $V(\phi)\mapsto V(2\phi_0-\phi)$. These redundancies must be fixed consistently before comparing parametric functions in field space. More generally, agreement with a reconstructed on-shell curve is necessary for that background branch but does not by itself prove that the fitted potential generates the complete target history when evolved from appropriate initial data.

This deliberately restricted inverse problem is also motivated by the broader underdetermination of scalar-field DE: present observations constrain only a limited set of effective combinations, while conclusions about dynamical single-field models remain sensitive to the adopted datasets, theory space, and priors \cite{Garcia-Garcia:2026nzy}. Our objective is therefore not to infer unique microphysics from a background curve, but to identify the necessary existence, invertibility, and regularity conditions before such an interpretation is attempted.

The inverse framework used here builds directly on Ref.~\cite{Adil:2026kfn}, which mapped prescribed DE density histories to $p_{\rm de}(z)$, the signed kinetic contribution $K(z)$, $\phi(z)$, and the on-shell curve $V(\phi)$, formulated the fixed-sign condition, and introduced a conditional potential-space comparison. That work considered the Chevallier--Polarski--Linder (CPL) form, a smooth sign-switching $\tanh$ history, and an emergent profile. Separately, Refs.~\cite{Akarsu:2025gwi,Akarsu:2026lva} studied forward homogeneous phantom dynamics for a bounded shifted-$\tanh$ potential. These two strands provide the methodological basis for the reconstruction and dynamical analyses developed below.

Our first step is to apply the inverse map specifically to the error-function ECDM, compact smooth-step SSCDM, and exact ladder-like L$\Lambda$CDM histories developed in Refs.~\cite{Bouhmadi-Lopez:2025ggl,Bouhmadi-Lopez:2025spo,Bouhmadi-Lopez:2026ckz}. We study representative profile families motivated by the phenomenological analyses \cite{Ibarra-Uriondo:2026zbp,Bouhmadi-Lopez:2026vyc}; the precise benchmark values are specified with the model definitions below, and no new data likelihood is evaluated. Our principal result is a regularity hierarchy hidden at the fluid level: ECDM admits an ordinary smooth on-shell reconstruction on every finite-redshift interval considered here; exact compact SSCDM gives a regular trajectory but a $C^1$, non-$C^2$ endpoint potential with the local scaling $V-V_e\propto|\phi-\phi_e|^{4/3}$ and hence a non-Lipschitz force. This is precisely what permits a nonunique delayed departure from an exactly frozen plateau. The exact Heaviside Ladder, by contrast, would require the squared field velocity to reproduce Dirac measures and therefore has no ordinary classical realization within the minimally coupled fixed-sign one-field action. This narrow obstruction does not rule out the Ladder as a phenomenological fluid, a smoothed transition, additional or noncanonical fields, interactions, or modified gravity. Under a mollifier of width $\epsilon$, the field excursion collapses as $\mathcal O(\sqrt{\epsilon})$ while the kinetic and potential peaks grow as $\mathcal O(\epsilon^{-1})$; its finite plotted spikes and field-space curve are consequently regulator-dependent fluid diagnostics.

In a second step, we quantify how efficiently five closed-form families represent the reconstructed curves in field space, for the ECDM target and for the retained transition-side interval of the $\Delta x=0.4$ SSCDM reconstruction: a generalized axion-like potential, a shifted-$\tanh$ form, a Gaussian feature, a regularized inverse-quadratic profile, and the sigmoid--Gaussian feature ansatz introduced here. The comparison is a controlled function-approximation experiment on synthetic potential-space data; its construction, and the precise sense in which its evidence scores are to be read, are specified in Sec.~\ref{sec:method}.

In a third step, we integrate representative regular members of the sigmoid--Gaussian and generalized axion-like families through the homogeneous Klein--Gordon--Friedmann equations, on a recombination-to-present background with an explicit massive neutrino. These integrations exhibit complete negative-to-positive crossings of the scalar energy density and distinguish the zero of the potential, $z_{\rm t}$, from the zero of the total scalar density, $z_\dagger$: the non-positive signed kinetic contribution separates the two events. The evolved parameter combinations are representative smooth members of the two families rather than the posterior-summary vectors of the potential-space comparison; in particular, the axion evolutions use the regular cosine case $n=1$. A closing subsection then evolves selected fitted representations from explicitly stated reconstructed or seeded initial data and quantifies how closely their unretuned forward evolutions reproduce the prescribed histories.

Relative to Refs.~\cite{Adil:2026kfn,Akarsu:2025gwi}, the new contributions are therefore the regularity classification of the ECDM--SSCDM--L$\Lambda$CDM trio, including the exact Ladder obstruction, compact-SSCDM endpoint nonanalyticity, and the unified plateau-endpoint criterion; conditional potential-space comparisons for the ECDM target and the retained SSCDM field interval, including the sigmoid--Gaussian feature family; and representative qualitative dynamical illustrations beyond the pure shifted-$\tanh$ case, together with a dynamical closure test connecting the three layers.

Three statements of scope apply throughout. First, the potential-space comparison is a function-approximation experiment on synthetic data drawn from the reconstructed curves: its evidence differences are conditional on the adopted field coordinate, sampling, noise prescription, and priors, and they are not observational Bayes factors, posterior odds between cosmologies, or constraints on cosmological parameters. Second, in the forward integrations $H_0$ is an input rather than an inferred quantity; no CMB, BAO, supernova, growth, or local-distance likelihood is evaluated anywhere in this paper, and no relief of the $H_0$ tension is claimed. Third, for $\xi=-1$ the two-derivative action is a quantum ghost, and every regularity statement below concerns the homogeneous background of this effective proxy; none amounts to perturbative stability or an ultraviolet completion. The reconstruction, the shape comparison, and the dynamical examples are three logically distinct layers of the analysis; they are kept separate throughout and are connected only by the controlled closure test of Sec.~\ref{subsec:closure}.

The paper is organized as follows. In Sec.~\ref{sec:background} we review the background cosmological equations, and in Sec.~\ref{sec:scalarfield} we develop the homogeneous scalar reconstruction and its fixed-sign consistency condition. In Sec.~\ref{sec:method} we define the three phenomenological histories, analyze their regularity, introduce the closed-form potential families, and specify the conditional potential-space comparison. The background reconstructions and the two within-target rankings are presented in Sec.~\ref{sec:results}. Section~\ref{sec:dynamics} studies representative Klein--Gordon--Friedmann evolutions for smooth members of the sigmoid--Gaussian and axion-like families, and closes with a dynamical test of the fitted potentials. We conclude in Sec.~\ref{sec:conclude}.

%%-----------------------------------------------------------------------------%%
\section{Background}
\label{sec:background}
%%-----------------------------------------------------------------------------%%

In the framework of general relativity, we consider a spatially flat FLRW spacetime, whose constant-cosmic-time hypersurfaces are homogeneous and isotropic. In comoving coordinates $(t,\mathbf{x})$, the corresponding line element is
\begin{equation}
    {\rm d}s^2=-{\rm d}t^2+a^2(t){\rm d}\mathbf{x}^2,
\end{equation}
where $a(t)$ is the scale factor and $t$ is cosmic time. Throughout this work, we normalize the scale factor at the present epoch by setting $a_0\equiv a(t_0)=1$, so that $1+z=a^{-1}$. For a spatially flat FLRW spacetime, Einstein's field equations reduce to the Friedmann equations

\begin{equation}
    3M_{\rm Pl}^2H^2=\sum_A \rho_A, \quad -2M_{\rm Pl}^2\dot H=\sum_A(\rho_A+p_A),
    \label{eq:friedmann}
\end{equation}
where $H\equiv\dot a/a$ is the Hubble parameter, an overdot denotes differentiation with respect to cosmic time, and $M_{\rm Pl}=(8\pi G)^{-1/2}$ is the reduced Planck mass. The sum runs over $A\in\{{\rm r,m,de}\}$, denoting radiation, non-relativistic matter, and DE, respectively. The allocation of a massive-neutrino component between the first two sectors must be specified consistently when the equations are integrated to recombination; this bookkeeping is made explicit in Sec.~\ref{sec:dynamics}. We adopt natural units with $c=\hbar=1$, while keeping $M_{\rm Pl}$ explicit. For the scalar-field reconstruction below, we use the dimensionless field $\widetilde\phi\equiv\phi/M_{\rm Pl}$.

Assuming that each component is independently conserved at the background level, $\nabla_\mu T^{\mu\nu}_{A} = 0$, one obtains the corresponding continuity equation
\begin{equation}
\dot{\rho}_A + 3H\left(\rho_A + p_A\right) = 0 \, .
\label{eq:continuity}
\end{equation}

The equation-of-state (EoS) parameter of each component is
\begin{equation}
    w_{\mathrm{r}}=\frac{p_{\mathrm{r}}}{\rho_{\mathrm{r}}}=\frac{1}{3}, \quad w_{\mathrm{m}}=\frac{p_{\mathrm{m}}}{\rho_{\mathrm{m}}}=0, \quad w_{\mathrm{de}}=\frac{p_{\mathrm{de}}}{\rho_{\mathrm{de}}}.
    \label{eq:eos_terms}
\end{equation}

On an interval on which $\rho_{\rm de}\neq0$ and $w_{\rm de}$ is locally integrable, an EoS parameterization of Eq.~\eqref{eq:continuity} gives
\begin{equation}
    \rho_{\mathrm{de}}(z)=\rho_{\mathrm{de}0}\exp\left[3\int_0^z\frac{1+w_{\mathrm{de}}(\tilde z)}{1+\tilde z}\,{\rm d}\tilde z\right].
\end{equation}
Starting from a nonzero density, this exponential representation cannot pass through zero. At a regular sign change, $w_{\rm de}$ becomes singular, and the density-based form of the continuity equation must instead be used directly. Matter and radiation evolve as $\rho_{\rm m}=\rho_{\rm m0}(1+z)^3$ and $\rho_{\rm r}=\rho_{\rm r0}(1+z)^4$. We introduce the present-day critical density $\rho_{\rm c0}\equiv3H_0^2M_{\rm Pl}^2$, the present-day density parameters $\Omega_{A0}\equiv\rho_{A0}/\rho_{\rm c0}$, and the densities normalized to $\rho_{\rm c0}$, $\widetilde\Omega_A(z)\equiv\rho_A(z)/\rho_{\rm c0}$; thus $\widetilde\Omega_A(0)=\Omega_{A0}$. Defining $E(z)\equiv H(z)/H_0$, the first Friedmann equation becomes

\begin{equation}
\begin{aligned}
    E^2(z)= \Omega_{\mathrm{r}0}(1+z)^4+\Omega_{\mathrm{m}0}(1+z)^3+\widetilde{\Omega}_{\mathrm{de}}(z).
    \end{aligned}
    \label{eq:friedmann_dimless}
\end{equation}
Spatial flatness implies
\begin{equation}
    1=\Omega_{\rm m0}+\Omega_{\rm r0}+\Omega_{\rm de0}.
\end{equation}

For the low-redshift background reconstruction arrays, we use the explicit photon-plus-massless-neutrino prescription below \cite{Planck:2018vyg}:
\begin{equation}
\begin{aligned}
 \omega_\gamma
 ={}&2.47297928\times10^{-5}
 \left(\frac{T_{\rm CMB}}{2.7255\,{\rm K}}\right)^4,\\
 \omega_{\rm r}\equiv\Omega_{\rm r0}h^2
 ={}&\omega_\gamma\left[1+\frac{7}{8}
 \left(\frac{4}{11}\right)^{4/3}N_{\rm eff}\right].
\end{aligned}
 \label{eq:radiation_background}
\end{equation}
Here $h\equiv H_0/(100\,{\rm km}\,{\rm s}^{-1}\,{\rm Mpc}^{-1})$, $T_{\rm CMB}=2.7255\,{\rm K}$ is the measured CMB monopole temperature \cite{Fixsen:2009ug}, and we adopt $N_{\rm eff}=3.046$ in Eq.~\eqref{eq:radiation_background}. The recombination-to-present formulation in Sec.~\ref{sec:dynamics} instead separates the single massive-neutrino species and follows its relativistic-to-nonrelativistic transition, avoiding double counting between the matter and radiation sectors. The photon normalization is common to the two numerical layers; only their treatment of the neutrino sector differs. Section~\ref{sec:dynamics} reports an end-to-end comparison with CLASS using identical photon, massless-neutrino, massive-neutrino, and matter inputs. For the low-redshift background reconstructions, Eq.~\eqref{eq:friedmann_dimless} then determines the expansion once $\widetilde\Omega_{\rm de}(z)$ is specified. In the following section, we map this phenomenological input to an effective homogeneous scalar-field description.

%%-----------------------------------------------------------------------------%%
\section{Effective scalar-field reconstruction}
\label{sec:scalarfield}
%%-----------------------------------------------------------------------------%%

\noindent In this section, we translate a phenomenological DE history specified by $\rho_{\rm de}(z)$ into an effective scalar-field description in a spatially flat FLRW spacetime. The correspondence is an on-shell mapping of the homogeneous background: by itself, it neither supplies a fundamental completion nor establishes perturbative or quantum stability.

%%-----------------------------------------------------------------------------%%
\subsection{Minimally coupled scalar field and Klein--Gordon equation}
%%-----------------------------------------------------------------------------%%

For definiteness, consider general relativity with separately conserved matter and radiation, and a minimally coupled scalar field,
\begin{equation}
\begin{aligned}
 \mathcal{S}={}&\int {\rm d}^4x\sqrt{-g}
 \left[\frac{M_{\rm Pl}^2}{2}R
 -\frac{\xi}{2}g^{\mu\nu}\partial_\mu\phi\partial_\nu\phi
 -V(\phi)\right] \\
 &+\mathcal{S}_{\rm m}+\mathcal{S}_{\rm r},
\end{aligned}
\label{eq:action}
\end{equation}
where $\xi=+1$ describes a canonical field \cite{Ratra:1987rm} and $\xi=-1$ a phantom field \cite{Caldwell:1999ew}. We define $X\equiv-\frac{1}{2}g^{\mu\nu}\partial_\mu\phi\,\partial_\nu\phi$, which reduces to $X=\dot\phi^2/2$ for a homogeneous configuration $\phi=\phi(t)$.

The corresponding energy density and pressure are
\begin{equation}
    \begin{aligned}
& \rho_{\phi}=\xi X+V(\phi)=\frac{\xi}{2} \dot{\phi}^2+V(\phi), \\
& p_{\phi}=\xi X-V(\phi)=\frac{\xi}{2} \dot{\phi}^2-V(\phi).
\end{aligned}\label{eq:rho_p}
\end{equation}

Equations~\eqref{eq:rho_p} give
\begin{equation}
    \begin{aligned}
& \rho_{\phi}+p_{\phi}=2\xi X=\xi \dot{\phi}^2\quad \textnormal{and}\quad \rho_{\phi}-p_{\phi}=2V(\phi).
\end{aligned}\label{eq:rhoplusp}
\end{equation}
The first combination determines the null energy condition (NEC): a real canonical field satisfies $\rho_\phi+p_\phi\geq0$, whereas a real phantom field satisfies $\rho_\phi+p_\phi\leq0$. For $\xi=-1$, the propagating scalar has a negative kinetic eigenvalue and is therefore a genuine ghost if Eq.~(\ref{eq:action}) is treated as a fundamental quantum field theory. The fact that the two-derivative model has unit rest-frame sound speed does not remove this ghost, and introducing a cutoff alone does not cure the negative-energy degree of freedom \cite{Carroll:2003st,Cline:2003gs}. Accordingly, throughout this work, the phantom field is used only as a formal homogeneous effective proxy for the prescribed background history; no ghost-free ultraviolet completion is claimed.

Varying Eq.~(\ref{eq:action}) with respect to the homogeneous field gives the Klein--Gordon (KG) equation
\begin{equation}
    \ddot{\phi}+3 H \dot{\phi} +{\xi} V_{, \phi}=0,
    \label{eq:KG}
\end{equation}
where $V_{,\phi}\equiv{\rm d}V/{\rm d}\phi$. Using ${\rm d}/{\rm d}t=-(1+z)H\,{\rm d}/{\rm d}z$, with a prime denoting ${\rm d}/{\rm d}z$, Eq.~(\ref{eq:KG}) becomes
\begin{equation}
    \phi''+\left(\frac{H'}{H}-\frac{2}{1+z}\right)\phi' +\frac{\xi}{H^2(1+z)^2} V_{, \phi}=0.
    \label{eq:KG_mod}
\end{equation}

%%-----------------------------------------------------------------------------%%
\subsection{Perfect-fluid--scalar-field mapping}
\label{subsec:mapping}
%%-----------------------------------------------------------------------------%%

At the background level, the DE component is described as a perfect fluid with density $\rho_{\rm de}(z)$ and pressure $p_{\rm de}(z)$. Rewriting its continuity equation in terms of redshift gives
\begin{equation}
    \rho_{\rm de}+p_{\rm de}=\frac{1+z}{3}\frac{\textrm{d}\rho_{\rm de}}{\textrm{d}z}.
    \label{eq:16}
\end{equation}
The combination $\rho_{\rm de}+p_{\rm de}$ is the enthalpy, or inertial mass density, and its sign determines whether the DE sector satisfies or violates the NEC.

Owing to the nature of the models under study, we must allow for both positive and negative energy densities. As a consequence, the phantom-divide line $w_{\rm de}=-1$ does not act as a universal boundary between canonical and phantom regimes. Although we occasionally display the quantity $w_{\rm de}=p_{\rm de}/\rho_{\rm de}$ for illustrative purposes, this parameter becomes ill defined at the zero-crossing $\rho_{\rm de}=0$, where it diverges for purely kinematic reasons. Therefore, while the combination $\rho_{\rm de}(1+w_{\rm de})$ may serve as a useful indicator when $w_{\rm de}$ is well defined, the fundamental physical quantity is the inertial mass density
\begin{equation}
\mathcal{I}_{\rm de} \equiv \rho_{\rm de} + p_{\rm de},
\label{eq:imd}
\end{equation}
whose vanishing defines the null energy condition boundary (NECB). $\mathcal{I}_{\rm de}$ remains regular at $\rho_{\rm de}=0$, and the locus $\mathcal{I}_{\rm de}=0$ therefore replaces the phantom-divide line as the branch-independent separator between quintessence-like and phantom-like behavior. Four distinct regimes can then be identified according to the signs of $\rho_{\rm de}$ and $\mathcal{I}_{\rm de}$: the quintessence-like regime, $\mathcal{I}_{\rm de}>0$, comprising the n- and p-quintessence branches for $\rho_{\rm de}<0$ and $\rho_{\rm de}>0$, respectively, and the phantom-like regime, $\mathcal{I}_{\rm de}<0$, comprising the corresponding n- and p-phantom branches, following the approach of Refs.~\cite{Akarsu:2025dmj,Akarsu:2026anp,Gokcen:2026pkq,Akarsu:2026pia}. The complementary combination 
\begin{equation}
\mathcal{M}_{\rm de}\equiv\rho_{\rm de}+3p_{\rm de},
\end{equation}
is the active gravitational mass density. It is likewise regular across the crossing and governs the DE sector's contribution to Raychaudhuri focusing. Thus, $\mathcal{M}_{\rm de}<0$ marks a gravitationally repulsive DE sector; the familiar criterion $w_{\rm de}<-1/3$ is equivalent only for $\rho_{\rm de}>0$ and reverses for $\rho_{\rm de}<0$ \cite{Akarsu:2026pia}.

\subsection{Reconstructing the field and single-field consistency}
\label{subsec:rec_phi}

To reconstruct the field, we first introduce the signed kinetic contribution
\begin{equation}
    K\equiv\frac{\xi}{2}\dot\phi^2.
\end{equation}
Identifying $\rho_\phi\leftrightarrow\rho_{\rm de}$ and $p_\phi\leftrightarrow p_{\rm de}$, Eqs.~(\ref{eq:rhoplusp}) and (\ref{eq:16}) imply
\begin{equation}
    \xi \dot{\phi}^2=\frac{1+z}{3}\frac{\textrm{d}\rho_{\rm de}}{\textrm{d}z}.
\end{equation}
In terms of the density normalized to $\rho_{\rm c0}$,
\begin{equation}
\label{eq:fracenergy}
    \xi \left(\frac{\dot{\phi}}{M_{\rm Pl}}\right)^2=(1+z)H_0^2\frac{\textrm{d}\widetilde\Omega_{\rm de}}{\textrm{d}z}.
\end{equation}
or, equivalently,
\begin{equation}
     \left(\frac{\phi^\prime}{M_{\rm Pl}}\right)^2=\frac{1}{\xi}\frac{1}{(1+z) E^2}\frac{\textrm{d}\widetilde\Omega_{\rm de}}{\textrm{d}z}.
     \label{eq:phi_dev}
\end{equation}
For a real field on an expanding redshift interval with $E^2>0$ and $z>-1$, the right-hand side must be non-negative. Since $\xi=\pm1$, a necessary fixed-sign single-field consistency condition is therefore
\begin{equation}
    \xi\,\frac{{\rm d}\widetilde\Omega_{\rm de}}{{\rm d}z}\geq0.
    \label{eq:single_field_condition}
\end{equation}
This inequality must hold throughout the reconstructed interval. If ${\rm d}\widetilde\Omega_{\rm de}/{\rm d}z$ changes sign, no single minimally coupled real scalar with fixed $\xi$ can reproduce the entire history. On a regular interval satisfying Eq.~(\ref{eq:single_field_condition}), define
\begin{equation}
\begin{aligned}
 \mathcal Q(z)&\equiv
 \left[\frac{1}{\xi(1+z)E^2(z)}
 \frac{{\rm d}\widetilde\Omega_{\rm de}}{{\rm d}z}\right]^{1/2},\\
 \frac{\phi_s(z)}{M_{\rm Pl}}
 &=\widetilde\phi_{\rm ref}
 +s\int_{z_{\rm ref}}^z\mathcal Q(\tilde z)\,{\rm d}\tilde z,
 \qquad s=\pm1.
\end{aligned}
\label{eq:phi_branches}
\end{equation}
The two branches are thereby written with a common field-speed magnitude. The integration constant $\widetilde\phi_{\rm ref}$ fixes the origin of the dimensionless field $\widetilde\phi=\phi/M_{\rm Pl}$. The two signs are physically equivalent under a field reflection accompanied by the corresponding reflected potential, $\widetilde\phi\to2\widetilde\phi_{\rm ref}-\widetilde\phi$ and $\widetilde V(\widetilde\phi)\to \widetilde V(2\widetilde\phi_{\rm ref}-\widetilde\phi)$; the KG equation does not select one of them. A definite origin and orientation must nevertheless be fixed before comparing parametric functions in potential space, because parameter priors need not be invariant under these coordinate choices.

Indeed, differentiating $\rho_\phi$ and using the continuity equation gives $\xi\dot\phi\,[\ddot\phi+3H\dot\phi+\xi V_{,\phi}]=0$. Thus, the KG equation follows wherever $\dot\phi\neq0$, and the smooth on-shell map is dynamically consistent along either reflected branch. On an open interval on which the field is exactly frozen, energy conservation alone reduces to an identity, and a scalar completion must additionally satisfy $V_{,\phi}(\phi_e)=0$. At each SSCDM endpoint reconstructed below, the potential is once continuously differentiable (class $C^1$) and has a vanishing limiting force. It therefore satisfies this condition on the plateau trajectory; its non-Lipschitz force nevertheless makes the corresponding initial-value problem nonunique. By contrast, constant nonzero $\mathcal I_{\rm de}$ implies $K=\mathcal I_{\rm de}/2\neq0$, so the field has no exactly frozen interval and the compact-plateau endpoint issue does not arise.

%%-----------------------------------------------------------------------------%%
\subsection{Reconstructing the potential, pressure, and kinetic term}
%%-----------------------------------------------------------------------------%%

The pressure associated with a DE density profile follows from Eq.~\eqref{eq:continuity}:
\begin{equation}
    p_{\rm de}(z)=-\rho_{\rm de}(z)+\frac{1+z}{3}\frac{\textrm{d}\rho_{\rm de}(z)}{\textrm{d}z}.
    \label{eq:pressure}
\end{equation}

The signed kinetic and potential contributions then follow from Eq.~\eqref{eq:rhoplusp}:
\begin{equation}
    \begin{aligned}
        K(z)&=\frac{1}{2}\left[\rho_{\rm de}(z)+p_{\rm de}(z)\right]=\frac{1+z}{6}\frac{\textrm{d}\rho_{\rm de}(z)}{\textrm{d}z},  \\
        V(z)&=\frac{1}{2}\left[\rho_{\rm de}(z)-p_{\rm de}(z)\right]=\rho_{\rm de}(z)-\frac{1+z}{6}\frac{\textrm{d}\rho_{\rm de}(z)}{\textrm{d}z}.
    \end{aligned}
    \label{eq:V_K_delta}
\end{equation}

These relations remain well defined when the DE density changes sign smoothly: the ratio $w_{\rm de}=p_{\rm de}/\rho_{\rm de}$ then diverges, but $p_{\rm de}$, $K$, and $V$ can remain finite. Where $\phi(z)$ is monotonic, eliminating $z$ between Eqs.~(\ref{eq:phi_dev}) and (\ref{eq:V_K_delta}) yields the reconstructed on-shell potential $V(\phi)$ along the background trajectory. This construction does not, by itself, establish a unique off-shell theory away from the field interval sampled by that trajectory.

%%-----------------------------------------------------------------------------%%
\subsection{Canonical versus phantom reconstruction}
\label{sec:quin_vs_phant}
%%-----------------------------------------------------------------------------%%
We now compare the two fixed-sign possibilities.

\textit{Canonical field ($\xi = +1$) --} For a canonical scalar field, Eqs.~(\ref{eq:continuity}) and (\ref{eq:rhoplusp}) imply $\dot{\rho}_\phi = -3H\dot{\phi}^2$. For an expanding Universe with $H > 0$, this ensures that the DE density decreases monotonically with time. Consequently, if $\rho_\phi$ is negative at earlier epochs, it cannot evolve toward and cross $\rho_\phi = 0$. The same conclusion follows from the continuity equation $\dot{\rho}_\phi + 3H\rho_\phi(1 + w_\phi) = 0$. In the $n$-quintessence regime, where $\rho_\phi < 0$ and $w_\phi < -1$, one again finds $\dot{\rho}_\phi < 0$, preventing any transition to a $p$-quintessence phase. It follows that such a scenario cannot be realized by a canonical scalar field.

\textit{Phantom field ($\xi=-1$) --} For a phantom field, Eqs.~(\ref{eq:continuity}) and (\ref{eq:rhoplusp}) imply $\dot\rho_\phi=3H\dot\phi^2$. Thus, in an expanding Universe with $H>0$, the energy density increases monotonically and can cross from negative to positive values. Both $p_\phi$ and $\rho_\phi$ may remain regular; the divergence of $w_\phi$ arises solely because the denominator $\rho_\phi$ passes through zero, which occurs at $V=X$. At a transverse crossing with finite $X>0$, as in the continuous targets studied below, $p_\phi=-2X=-2V<0$ remains finite and $\dot\rho_\phi=-3Hp_\phi=6HX>0$. To display the limiting behavior, write $V=X\pm\delta$, with $\delta\geq0$. For $V=X-\delta$, one has $\rho_\phi=-\delta$ and $w_\phi=-1+2X/\delta\to+\infty$ as $\delta\to0^+$. For $V=X+\delta$, one has $\rho_\phi=\delta$ and $w_\phi=-1-2X/\delta\to-\infty$. If instead $X=0$ at a higher-order density zero, these strict limiting statements need not follow. Away from the crossing, when $X\ll|V|$, the EoS approaches the cosmological-constant value. The pole in $w_\phi$ is therefore a ratio singularity rather than a singularity of the homogeneous density or pressure \cite{Dabrowski:2009kg,Bouhmadi-Lopez:2009ggt,Bouhmadi-Lopez:2013tua,Bouhmadi-Lopez:2019zvz,Ozulker:2022slu}. This regular background behavior should not, however, be confused with quantum stability: the negative kinetic term remains ghost-like as discussed above.

%%-----------------------------------------------------------------------------%%
\section{Methodology}\label{sec:method}
%%-----------------------------------------------------------------------------%%

Following the methodology developed in Ref.~\cite{Adil:2026kfn}, we apply the background reconstruction of Sec.~\ref{sec:scalarfield} to three prescribed DE density histories, $\rho_{\rm de}(z)$, or equivalently $\widetilde\Omega_{\rm de}(z)\equiv\rho_{\rm de}(z)/\rho_{\rm c0}$. Their differing degrees of regularity allow us to compare a smooth history, a compact smooth-step history, and a distributional ladder. For ECDM and the retained transition-side interval of SSCDM, we subsequently compare the reconstructed on-trajectory potentials with several closed-form ans\"atze in field space. These are conditional potential-shape comparisons based on synthetic points, rather than likelihood analyses of cosmological observations.

%%-----------------------------------------------------------------------------%%
\subsection{Phenomenological sign-switching DE histories}\label{subsec:dark_energy_profile}
%%-----------------------------------------------------------------------------%%

We consider three phenomenological histories for the normalized DE density, $\widetilde\Omega_{\rm de}(z)\equiv\rho_{\rm de}(z)/\rho_{\rm c0}$, with $\widetilde\Omega_{\rm de}(0)=\Omega_{\rm de0}$. These histories are used as inputs to the reconstruction developed in Sec.~\ref{sec:scalarfield}. Two of them, SSCDM and ECDM, interpolate continuously between negative and positive DE densities, whereas L$\Lambda$CDM is deliberately defined through a finite sequence of discontinuous steps. This distinction will prove important for the mathematical status of the corresponding scalar-field reconstruction. Throughout this discussion, ``AdS-like'' and ``dS-like'' refer only to the negative- and positive-vacuum-energy-like regimes of the DE sector. They do not imply that the complete matter--radiation--DE FLRW spacetime is an exact anti-de Sitter or de Sitter geometry.

\textit{1. L$\mathit{\Lambda}$CDM --} The first scenario is the L$\Lambda$CDM model, in which the DE density evolves through a ladder-like transition function introduced in Ref.~\cite{Bouhmadi-Lopez:2025ggl} and subsequently studied in Refs.~\cite{Bouhmadi-Lopez:2025spo,Ibarra-Uriondo:2026zbp}. It provides a discretized generalization of abrupt $\Lambda_{\rm s}$CDM \cite{Akarsu:2019hmw,Akarsu:2021fol,Akarsu:2022typ,Akarsu:2023mfb}: rather than undergoing a single sign reversal, the DE density changes by equal increments at a finite set of redshifts. Throughout this work we take an even number of steps and fix $N=8$. The normalized DE density is
\begin{equation}
\widetilde\Omega_{\rm de}(z)=\Omega_{\rm de0}\left[1-\frac{2}{N}\sum_{n=1}^{N}\mathcal H(z-z_n)\right],
\label{densityladder}
\end{equation}
where $\mathcal H$ denotes the Heaviside function. Because an even-step ladder has a finite zero-density plateau rather than a unique zero-crossing event, we use $z_\dagger$ to denote the midpoint of that plateau. The midpoint-centered transition grid used in all numerical arrays and figures in this work is
\begin{equation}
z_n=z_\dagger+\left(n-\frac{N+1}{2}\right)\Delta z,\qquad n=1,\ldots,N,
\label{eq:ladder_grid}
\end{equation}
or, equivalently,
\begin{equation}
\begin{aligned}
z_{\rm lo}\equiv z_1&=z_\dagger-\frac{N-1}{2}\Delta z,\\
z_{\rm hi}\equiv z_N&=z_\dagger+\frac{N-1}{2}\Delta z,\\
z_n&=z_{\rm lo}+(n-1)\Delta z.
\end{aligned}
\label{eq:ladder_grid_endpoints}
\end{equation}
Here $\Delta z$ is the separation between adjacent transitions. We require $z_{\rm lo}>0$, so that all steps precede the present epoch and $\widetilde\Omega_{\rm de}(0)=\Omega_{\rm de0}$. The factor $N+1$ in Eq.~\eqref{eq:ladder_grid} follows from centering $N$ indexed points symmetrically about $z_\dagger$; it does not introduce an additional step. Since redshift decreases with cosmic time, the transition begins at $z_{\rm hi}$ and ends at $z_{\rm lo}$. An edge-anchored convention is related to Eqs.~\eqref{eq:ladder_grid}--\eqref{eq:ladder_grid_endpoints} by a half-step redefinition of the parameter called $z_\dagger$; keeping the same numerical $z_\dagger$ in the two conventions would instead shift every transition by $\Delta z/2$.

For the benchmark $(N,\Delta z,z_\dagger)=(8,0.15,1.8)$, the transition redshifts are
\begin{equation}
\begin{aligned}
\{z_n\}_{n=1}^{8}={}&\{1.275,\ 1.425,\ 1.575,\ 1.725,\\
&\phantom{\{}1.875,\ 2.025,\ 2.175,\ 2.325\}.
\end{aligned}
\label{eq:ladder_benchmark_grid}
\end{equation}
Thus, away from the jump points, the central zero-density plateau is the open interval $1.725<z<1.875$, whose midpoint is $z_\dagger=1.8$. More generally, for even $N$ it is $z_{N/2}<z<z_{N/2+1}$ and has width $\Delta z$. For odd $N$ there is no finite zero-density plateau; the central step changes the sign directly. We take $\mathcal H(0)=1/2$ to fix the isolated values at the jumps. This convention has no effect on integrated quantities or on the distributional derivative below. Read forward in cosmic time, Eq.~\eqref{densityladder} therefore evolves monotonically from $-\Omega_{\rm de0}$ at high redshift to $+\Omega_{\rm de0}$ at low redshift. In the strict Heaviside model, differentiation gives
\begin{equation}
\frac{{\rm d}\rho_{\rm de}}{{\rm d}z}=-\rho_{\rm c0}\Omega_{\rm de0}\frac{2}{N}\sum_{n=1}^{N}\delta(z-z_n).
\label{eq:ladder_delta}
\end{equation}
The identity holds in the sense of distributions and is independent of the chosen value of $\mathcal H(0)$. It also exposes an important limitation of the scalar mapping. On the phantom branch, $K=(1+z)\rho'_{\rm de}/6=-\dot\phi^2/2$ would require the non-negative kinetic measure $\dot\phi^2\,{\rm d}t$ to contain Dirac mass. For every ordinary scalar configuration $\phi\in W^{1,2}_{\rm loc}$, however, $|\dot\phi|^2\,{\rm d}t$ is absolutely continuous with respect to Lebesgue measure and cannot equal a singular Dirac measure. Promoting $\dot\phi$ itself to a Schwartz distribution does not help, because the stress tensor would then require an undefined nonlinear product, heuristically written as the square of $\sqrt{\delta}$ \cite{Schwartz:1954multiplication}. Formal use of Eq.~\eqref{eq:phi_dev} additionally requires multiplying a Dirac measure by $E^{-2}$, which is discontinuous at the same jump.

The exact L$\Lambda$CDM history therefore has no realization in the ordinary Sobolev configuration space of the minimally coupled one-field action \eqref{eq:action}. This is a deliberately narrow obstruction. It neither invalidates the Ladder as an effective fluid history nor excludes noncanonical or multiple fields, interacting sectors, modified gravity, a specified generalized-function algebra, or a family of finite-width regulated theories. Such constructions require additional prescriptions and do not furnish a regulator-independent ordinary potential $V(\phi)$ in the strict Heaviside limit.

Correspondingly, $p_{\rm de}$, $K$, and $V$ contain impulsive contributions at the steps. The algebraic Friedmann constraint makes $H^2$ discontinuous; on the expanding branch with $H>0$ and $E^2>0$, $H$ therefore jumps. The scale factor can remain continuous, but its extrinsic curvature is discontinuous, so $\dot H$ and the FLRW Ricci curvature acquire impulsive terms in the corresponding weak or mollified description. The strict geometry should therefore be regarded only as an idealized weak limit; products involving the jumps and impulses require a specified regularization. To make the regulator dependence explicit, replace each delta function with a fixed-shape, non-negative, normalized mollifier $\delta_\epsilon(u)=\epsilon^{-1}g(u/\epsilon)$, with $\int g(u)\,{\rm d}u=1$ and $\epsilon\ll\Delta z$. Provided that the correspondingly smoothed background has finite positive $E^2$, $|\phi'|=\mathcal O(\epsilon^{-1/2})$ across a step and hence $\Delta\phi=\mathcal O(\sqrt{\epsilon})$, whereas the pointwise peaks in $K$ and $V$ scale as $\mathcal O(\epsilon^{-1})$. The limiting impulse weights in redshift space for $K$ and for the impulsive part of $V$ are fixed by the density jump and are independent of the normalized profile $g$. By contrast, their finite-$\epsilon$ shapes and peak heights, the coefficient of $\Delta\phi=\mathcal O(\sqrt{\epsilon})$, and the resulting field-space curve depend on the regulator. The L$\Lambda$CDM plots below should therefore be read as finite-resolution representations of a distributional fluid benchmark, not as the strict potential of a regular scalar field.

The discrete construction nevertheless has an interesting, albeit heuristic, connection with causal-set ideas. In ``everpresent $\Lambda$'' scenarios \cite{Ahmed:2002mj,Zwane:2017xbg,Das:2023hbw}, the number of causal-set elements is proportional to the spacetime four-volume, $N_{\rm cs}\propto\mathcal V$, and Poisson statistics give $\Delta N_{\rm cs}\sim\sqrt{N_{\rm cs}}$. Together with the conjugacy estimate $\Delta\Lambda\,\Delta\mathcal V\sim1$, this leads in Planck units to $|\Lambda|_{\rm rms}\sim\mathcal V^{-1/2}$. For a Hubble four-volume, $\mathcal V\sim H^{-4}$, one therefore obtains $|\Lambda|_{\rm rms}\sim H^2$, corresponding to an effective vacuum-energy density of the order of the critical density. The effective cosmological term may fluctuate in both magnitude and sign as the causal set grows. The staircase in Eq.~\eqref{densityladder} is suggestive of this discrete, stepwise picture and offers a useful contrast with the continuous SSCDM and ECDM histories.

The analogy should not be interpreted as a derivation or projection of causal-set dynamics. Everpresent-$\Lambda$ models are stochastic and generally undergo repeated, realization-dependent sign changes, whereas L$\Lambda$CDM is deterministic and monotonic, with uniformly spaced transitions and a single progression from negative to positive DE density in cosmic time. Thus, causal-set discreteness supplies a possible motivation for studying stepwise vacuum-energy histories, while Eq.~\eqref{densityladder} is a deterministic phenomenological foil inspired by that possibility. Its failure to admit a regular scalar representation is not a flaw in the fluid benchmark; rather, it sharply distinguishes an idealized discontinuous history from the continuous single-field constructions considered below.

\textit{2. SSCDM --} The second model is a two-parameter extension of $\Lambda$CDM in which the DE density is constant outside a finite transition interval and is joined inside that interval by a high-order smooth-step polynomial, introduced in Ref.~\cite{Bouhmadi-Lopez:2025ggl} and subsequently studied in Refs.~\cite{Bouhmadi-Lopez:2025spo,Ibarra-Uriondo:2026zbp}. In terms of $x=\ln a=-\ln(1+z)$, the normalized density is
\begin{equation}
\widetilde\Omega_{\rm de}(x)=\Omega_{\rm de0}
\begin{cases}
-1, & x\leq x_i,\\[4pt]
1-2f(t), & x_i<x<x_f,\\[4pt]
1, & x\geq x_f,
\end{cases}
\label{eq:density_sscdm}
\end{equation}
where
\begin{equation}
\begin{aligned}
f(t)&=126t^5-420t^6+540t^7-315t^8+70t^9,\\
t&=\frac{x_f-x}{x_f-x_i}.
\end{aligned}
\end{equation}
The polynomial obeys $f(0)=0$, $f(1)=1$, and $f(1-t)=1-f(t)$. Consequently, the zero of the DE density occurs at $t=1/2$, and the transition is symmetric in $x$, not in redshift. Defining $\Delta x\equiv x_f-x_i>0$ and $x_\dagger\equiv-\ln(1+z_\dagger)$, the correct boundaries are
\begin{equation}
x_i=x_\dagger-\frac{\Delta x}{2},\qquad x_f=x_\dagger+\frac{\Delta x}{2},
\label{eq:sscdm_x_boundaries}
\end{equation}
or, in terms of redshift,
\begin{equation}
1+z_i=(1+z_\dagger)e^{\Delta x/2},\qquad 1+z_f=(1+z_\dagger)e^{-\Delta x/2}.
\label{eq:sscdm_z_boundaries}
\end{equation}
Equivalently, $(1+z_\dagger)^2=(1+z_i)(1+z_f)$; in general, $z_\dagger$ is not the arithmetic midpoint of $z_i$ and $z_f$. The present-day normalization assumes $x_f\leq0$, or equivalently $\Delta x\leq2\ln(1+z_\dagger)$, so that the positive-density plateau has been reached by $z=0$. All SSCDM benchmarks used here satisfy this condition.

Since $f'(t)=630t^4(1-t)^4$, one has ${\rm d}\widetilde\Omega_{\rm de}/{\rm d}z<0$ inside the transition and equality on the two plateaus. SSCDM therefore satisfies the fixed-sign condition \eqref{eq:single_field_condition} for $\xi=-1$, with the effective field frozen wherever the DE density is constant.

Although Eq.~\eqref{eq:density_sscdm} is smooth enough for the background quantities used here, its exact compact support has a nontrivial implication for the reconstructed potential. The first four derivatives of the polynomial match the constant plateaus at each endpoint, so the piecewise density is $C^4$ in $x$. Let $\ell$ denote the distance from either endpoint measured inside the transition. Then
\begin{equation}
\widetilde\Omega_{\rm de}-\widetilde\Omega_e=\mathcal O(\ell^5),\qquad \frac{{\rm d}\widetilde\Omega_{\rm de}}{{\rm d}x}=\mathcal O(\ell^4).
\end{equation}
Using Eqs.~\eqref{eq:phi_dev} and \eqref{eq:V_K_delta} for $\xi=-1$ and assuming finite $E_e^2>0$ at the endpoint gives
\begin{equation}
\begin{aligned}
\widetilde\phi-\widetilde\phi_e&=\mathcal O(\ell^3),\\
\widetilde V-\widetilde V_e&=C_e|\widetilde\phi-\widetilde\phi_e|^{4/3}\\
&\quad+\mathcal O\!\left(|\widetilde\phi-\widetilde\phi_e|^{5/3}\right),
\end{aligned}
\label{eq:sscdm_endpoint_scaling}
\end{equation}
where $\widetilde V\equiv V/\rho_{\rm c0}$ and the finite coefficient $C_e>0$ depends on the endpoint. Thus, the exact reconstructed $V(\phi)$ is $C^1$ but not $C^2$ at the endpoints: $V_{,\phi}$ is continuous but non-Lipschitz, while $V_{,\phi\phi}$ diverges as $|\phi-\phi_e|^{-2/3}$. Locally, $V_{,\phi}\propto{\rm sgn}(\phi-\phi_e)|\phi-\phi_e|^{1/3}$.

The loss of uniqueness can be seen directly. At a departure endpoint, reflect the field if necessary and define $\psi\equiv\widetilde\phi-\widetilde\phi_e\geq0$. Using $\widetilde V-\widetilde V_e=C_e \psi^{4/3}+\cdots$, the phantom Klein--Gordon equation becomes locally
\begin{equation}
\ddot \psi+3H_e\dot \psi-4C_eH_0^2\psi^{1/3}=o(\psi^{1/3}),
\end{equation}
where $H_e$ is the finite Hubble rate at departure. Besides the solution $\psi=0$, any departure time $t_d$ within the regular expanding domain of the frozen background admits a local departing branch
\begin{equation}
\psi(t)=
\begin{cases}
0, & t\leq t_d,\\[3pt]
A_e(t-t_d)^3+\mathcal O\!\left((t-t_d)^4\right), & t>t_d,
\end{cases}
\label{eq:sscdm_waiting_solution}
\end{equation}
where $A_e\equiv(2C_eH_0^2/3)^{3/2}$. Indeed, $\ddot \psi$ and the non-Lipschitz force are both linear in $t-t_d$ at leading order, whereas Hubble friction begins at $\mathcal O((t-t_d)^2)$. All these branches share $(\psi,\dot \psi)=(0,0)$ at departure. The prescribed SSCDM history selects one particular $t_d$, whereas the reconstructed departure-side force and the plateau data admit the full waiting-time family. This statement is one-sided and on shell: the inverse map fixes the force only on the traversed field-space branch and does not select a unique off-shell continuation through $\phi_e$. Any $C^1$ completion that agrees with the reconstructed transition-side force retains the stationary solution and the departing branches displayed above. The nonuniqueness is therefore explicit, rather than merely a failure of a sufficient uniqueness criterion. A conventional analytic potential with Lipschitz force and initial data $\dot\phi=V_{,\phi}=0$ would instead remain frozen by uniqueness. Accordingly, the closed-form templates fitted below approximate the SSCDM reconstruction over the retained field interval; they do not reproduce its exact compact plateaus.

\textit{3. ECDM --} Finally, we consider a smooth sign-switching DE model governed by an error-function profile, introduced in Ref.~\cite{Bouhmadi-Lopez:2025ggl} and subsequently studied in Refs.~\cite{Bouhmadi-Lopez:2025spo,Ibarra-Uriondo:2026zbp}. The error function is $\operatorname{erf}(y)=2\pi^{-1/2}\int_0^y e^{-u^2}\,{\rm d}u$, and the normalized DE density is parameterized as
\begin{equation}
\widetilde\Omega_{\rm de}(x)=\Omega_{\rm de0}\frac{\operatorname{erf}\!\left[\eta(x-x_\dagger)\right]}{\operatorname{erf}\!\left(-\eta x_\dagger\right)},
\label{eq:pheno_model}
\end{equation}
where $x=-\ln(1+z)$ and $x_\dagger=-\ln(1+z_\dagger)$. The parameter $\eta$ controls the sharpness of the transition. By construction, $\widetilde\Omega_{\rm de}(0)=\Omega_{\rm de0}$ and $\widetilde\Omega_{\rm de}(x_\dagger)=0$. Writing $D\equiv\operatorname{erf}(-\eta x_\dagger)>0$, one has
\begin{equation}
\frac{{\rm d}\widetilde\Omega_{\rm de}}{{\rm d}x}=\frac{2\eta\Omega_{\rm de0}}{\sqrt{\pi}D}\exp\!\left[-\eta^2(x-x_\dagger)^2\right]>0,
\label{eq:ecdm_density_derivative}
\end{equation}
and hence ${\rm d}\widetilde\Omega_{\rm de}/{\rm d}z<0$ for $\eta>0$ and $z_\dagger>0$. On every finite-redshift interval on which $E^2>0$, this derivative is nonzero, the reconstructed field is monotonic, and the inverse-function theorem gives a smooth on-shell potential $V(\phi)$. In particular, $\rho_{\rm de}$, $p_{\rm de}$, $K$, $V$, and $\phi$ remain regular at the finite-redshift density crossing; the pole in $w_{\rm de}=p_{\rm de}/\rho_{\rm de}$ is only a ratio singularity. The fixed-sign condition \eqref{eq:single_field_condition} selects the phantom branch, $\xi=-1$.

This finite-interval statement should not be promoted to global analyticity on the closed completed field range. In the asymptotic future, let
\begin{equation}
\begin{aligned}
y&\equiv\eta(x-x_\dagger)\longrightarrow+\infty,\\
\widetilde\Omega_\infty&\equiv\frac{\Omega_{\rm de0}}{D},\\
\Delta\widetilde\phi&\equiv|\widetilde\phi_\infty-\widetilde\phi|.
\end{aligned}
\end{equation}
Using $E^2\to\widetilde\Omega_\infty$ and $\widetilde V=\widetilde\Omega_{\rm de}+({\rm d}\widetilde\Omega_{\rm de}/{\rm d}x)/6$ gives
\begin{equation}
\begin{aligned}
\Delta\widetilde\phi&\sim\left(\frac{2}{\eta\sqrt{\pi}}\right)^{1/2}\frac{e^{-y^2/2}}{y},\\
\widetilde V-\widetilde\Omega_\infty&\sim\frac{\eta\widetilde\Omega_\infty}{3\sqrt{\pi}}e^{-y^2}\\
&\sim\frac{\eta^2\widetilde\Omega_\infty}{3}(\Delta\widetilde\phi)^2\ln\!\left(\frac{1}{\Delta\widetilde\phi}\right),
\end{aligned}
\label{eq:ecdm_asymptotic_endpoint}
\end{equation}
up to subleading logarithms. Thus, the force tends to zero while the field-space curvature grows logarithmically at the completed asymptotic endpoint. This endpoint is reached only as $x\to+\infty$ ($z\to-1$), rather than at a finite transition time, and therefore does not produce the finite-time waiting-solution ambiguity of exact compact SSCDM. Our claim of a smooth ECDM reconstruction refers to the finite-redshift intervals used in the analysis.

\paragraph{Endpoint stability criterion.} The SSCDM and ECDM endpoint behaviors are two instances of one elementary criterion. Consider a phantom-branch history near a constant plateau of value $\widetilde\Omega_e$, and let $\varepsilon\equiv|\widetilde\Omega_{\rm de}-\widetilde\Omega_e|$. Because ${\rm d}\widetilde\Omega_{\rm de}/{\rm d}x\geq0$ on this branch, $\varepsilon$ grows from zero at a departure plateau and decays to zero at an arrival plateau, and $\widetilde V=\widetilde\Omega_{\rm de}+({\rm d}\widetilde\Omega_{\rm de}/{\rm d}x)/6$ gives
\begin{equation}
\widetilde V-\widetilde V_e=
\begin{cases}
\varepsilon+\dfrac{1}{6}\dfrac{{\rm d}\varepsilon}{{\rm d}x}>0, & \text{departure},\\[8pt]
\dfrac{1}{6}\left(\left|\dfrac{{\rm d}\varepsilon}{{\rm d}x}\right|-6\,\varepsilon\right), & \text{arrival}.
\end{cases}
\label{eq:endpoint_sign}
\end{equation}
A departure endpoint is therefore always a one-sided minimum of the on-shell potential. At an arrival endpoint, define the local settling rate $k_{\rm eff}\equiv-{\rm d}\ln\varepsilon/{\rm d}x$. Equation~\eqref{eq:endpoint_sign} then gives $\widetilde V-\widetilde V_e=(k_{\rm eff}-6)\varepsilon/6$. Provided that the sign of $k_{\rm eff}-6$ is fixed sufficiently close to the endpoint, the endpoint is a one-sided minimum when $k_{\rm eff}>6$ and a one-sided maximum when $k_{\rm eff}<6$; $k_{\rm eff}=6$ is marginal, and subleading terms determine the classification. The threshold is thus the $a^{-6}$ dilution law of free phantom kinetic energy, $|\rho_{\rm de}+p_{\rm de}|=\dot\phi^2\propto a^{-6}$. Since the homogeneous stability criterion is reversed for $\xi=-1$ (Sec.~\ref{sec:dynamics}), a minimum is unstable and a maximum stable under homogeneous phantom dynamics. The correspondence is exact for an exponential approach, $\varepsilon\propto a^{-k}$, on a plateau-dominated background with $E^2\to E_e^2$ and $H\to H_e$: Eqs.~\eqref{eq:phi_dev} and \eqref{eq:V_K_delta} then give $|\widetilde\phi-\widetilde\phi_e|\propto a^{-k/2}$ and
\begin{equation}
\widetilde V-\widetilde V_e\simeq\frac{k\,(k-6)}{24}\,E_e^2\,\big(\widetilde\phi-\widetilde\phi_e\big)^2,
\label{eq:endpoint_curvature}
\end{equation}
while linearizing the phantom Klein--Gordon equation about the frozen endpoint---where $\delta\rho_\phi$ vanishes at first order, so $H$ is unperturbed---gives the characteristic exponents $\lambda=-kH_e/2$ and $\lambda=(k-6)H_e/2$. For $k\neq3$, the first follows the reconstructed approach and the second is its companion; at $k=3$ the exponents coincide and the second independent solution acquires the usual factor of cosmic time. The companion exponent is positive precisely for $k>6$.

Both continuous targets lie on the unstable side. The ECDM arrival is Gaussian, $-{\rm d}\ln\varepsilon/{\rm d}x\to\infty$, reproducing the quadratic-with-logarithm law and logarithmically divergent curvature of Eq.~\eqref{eq:ecdm_asymptotic_endpoint}; exact compact SSCDM reaches its plateau at finite $x$, the extreme case, which sharpens the unstable minimum into the non-Lipschitz $|\widetilde\phi-\widetilde\phi_e|^{4/3}$ law of Eq.~\eqref{eq:sscdm_endpoint_scaling}. The nonuniqueness of Eq.~\eqref{eq:sscdm_waiting_solution} then acquires a simple physical reading. Departure from the high-redshift plateau is departure from an unstable equilibrium: dynamically natural, and rendered nonunique even from exactly frozen data by the non-Lipschitz force. Arrival at the late-time plateau is instead a fine-tuned approaching branch of an unstable equilibrium: the dS-like state of the reconstruction is not an attractor of its own scalar dynamics. For compact SSCDM, this instability is realized explicitly---any $C^1$ completion preserving the transition-side branch admits spontaneous re-departure from the late-time plateau after an arbitrary waiting time, by the mechanism of Eq.~\eqref{eq:sscdm_waiting_solution} applied at that endpoint---while for ECDM the corresponding endpoint lies at infinite time and is unstable to arbitrarily small perturbations of the arrival data. These statements are one-sided and on shell: they concern completions that preserve the reconstructed branch.

The three histories therefore do not stand on identical theoretical footing. ECDM yields an ordinary smooth on-shell scalar reconstruction on every finite interval considered here; compact SSCDM yields a regular homogeneous trajectory but a nonanalytic endpoint potential; and exact L$\Lambda$CDM remains a distributional fluid benchmark with no regular one-field realization within Eq.~\eqref{eq:action}. This last statement is not a no-go theorem for sign-switching cosmology itself. The parametric comparison below treats ECDM and the retained field interval of the compact SSCDM target separately; neither finite-interval fit alters the analytic endpoint classification.

\paragraph{Benchmark inputs and numerical rendering.} For a uniform comparison, the three prescribed histories use the same background normalization and characteristic redshift. The common numerical parameters are
\begin{equation*}
\begin{aligned}
H_0&=70\ {\rm km\,s^{-1}\,Mpc^{-1}}, & h&=0.7,\\
\Omega_{\rm m0}&=0.31, & \Omega_{\rm r0}&=8.53818\times10^{-5},\\
\Omega_{\rm de0}&=1-\Omega_{\rm m0}-\Omega_{\rm r0}\\[-2pt]
&\simeq0.6899146.
\end{aligned}
\end{equation*}
The radiation value follows from Eq.~\eqref{eq:radiation_background} with $T_{\rm CMB}=2.7255\,{\rm K}$ and $N_{\rm eff}=3.046$. The profile-specific benchmark inputs are
\begin{equation*}
\begin{aligned}
{\rm ECDM}:&\quad z_\dagger=1.8,\qquad \eta=5.0,\\
{\rm SSCDM}:&\quad z_\dagger=1.8,\qquad \Delta x=0.4,\\[-2pt]
&\quad z_i=2.4199277,\qquad z_f=1.2924461,\\
{\rm L}\Lambda{\rm CDM}:&\quad z_\dagger=1.8,\qquad \Delta z=0.15,\qquad N=8.
\end{aligned}
\end{equation*}
The SSCDM endpoints obey $(1+z_\dagger)^2=(1+z_i)(1+z_f)$ and therefore implement a transition symmetric in $x=\ln a$, consistently with Eqs.~\eqref{eq:sscdm_x_boundaries}--\eqref{eq:sscdm_z_boundaries}. For L$\Lambda$CDM, these parameters generate the centered transition array in Eq.~\eqref{eq:ladder_benchmark_grid}.

The source arrays contain $N_z=30000$ uniformly spaced redshift samples over $0\leq z\leq100$, ordered from high to low redshift, with $|\delta z_{\rm grid}|=100/29999=3.33344\times10^{-3}$. All displayed background quantities, scalar reconstructions, and potential fits use $0\leq z\leq5$. The extension to $z>5$ is retained only for source-array compatibility and a grid-level positivity diagnostic; it is not physically interpreted with the low-redshift massless-neutrino approximation. For all three benchmarks, $E^2(z)$ remains positive on the stored grid, with the minimum sampled value $E^2_{\min}=1$ at $z=0$. The recombination-to-present shooting calculations of Secs.~\ref{subsec:append_sig_bump} and \ref{subsec:append_axion} instead evolve the explicit massive-neutrino background from $z_*=1090$; the closure test of Sec.~\ref{subsec:closure} returns to the low-redshift reconstruction background over $0\leq z\leq5$.

The pressure is evaluated from Eq.~\eqref{eq:pressure}. Derivatives are computed on the same redshift-coordinate array with the second-order finite-difference prescription \texttt{numpy.gradient} \cite{Harris:2020xlr} and \texttt{edge\_order=2}, without interpolation before differentiation. No analytic mollifier or smoothing kernel is applied to the exact L$\Lambda$CDM density. Its finite plotted spikes arise solely from applying the finite-difference stencil across the sampled Heaviside discontinuities; they occupy approximately one to two grid cells and have resolution-dependent heights and shapes. They are therefore numerical renderings of distributional impulses, not regular transitions of physical width $|\delta z_{\rm grid}|$.

%%-----------------------------------------------------------------------------%%
\subsection{Parametric scalar-field potential ans\"atze}
%%-----------------------------------------------------------------------------%%

To compare closed-form potential shapes with reconstructed DE histories, we consider the five parametric ans\"atze listed below. They are fitted separately to the ECDM on-shell potential and to the retained transition-side interval of the SSCDM target. Agreement on either interval neither selects a unique off-shell completion nor establishes the global dynamics of the potential. Although several of these forms are inspired by canonical quintessence models, here they are used as potential-shape families for the fixed-sign phantom reconstruction; canonical tracker or stability properties, therefore, do not follow automatically \cite{Copeland:2006wr,Tsujikawa:2013fta}.

We use the dimensionless field coordinate $\widetilde\phi\equiv\phi/M_{\rm Pl}$, and all field locations and widths below are expressed in this coordinate. The physical potential $V$ has dimensions of energy density. Thus, $V_0$, $A$, the quantities denoted by $\Lambda$ in the shifted-$\tanh$ and sigmoid--Gaussian models, and $\lambda^2$ have dimensions of energy density, whereas the axion parameter $\Lambda$ has dimensions of mass and $\Lambda^4$ has dimensions of energy density. In the potential-space inference, $V$ and all energy-density amplitudes are expressed in units of $\rho_{\rm c0}$; the corresponding reported amplitudes are therefore dimensionless ratios. The symbol $\Lambda$ is model dependent and does not have the same physical dimension in all five ans\"atze.

For each target, the additive origin and orientation of $\widetilde\phi$ are fixed before the inference and held constant for all five families. In particular, the axion form below is a zero-phase model, so its quoted evidence is conditional on this field-coordinate convention. Introducing an independent axion phase would define a different model and require a new evidence calculation.

\paragraph{Generalized axion-like potential.}

A generalized axion-like potential of a form used in late-time scalar-field studies \cite{Kamionkowski:2014zda,Emami:2016mrt,Boiza:2024fmr,Chiang:2025qxg} is
\begin{equation}
V(\widetilde\phi)=\Lambda^4\left[1-\cos\left(\frac{\widetilde\phi}{\eta}\right)\right]^n+V_0.
\label{eq:axion_template}
\end{equation}
Here $\eta>0$ fixes the period $2\pi\eta$ in $\widetilde\phi$, $n>0$ controls the shape, and $V_0$ is a constant offset. For $\Lambda^4>0$, the potential ranges from $V_0$ to $V_0+2^n\Lambda^4$ and therefore takes both signs only when $-2^n\Lambda^4<V_0<0$. The axion field-space scale $\eta$ is unrelated to the ECDM transition sharpness denoted by the same symbol in Eq.~\eqref{eq:pheno_model}.

For noninteger $n$, Eq.~\eqref{eq:axion_template} is a closed-form but not generally a globally analytic potential. Near a minimum, $\widetilde\phi=2\pi k\eta+\delta\widetilde\phi$, one has $V-V_0\propto|\delta\widetilde\phi|^{2n}$. Its gradient diverges for $n<1/2$, has a cusp at $n=1/2$, and is continuous while the curvature diverges for $1/2<n<1$. Any subsequent dynamical use must therefore verify that the trajectory does not encounter these non-regular minima.

\paragraph{Shifted-$\tanh$ potential.} To model a smooth transition between two field-space plateaus, we use
\begin{equation}
V(\widetilde\phi)=\frac{\Lambda(\xi_1+1)}{2}-\frac{\Lambda(\xi_1-1)}{2}\tanh\left[\nu\left(\widetilde\phi-\widetilde\phi_{\rm c}\right)\right].
\label{eq:tanh_template}
\end{equation}
Its two asymptotes are $V(-\infty)=\Lambda\xi_1$ and $V(+\infty)=\Lambda$. Thus, for $\Lambda\neq0$, $\xi_1$ is the ratio of the low-field to high-field plateau; it is unrelated to the kinetic-sign parameter $\xi$ in Eq.~\eqref{eq:action}. The two plateaus have opposite signs precisely when $\xi_1<0$. The parameter $\nu>0$ controls the inverse transition width and $\widetilde\phi_{\rm c}$ its center. At $\Lambda=0$ the entire model collapses to zero, while at $\xi_1=1$ the transition disappears and $\nu$ and $\widetilde\phi_{\rm c}$ become unidentifiable. These nested limits contribute to the prior dependence of the evidence.

\paragraph{Gaussian feature potential.} As a deliberately restrictive one-feature baseline, we consider
\begin{equation}
V(\widetilde\phi)=A\exp\left[-\frac{(\widetilde\phi-\widetilde\phi_{\rm c})^2}{2w^2}\right].
\label{eq:gaussian_template}
\end{equation}
Here $A$ is a signed amplitude, $\widetilde\phi_{\rm c}$ is the feature center, and $w>0$ is its standard deviation. This ansatz approaches zero at both field-space asymptotes and has the sign of $A$ at every finite field value. It therefore cannot by itself represent two nonzero plateaus or a sign-changing potential.

\paragraph{Regularized inverse-quadratic potential.} Inspired by inverse-power-law quintessence potentials \cite{Ratra:1987rm}, we use the regularized fitting form
\begin{equation}
V(\widetilde\phi)=\frac{\lambda^2}{(\widetilde\phi-\widetilde\phi_{\rm c})^2+\epsilon^2}+V_0.
\label{eq:inverse_template}
\end{equation}
Here $\epsilon>0$ regularizes the inverse-quadratic profile and $V_0$ is its asymptotic value. For finite $\epsilon$ the potential has no pole: its peak above the offset is $\lambda^2/\epsilon^2$, and its half-width at half-maximum is $\epsilon$. Consequently, $\lambda$ and $\epsilon$ jointly determine the peak height and are structurally correlated. The profile takes both signs when $V_0<0<V_0+\lambda^2/\epsilon^2$. The canonical tracker motivation of the unregularized inverse-power-law form should not be interpreted as establishing tracker behavior for the offset, regularized phantom template used here.

\paragraph{Sigmoid--Gaussian feature potential.} As a phenomenological ansatz combining an asymptotic transition with a localized feature, we introduce
\begin{equation}
\begin{aligned}
\frac{V(\widetilde\phi)}{\Lambda}&=\tanh\!\left(\frac{\widetilde\phi-\widetilde\phi_{\rm a}}{\sigma_{\rm a}}\right)\\
&\quad+\alpha\exp\!\left[-\frac{(\widetilde\phi-\widetilde\phi_{\rm b})^2}{\sigma_{\rm b}^2}\right].
\end{aligned}
\label{eq:sigmoid_template}
\end{equation}
For $\sigma_{\rm a},\sigma_{\rm b}>0$, the asymptotic values are fixed to $V(-\infty)=-\Lambda$ and $V(+\infty)=+\Lambda$; hence, the global plateaus have equal magnitude and opposite sign. The localized contribution has an amplitude of $\Lambda\alpha$, so $\alpha$ is a relative amplitude and is structurally correlated with $\Lambda$. Its sign produces an enhancement or suppression according to the sign of the product $\Lambda\alpha$, not $\alpha$ alone. With the convention in Eq.~\eqref{eq:sigmoid_template}, $\sigma_{\rm b}$ is the $1/e$ half-width and the corresponding Gaussian standard deviation is $\sigma_{\rm b}/\sqrt{2}$.

The inference below restricts $\alpha\geq0$. Negative-$\alpha$ deformations are therefore not included, and the evidence for this family is conditional on the adopted field orientation. At $\alpha=0$, $\widetilde\phi_{\rm b}$ and $\sigma_{\rm b}$ become unidentifiable, while at $\Lambda=0$ all shape parameters are unidentifiable.

For positive widths and $\epsilon$, the shifted-$\tanh$, Gaussian, regularized inverse-quadratic, and sigmoid--Gaussian forms are real analytic on the real field axis. The generalized axion potential requires the regularity qualification stated above. None of these families reproduces the exact compact endpoints of SSCDM globally. Their regular equilibria can also have the opposite homogeneous stability from the reconstructed late-time endpoints. In the parameter branches relevant here, the shifted-$\tanh$ and sigmoid--Gaussian plateaus are approached from below and can act as phantom attractors. The positive-amplitude maxima of the Gaussian and regularized inverse-quadratic templates, and the maxima of axion members with a regular force, such as those with $n\geq1$, are likewise locally stable under homogeneous phantom dynamics. Both reconstructed targets instead approach their late-time plateaus from above, as unstable equilibria. Agreement in sampled potential values over a finite field interval can therefore coexist with qualitatively different late-time stability, a distinction that the function-value comparison below does not probe. Accordingly, the SSCDM comparison below is restricted to the retained transition-side field interval, which includes the high-redshift compact endpoint but not the short tail approaching the late-time endpoint.

%%-----------------------------------------------------------------------------%%
\subsection{Conditional template comparison in potential space}
\label{subsec:bayes}
%%-----------------------------------------------------------------------------%%

To quantify how efficiently these potential families represent a fixed reconstructed target, we perform a conditional Bayesian fit directly in potential space. We write $\widetilde\phi\equiv\phi/M_{\rm Pl}$ and $\widetilde V\equiv V/\rho_{\rm c0}$, and describe a family $M$ by $\widetilde V(\widetilde\phi;\boldsymbol\theta,M)$. This construction is not an observational likelihood and does not compare the cosmological viability of the underlying DE histories. It compares parametric functions for a prescribed target curve, field interval, sampling rule, synthetic noise prescription, and set of parameter priors.

The reconstruction fixes the field only up to a reflection and an additive constant. We therefore adopt one field orientation and origin before constructing the potential-space dataset and use the same convention for every family. These choices are part of the comparison because the ans\"atze and their finite priors need not be invariant under a translation or reflection. Likewise, sampling uniformly in $\widetilde\phi$ and sampling uniformly in $z$ before mapping to $\widetilde\phi$ assign different weights along the target curve.

For each target, we fix one common set of abscissae $\{\widetilde\phi_i\}_{i=1}^{N}$ for all five families and construct one synthetic realization,
\begin{equation}
\widetilde V_i^{\rm(mock)}=\widetilde V_{\rm tar}(\widetilde\phi_i)+\delta\widetilde V_i,\qquad \delta\widetilde V_i\sim\mathcal N(0,\sigma_i^2),
\label{eq:data_mock}
\end{equation}
where
\begin{equation}
\sigma_i^2=\left(\sigma_{\rm rel}|\widetilde V_{\rm tar}(\widetilde\phi_i)|\right)^2+\sigma_{\rm abs}^2.
\label{eq:mock_variance}
\end{equation}
Here both $\sigma_i$ and $\sigma_{\rm abs}$ use the same $\rho_{\rm c0}$ normalization as $\widetilde V$, while $\sigma_{\rm rel}$ is dimensionless. The resulting dataset is
\begin{equation}
\mathcal D\equiv\left\{(\widetilde\phi_i,\widetilde V_i^{\rm(mock)},\sigma_i)\right\}_{i=1}^{N}.
\end{equation}

For the ECDM comparison reported here, we set the field origin at the high-redshift endpoint $z_{\rm hi}=5$ and select the branch $s=-1$ in Eq.~\eqref{eq:phi_branches}. With $\widetilde\phi_{\rm ref}=0$ and $z_{\rm ref}=z_{\rm hi}$, this convention is
\begin{equation}
\begin{aligned}
\widetilde\phi(z)&=\int_z^{z_{\rm hi}}\mathcal Q(\bar z)\,{\rm d}\bar z, & \widetilde\phi'(z)&=-\mathcal Q(z),\\
\dot{\widetilde\phi}&=(1+z)H\mathcal Q(z)\geq0.
\end{aligned}
\label{eq:field_orientation_fit}
\end{equation}
Thus, $\widetilde\phi=0$ at $z=5$ and the field increases forward in cosmic time.

The reconstruction arrays are initially sampled uniformly in redshift. Before constructing the fitting datasets, the $(\widetilde\phi,\widetilde V_{\rm tar})$ pairs are sorted by $\widetilde\phi$, repeated field coordinates are removed, and the target is transferred by piecewise-linear interpolation, implemented with \texttt{numpy.interp} \cite{Harris:2020xlr}, to $N=100$ field points uniformly spaced over the retained interval, including both endpoints. For ECDM, the full reconstructed trajectory covers $0\leq z\leq5$, with $\widetilde\phi(z_{\rm hi}=5)=0$, and the complete trajectory has $\Delta\widetilde\phi_{\rm ECDM}\simeq0.38369$ at $z=0$. The potential comparison uses $0\leq\widetilde\phi\leq0.38$ and $\widetilde\phi_{\rm sc}=0.38$; the upper fitting endpoint corresponds to $z\simeq0.676$. Thus, the comparison omits only the final field-space segment $0.38<\widetilde\phi\lesssim0.38369$. The plotted horizontal range extends to $1.05\max_i\widetilde\phi_i=0.399\simeq0.40$; no endpoint clamping is used.

For the $x$-symmetric SSCDM benchmark $(z_\dagger,\Delta x)=(1.8,0.4)$, the reconstruction is likewise evaluated over $0\leq z\leq5$, with $\widetilde\phi(z=5)=0$. The complete transition-side branch terminates at $(\widetilde\phi_{\rm end},z_f)\simeq(0.23744185,1.292)$. The five-family comparison uses the common interval $0\leq\widetilde\phi\leq0.23$, whose upper endpoint corresponds to $z\simeq1.434$, sampled at 100 uniformly spaced field points. All five SSCDM likelihoods and marginal likelihoods therefore refer to the same fixed synthetic dataset. As a sensitivity test, repeating the generalized axion-like fit over $0\leq\widetilde\phi\leq0.23744$ shifts the direct weighted posterior medians from $(\Lambda^4/\rho_{\rm c0},\eta,n,V_0/\rho_{\rm c0})=(1.72610,0.0469695,0.428204,-0.811441)$ to $(1.72931,0.0471316,0.424007,-0.814893)$. Each shift is smaller than $0.35$ times the corresponding combined posterior standard deviation. The axion-potential shape and its marginal parameter summaries are therefore insensitive to the omitted tail at the resolution of this experiment. Comparable full-endpoint tests have not been performed for the other four families, so the reported SSCDM ranking applies specifically to $0\leq\widetilde\phi\leq0.23$.

For both targets, the reported synthetic-noise parameters are
\begin{equation}
\sigma_{\rm rel}=0.10,\qquad \sigma_{\rm abs}=0.05,
\end{equation}
in the normalization of Eq.~\eqref{eq:mock_variance}. The black points for each target constitute one Gaussian mock realization rather than noiseless target values carrying design weights, and the same target-specific mock is fitted by all five families. The NumPy generator \texttt{numpy.random.default\_rng(10)} is initialized separately when generating the ECDM and SSCDM mocks. Resetting the same seed reuses the same underlying standard-normal pseudorandom sequence, although the two physical mocks differ because their means and pointwise scales differ; they should therefore not be regarded as statistically independent. No cross-target evidence comparison is made. Each family is then analyzed in a separate, non-resumed sampler instance, with no explicit UltraNest random seed imposed.

The diagonal covariance in Eq.~\eqref{eq:mock_variance} is an adopted weighting rule. The points sample a common deterministic reconstructed curve; they are not independent cosmological measurements. We do not propagate covariance arising from the phenomenological target parameters, numerical reconstruction, or interpolation, and the coordinates $\widetilde\phi_i$ are treated as exact. The posterior widths quantify parameter uncertainty within this synthetic prescription, and the marginal likelihoods quantify prior-weighted representational performance.

For each family, we use the finite proper priors specified in Sec.~\ref{subsec:priors}. The potential scale is fixed from the noiseless interpolated target before the Gaussian mock is drawn:
\begin{equation}
\widetilde V_{\rm sc}=\max\left\{\max_i\left|\widetilde V_{{\rm tar},i}\right|,10^{-3}\right\}.
\label{eq:target_prior_scale}
\end{equation}
With the common photon normalization of Eq.~\eqref{eq:radiation_background}, the corresponding target scales are
\begin{equation}
\widetilde V_{\rm sc}^{\rm ECDM}=0.86921809,\qquad \widetilde V_{\rm sc}^{\rm SSCDM}=1.53900555.
\end{equation}
These scales were constructed from the noiseless targets rather than from the realized mocks. The priors are therefore target- and coordinate-dependent but are not adaptive to the particular Gaussian mock realization. The potential-space inputs are insensitive to the radiation normalization at the level relevant here: replacing $\omega_\gamma=2.47297928\times10^{-5}$, the value associated with $T_{\rm CMB}=2.7255\,{\rm K}$ in Eq.~\eqref{eq:radiation_background}, by the older value $\omega_\gamma=2.469\times10^{-5}$ associated with $T_{\rm CMB}=2.725\,{\rm K}$, which lowers $\Omega_{\rm r0}$ by $1.4\times10^{-7}$ at fixed $N_{\rm eff}=3.046$ and $h=0.7$, changes the target values on the $N=100$ fitting grids by at most $1.37681\times10^{-5}\sigma_i$ for ECDM and $2.66252\times10^{-5}\sigma_i$ for SSCDM. These shifts are negligible compared with the adopted synthetic uncertainties and do not alter the reported potential-space comparison at the quoted precision. Treating the $\sigma_i$ as fixed design weights, the adopted factorized Gaussian log-likelihood is
\begin{equation}
\begin{aligned}
\log\mathcal L(\mathcal D|\boldsymbol\theta,M)=-\frac{1}{2}\sum_{i=1}^{N}\Bigg\{&\frac{\left[\widetilde V_i^{\rm(mock)}-\widetilde V(\widetilde\phi_i;\boldsymbol\theta,M)\right]^2}{\sigma_i^2}\\
&+\log\left(2\pi\sigma_i^2\right)\Bigg\}.
\end{aligned}
\end{equation}
The normalized posterior is
\begin{equation}
p(\boldsymbol\theta|\mathcal D,M)=\frac{\mathcal L(\mathcal D|\boldsymbol\theta,M)\pi(\boldsymbol\theta|M)}{\mathcal Z(\mathcal D|M)},
\end{equation}
where
\begin{equation}
\mathcal Z(\mathcal D|M)=\int\mathcal L(\mathcal D|\boldsymbol\theta,M)\,\pi(\boldsymbol\theta|M)\,\dd\boldsymbol\theta
\end{equation}
is estimated using nested sampling \cite{Skilling:2004pqw}. Because the field convention, synthetic dataset, and target-dependent prior prescription form part of the definition of each experiment, the resulting evidence differences are template scores internal to that experiment.

\begin{table*}
\centering
\caption{Proper prior distributions used in the reported potential-space fits. Each cell lists, from top to bottom, the sampled parameter, prior type, and support. For each target, $\widetilde\phi_{\rm sc}$ and $\widetilde V_{\rm sc}$ were held fixed during sampling, with $\widetilde V_{\rm sc}$ calculated from the noiseless interpolated target through Eq.~\eqref{eq:target_prior_scale}. These ranges define a target-specific, coordinate-dependent prior prescription (Sec.~\ref{subsec:bayes}). Dashes pad the columns of templates with fewer sampled parameters.}
\label{tab:priors}
{\footnotesize
\renewcommand{\arraystretch}{1.35}
\setlength{\tabcolsep}{3pt}
\newcommand{\priorentry}[3]{%
  \shortstack{$#1$\\[-2pt]\textup{#2}\\[-2pt]$[#3]$}}
\begin{tabular}{@{}ccccc@{}}
\toprule
\shortstack{\emph{Generalized}\\[-1pt]\emph{axion-like}}
&
\shortstack{\emph{Shifted-}\\[-1pt]$\tanh$}
&
\shortstack{\emph{Gaussian}\\[-1pt]\emph{feature}}
&
\shortstack{\emph{Regularized}\\[-1pt]\emph{inverse-quadratic}}
&
\shortstack{\emph{Sigmoid--Gaussian}\\[-1pt]\emph{feature}}
\\
\midrule

\priorentry{\Lambda^4/\rho_{\rm c0}}{Log-uniform}
  {10^{-3}\widetilde V_{\rm sc},\,3\widetilde V_{\rm sc}}
&
\priorentry{\Lambda/\rho_{\rm c0}}{Uniform}
  {-10\widetilde V_{\rm sc},\,10\widetilde V_{\rm sc}}
&
\priorentry{A/\rho_{\rm c0}}{Uniform}
  {-10\widetilde V_{\rm sc},\,10\widetilde V_{\rm sc}}
&
\priorentry{\lambda/\sqrt{\rho_{\rm c0}}}{Log-uniform}
  {10^{-4}\sqrt{\widetilde V_{\rm sc}},\,
   10\sqrt{\widetilde V_{\rm sc}}}
&
\priorentry{\Lambda/\rho_{\rm c0}}{Uniform}
  {-10\widetilde V_{\rm sc},\,10\widetilde V_{\rm sc}}
\\
\addlinespace[2pt]

\priorentry{\eta}{Log-uniform}
  {0.02\widetilde\phi_{\rm sc},\,1.5\widetilde\phi_{\rm sc}}
&
\priorentry{\xi_1}{Uniform}{-5,\,5}
&
\priorentry{\widetilde\phi_{\rm c}}{Uniform}
  {-3\widetilde\phi_{\rm sc},\,3\widetilde\phi_{\rm sc}}
&
\priorentry{\widetilde\phi_{\rm c}}{Uniform}
  {-3\widetilde\phi_{\rm sc},\,3\widetilde\phi_{\rm sc}}
&
\priorentry{\widetilde\phi_{\rm a}}{Uniform}
  {-0.5\widetilde\phi_{\rm sc},\,0.5\widetilde\phi_{\rm sc}}
\\
\addlinespace[2pt]

\priorentry{n}{Uniform}{0.05,\,2.0}
&
\priorentry{\nu}{Log-uniform}{10^{-2},\,10^{3}}
&
\priorentry{w}{Log-uniform}
  {10^{-3}\widetilde\phi_{\rm sc},\,10\widetilde\phi_{\rm sc}}
&
\priorentry{\epsilon}{Log-uniform}
  {10^{-5}\widetilde\phi_{\rm sc},\,\widetilde\phi_{\rm sc}}
&
\priorentry{\widetilde\phi_{\rm b}}{Uniform}
  {0,\,\widetilde\phi_{\rm sc}}
\\
\addlinespace[2pt]

\priorentry{V_0/\rho_{\rm c0}}{Uniform}
  {-10\widetilde V_{\rm sc},\,10\widetilde V_{\rm sc}}
&
\priorentry{\widetilde\phi_{\rm c}}{Uniform}
  {-3\widetilde\phi_{\rm sc},\,3\widetilde\phi_{\rm sc}}
&
--
&
\priorentry{V_0/\rho_{\rm c0}}{Uniform}
  {-10\widetilde V_{\rm sc},\,10\widetilde V_{\rm sc}}
&
\priorentry{\sigma_{\rm a}}{Log-uniform}
  {0.03\widetilde\phi_{\rm sc},\,0.80\widetilde\phi_{\rm sc}}
\\
\addlinespace[2pt]

-- & -- & -- & --
&
\priorentry{\sigma_{\rm b}}{Log-uniform}
  {0.03\widetilde\phi_{\rm sc},\,0.60\widetilde\phi_{\rm sc}}
\\
\addlinespace[2pt]

-- & -- & -- & --
&
\priorentry{\alpha}{Uniform}{0,\,3}
\\
\bottomrule
\end{tabular}
}
\end{table*}

We use \texttt{UltraNest} version \texttt{4.5.0} for the posterior and evidence calculations \cite{Buchner:2014qmx,Buchner2,Buchner:2021cql,Buchner:2021kpm}. The reported runs use \texttt{ReactiveNestedSampler} with a vectorized likelihood, \texttt{min\_num\_live\_points}$=250$, and the termination tolerance \texttt{dlogz}$=0.05$. No likelihood-call cap is imposed. Each family is sampled once in a separate sampler instance using overwrite rather than resume mode, and no separate UltraNest random seed is explicitly passed. Parameter summaries are weighted medians with weighted 16th and 84th percentiles calculated directly from the original nested-sampling points. The model-specific likelihood-call counts are reported in Table~\ref{tab:model_comparison_combined}. For a fixed target, we define
\begin{equation}
\Delta\log\mathcal Z_M=\log\mathcal Z_M-\max_{M'}\log\mathcal Z_{M'},
\label{eq:delta_logz}
\end{equation}
so that the best-ranked family has $\Delta\log\mathcal Z=0$. Only differences between families fitted to the same dataset are meaningful; in particular, absolute evidence values cannot be compared between ECDM and SSCDM, as their target curves, datasets, field intervals, and prior scales differ.

The curve summaries evaluate the deterministic templates on a 1000-point field grid. At each grid point, we calculate direct weighted equal-tail quantiles at $q=(0.025,0.16,0.50,0.84,0.975)$ from the saved UltraNest posterior samples, with the normalized posterior weights applied exactly once. No posterior resampling or display smoothing is used, and residuals are calculated relative to the pointwise weighted median. The resulting summaries are checked to satisfy
\begin{equation}
V_{0.025}\leq V_{0.16}\leq V_{0.50}\leq V_{0.84}\leq V_{0.975}.
\end{equation}
The same construction and nesting check are used for ECDM and SSCDM. These pointwise regions describe only the parameter-induced spread of the deterministic templates. They are neither uncertainty bands on $\widetilde V_{\rm tar}$ nor observational posterior-predictive intervals. Parameter corner plots are produced using \texttt{ChainConsumer} version \texttt{0.34.0} \cite{2016JOSS....1...45H}.

%%-----------------------------------------------------------------------------%%
\subsection{Priors}
\label{subsec:priors}
%%-----------------------------------------------------------------------------%%

For each synthetic experiment, the reported prior bounds use a target-dependent coordinate scale and a positive potential scale,
\begin{equation}
 \widetilde\phi_{\rm sc}\equiv\max_i|\widetilde\phi_i|,\qquad \widetilde V_{\rm sc}>0.
 \label{eq:prior_scales}
\end{equation}
The resulting bounds are held fixed throughout each nested-sampling run. Equation~\eqref{eq:target_prior_scale} specifies the pre-mock construction used for both reported experiments. The resulting ranges are empirical, target-specific priors rather than parameter-independent physical priors.

The scale $\widetilde\phi_{\rm sc}$ and several location priors depend on the field origin and orientation. Consequently, the field convention specified in Sec.~\ref{subsec:bayes} forms part of the definition of the comparison. In particular, the zero-phase axion model and the positive-only prior for $\widetilde\phi_{\rm b}$ are not translation invariant, while the restriction $\alpha\geq0$ means that the sigmoid--Gaussian prior family is not closed under the physically equivalent field reflection. Fixing the coordinate convention is therefore essential for interpreting the reported ranking.

All priors are proper distributions with the finite bounds listed in Table~\ref{tab:priors}. Log-uniform priors are used for the strictly positive scale parameters indicated there, whereas signed amplitudes, offsets, and locations receive uniform priors. The energy-density parameters in the table are expressed in the normalization actually sampled, namely units of $\rho_{\rm c0}$. The reported sigmoid--Gaussian runs impose $\alpha\in[0,3]$ and therefore do not sample negative values of the relative feature coefficient.

Several parameterizations contain nested or weakly identified limits. For example, the Gaussian center and width become irrelevant as $A\to0$; the sigmoid feature center and width become irrelevant as $\alpha\to0$; and the shifted-$\tanh$ shape parameters become irrelevant as $\Lambda\to0$. These regions are retained in the evidence integral, so their prior-volume weight is part of the reported Occam penalty.

%%-----------------------------------------------------------------------------%%
\section{Results}
\label{sec:results}
%%-----------------------------------------------------------------------------%%

In this section, we first compare the prescribed DE density histories and the background quantities reconstructed from them. We then present separate conditional potential-space comparisons for ECDM and the retained transition-side interval of SSCDM. For both targets, the prior scale was fixed from the noiseless reconstruction before mock generation, and the scores are to be read in the sense specified in Sec.~\ref{subsec:bayes}.

\begin{figure*}
    \centering
    \includegraphics[width=0.32\linewidth]{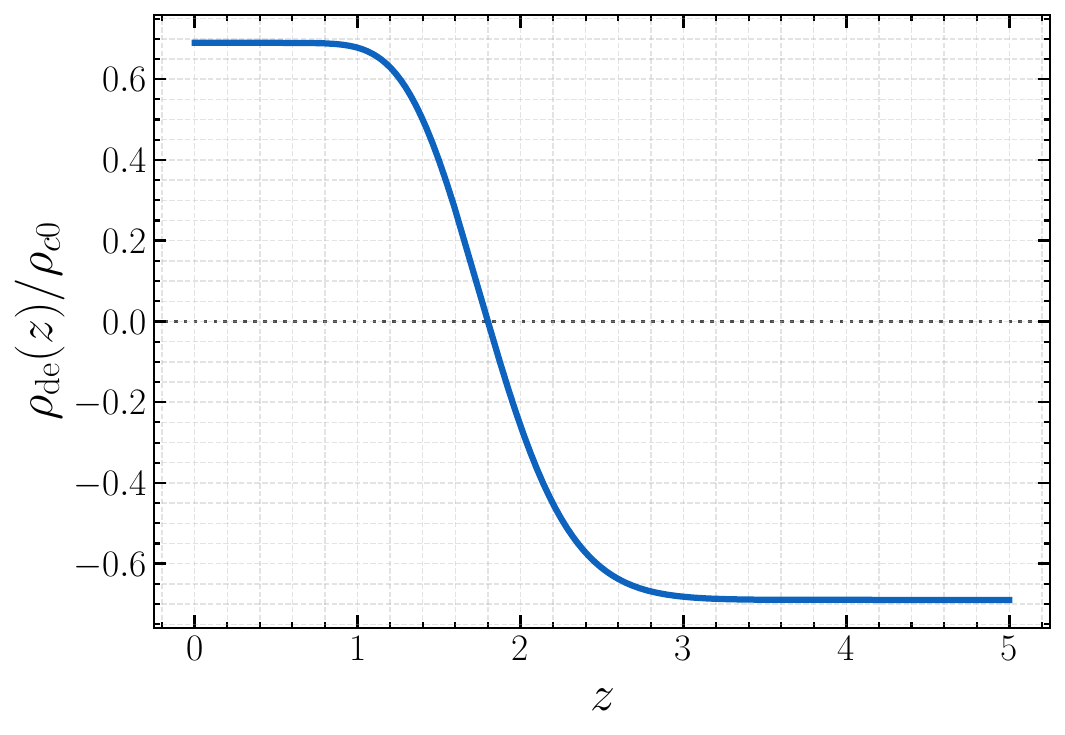}
    \includegraphics[width=0.32\linewidth]{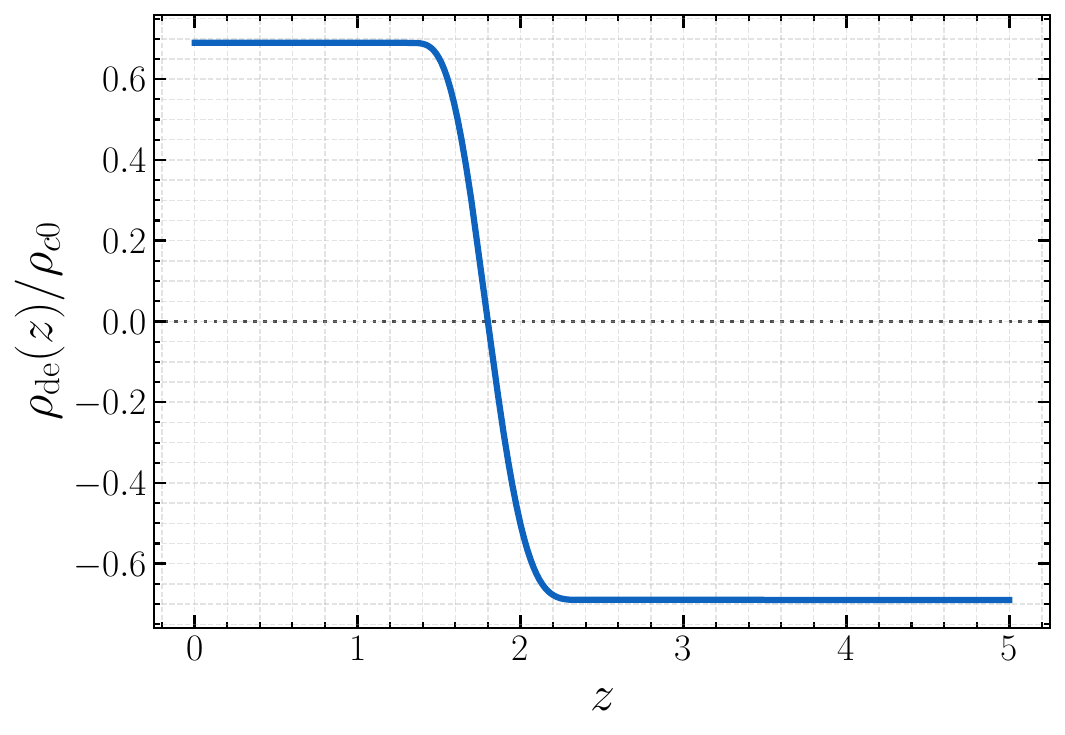}
    \includegraphics[width=0.32\linewidth]{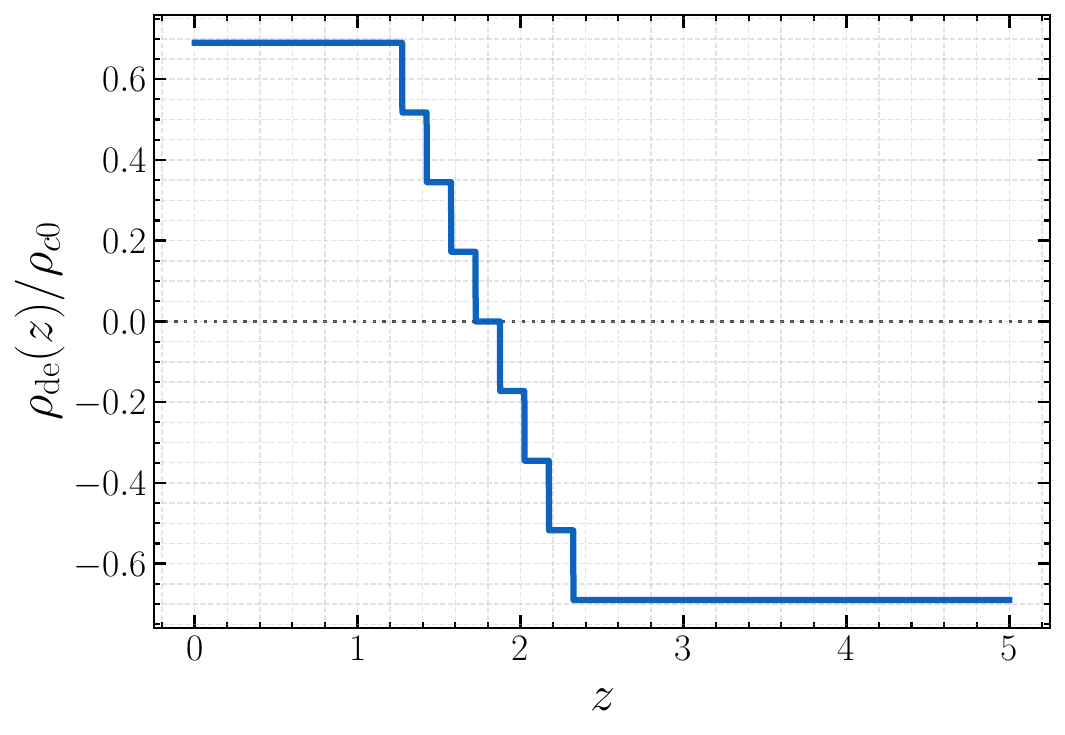}
    \caption{Evolution of the normalized DE density, $\widetilde\Omega_{\rm de}(z)=\rho_{\rm de}(z)/\rho_{\rm c0}$. The left, middle, and right panels show the ECDM, SSCDM, and L$\Lambda$CDM histories, respectively. Read forward in cosmic time, from high to low redshift, all three histories evolve from negative to positive DE density, although with different transition regularity. The L$\Lambda$CDM panel uses the eight transition redshifts in Eq.~\eqref{eq:ladder_benchmark_grid}; its central zero-density plateau is $1.725<z<1.875$.}
    \label{fig:rho_background}
\end{figure*}

\begin{figure*}
    \centering
    \includegraphics[width=0.32\linewidth]{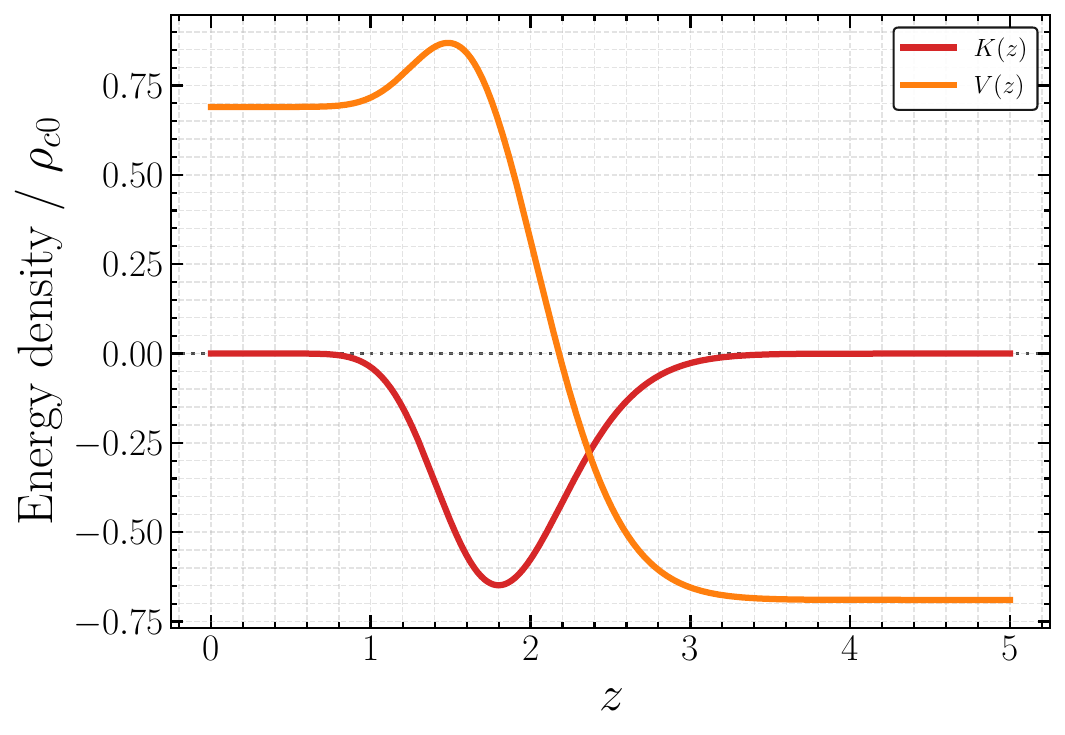}
    \includegraphics[width=0.32\linewidth]{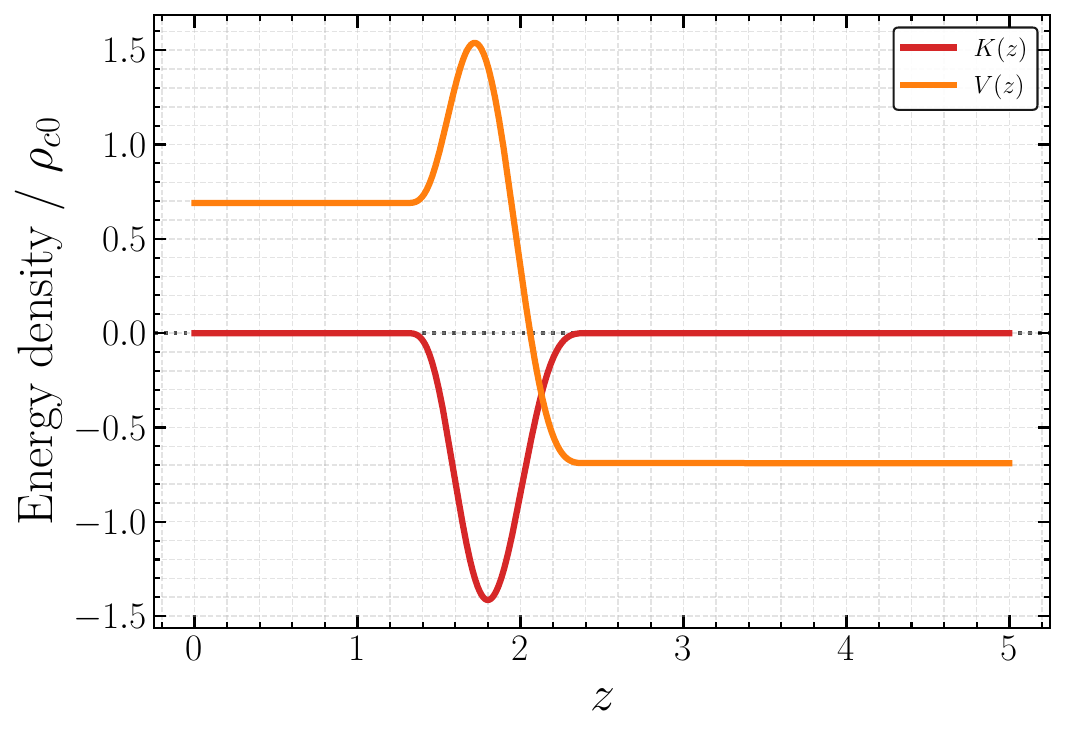}
    \includegraphics[width=0.32\linewidth]{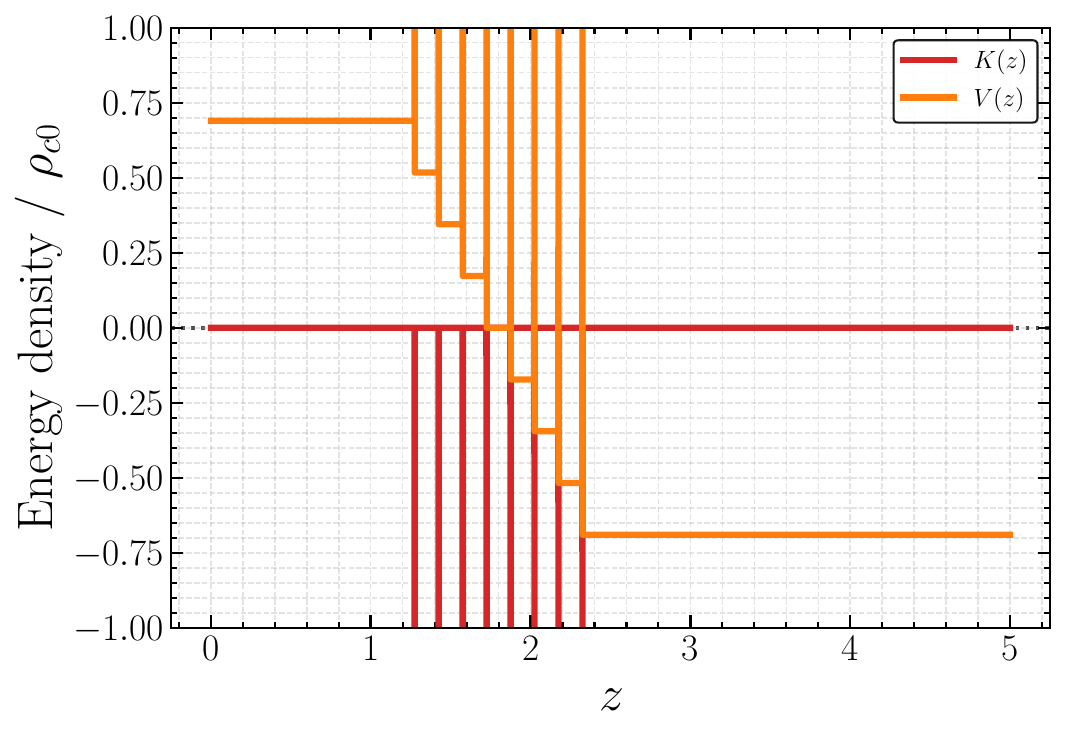}
    \caption{Evolution of the normalized potential contribution $\widetilde\Omega_V=V/\rho_{\rm c0}$ and signed kinetic contribution $\widetilde\Omega_K=K/\rho_{\rm c0}$ for the ECDM, SSCDM, and L$\Lambda$CDM histories in the left, middle, and right panels, respectively. For ECDM and SSCDM these quantities follow from the on-trajectory background reconstruction, with $K\leq0$ and with $K=0$ on the exact SSCDM plateaus. For the strict Heaviside ladder, $K$ and $V$ contain impulsive terms at the jumps; the finite vertical features in the right panel are finite-resolution representations whose heights and widths are not regulator independent. The impulses are clipped at the displayed vertical limits.}
    \label{fig:KV_background}
\end{figure*}

%%-----------------------------------------------------------------------------%%
\subsection{Background reconstruction}
\label{subsec:res_background}
%%-----------------------------------------------------------------------------%%

% ============================
%   Background Dynamics
% ============================

Figure~\ref{fig:rho_background} compares the three prescribed sign-switching DE histories. Here and below, the terms AdS-like and dS-like refer only to the negative- and positive-vacuum-energy-like regimes of the DE sector; they do not imply that the complete matter--radiation--DE spacetime is an exact anti-de Sitter or de Sitter geometry. The ECDM history (left panel) crosses $\rho_{\rm de}=0$ smoothly through its error-function profile. The SSCDM history (middle panel) joins two constant-density plateaus over a finite interval using the compact smooth-step interpolation. The L$\Lambda$CDM history (right panel) instead consists of a finite sequence of discontinuous jumps and, for the even value of $N$ used here, contains a central zero-density plateau. In every case, $\rho_{\rm de}(0)=\Omega_{\rm de0}\rho_{\rm c0}>0$, while the DE density is negative at sufficiently high redshift. Thus, read in the physical direction of cosmic evolution, the histories proceed from an early negative-density AdS-like regime to a late positive-density dS-like regime in the DE sector.

Figure~\ref{fig:KV_background} shows the reconstructed potential contribution $V(z)$ and signed kinetic contribution $K(z)$. For the continuous ECDM and SSCDM targets, ${\rm d}\widetilde\Omega_{\rm de}/{\rm d}z\leq0$ and hence $K\leq0$. Wherever the density varies, the fixed-sign scalar reconstruction therefore selects the phantom branch, $\xi=-1$; equality holds wherever the density is constant. In the ECDM case (left panel), the error-function transition produces a broad negative feature in $K$ and a corresponding smooth enhancement in $V$. Their sum reproduces the prescribed DE density. For SSCDM (middle panel), the evolution is confined to its compact transition interval: $K$ develops a narrow negative feature and $V$ a pronounced positive feature, while outside that interval $K=0$ and $V=\rho_{\rm de}$ exactly.

The L$\Lambda$CDM panel must be interpreted differently. Substitution of Eq.~\eqref{eq:ladder_delta} into Eq.~\eqref{eq:V_K_delta} shows that the strict ladder has negative Dirac impulses in $K$ and positive Dirac impulses in $V$, with weights proportional to $1+z_n$. Between successive jumps, $K=0$ and $V=\rho_{\rm de}$. Consequently, the finite spikes or vertical segments in the displayed curves should not be assigned regulator-independent amplitudes or widths: they represent a particular finite-resolution rendering of the distributional fluid history. The comparison between ECDM and SSCDM illustrates the contrast between a broadly smooth transition and a compactly localized one, but the exact ladder belongs to a different mathematical category and does not define a regular minimally coupled single-field background.

\begin{figure*}
    \centering
    \includegraphics[width=0.32\linewidth]{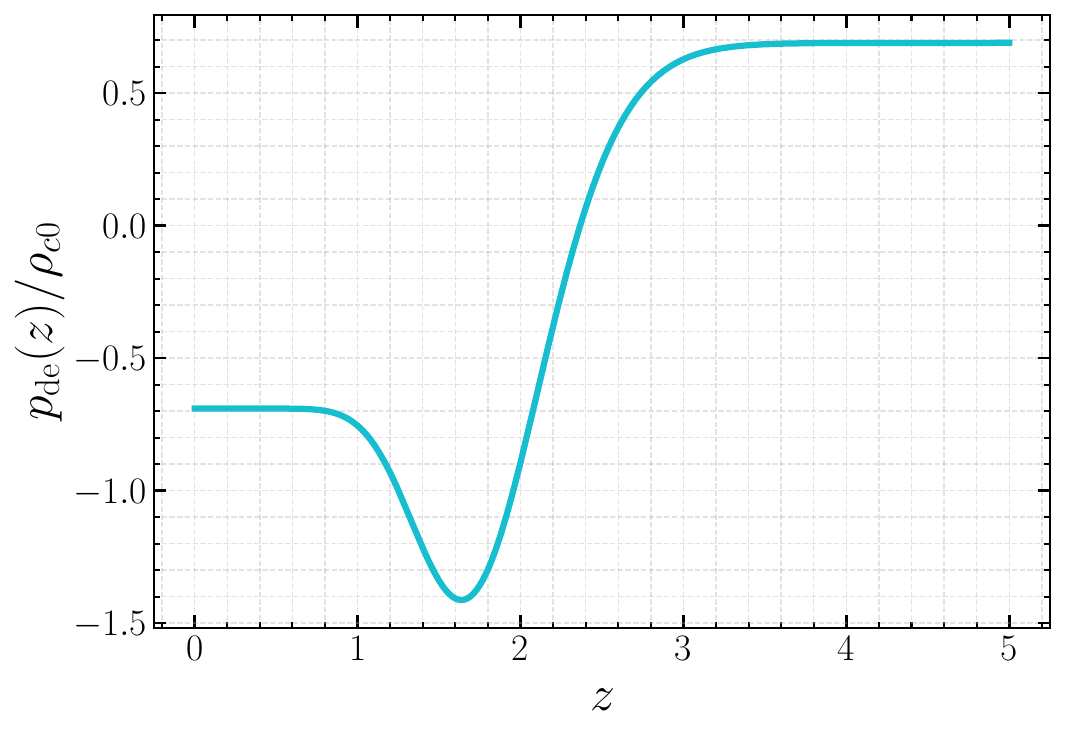}
    \includegraphics[width=0.32\linewidth]{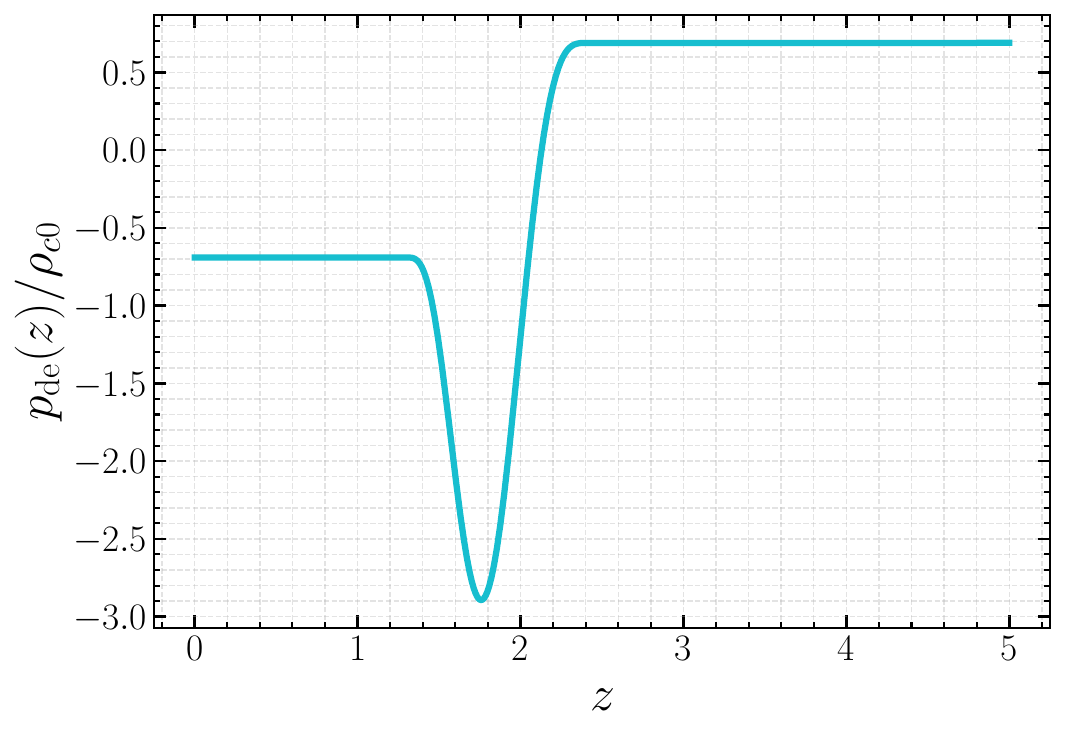}
    \includegraphics[width=0.32\linewidth]{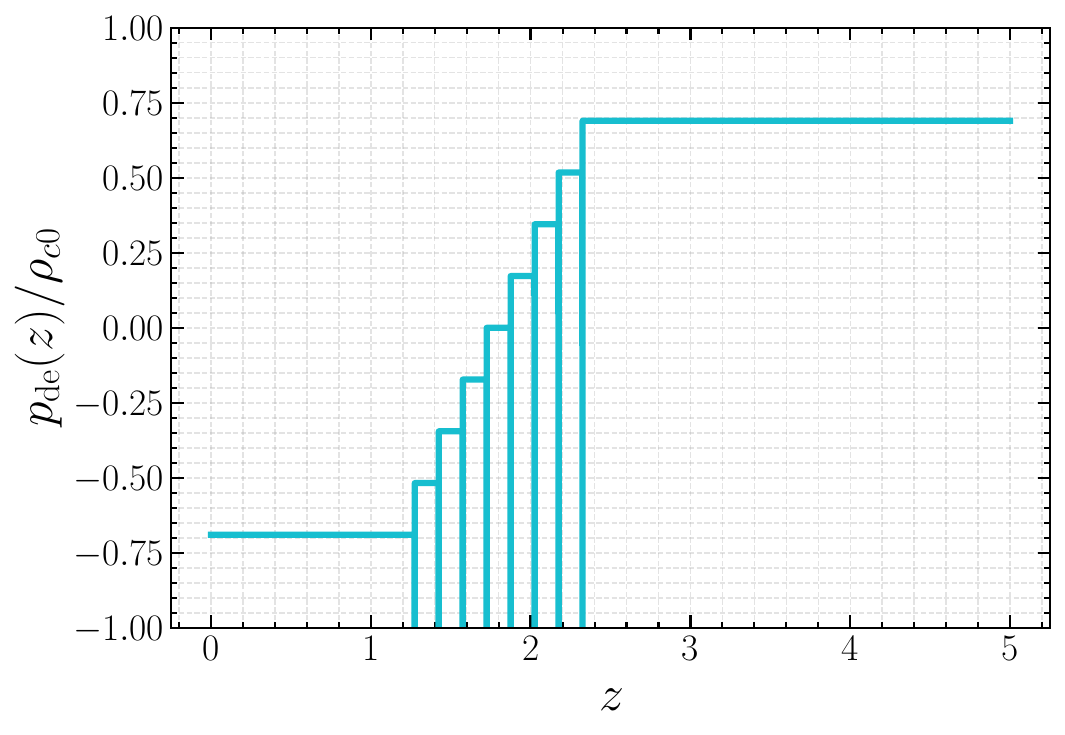}
    \caption{Evolution of the normalized DE pressure, $p_{\rm de}(z)/\rho_{\rm c0}$, for the ECDM, SSCDM, and L$\Lambda$CDM histories in the left, middle, and right panels, respectively. The continuous histories produce a broad ECDM feature and a compact SSCDM feature. In the strict ladder model, the pressure consists of a stepwise regular part together with Dirac impulses at the discontinuities; an ordinary line plot can display these impulses only through a finite-resolution prescription, clipped here at the displayed vertical limits.}
    \label{fig:p_background}
\end{figure*}

\begin{figure*}
    \centering
    \includegraphics[width=0.32\linewidth]{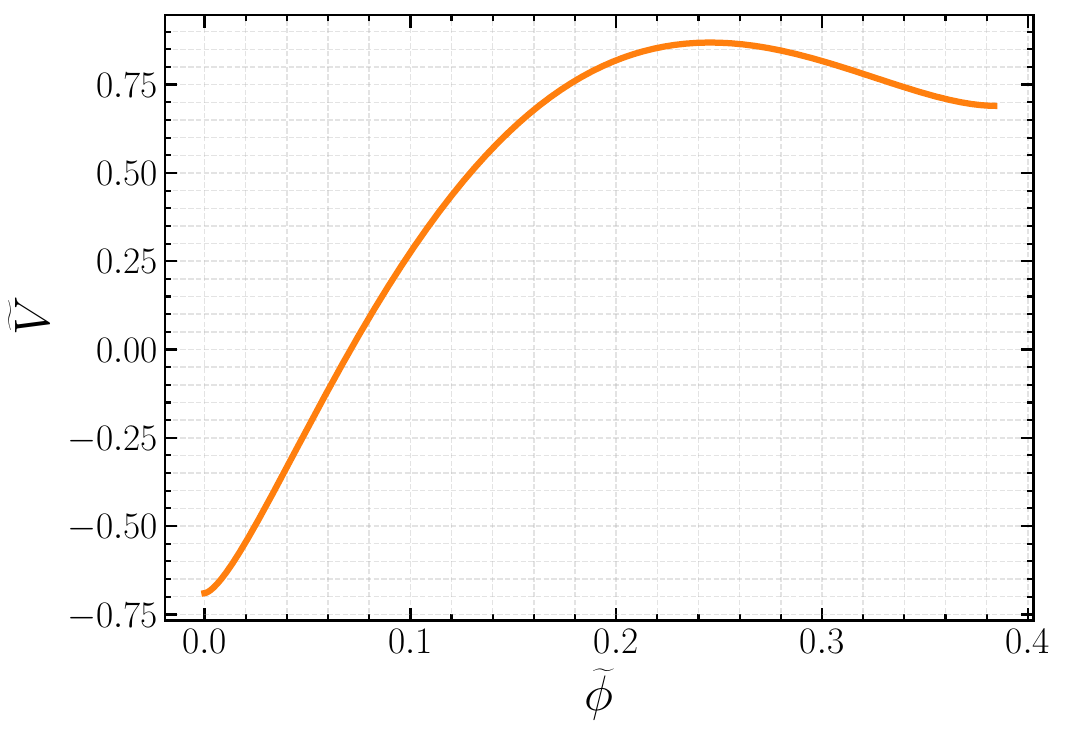}
    \includegraphics[width=0.32\linewidth]{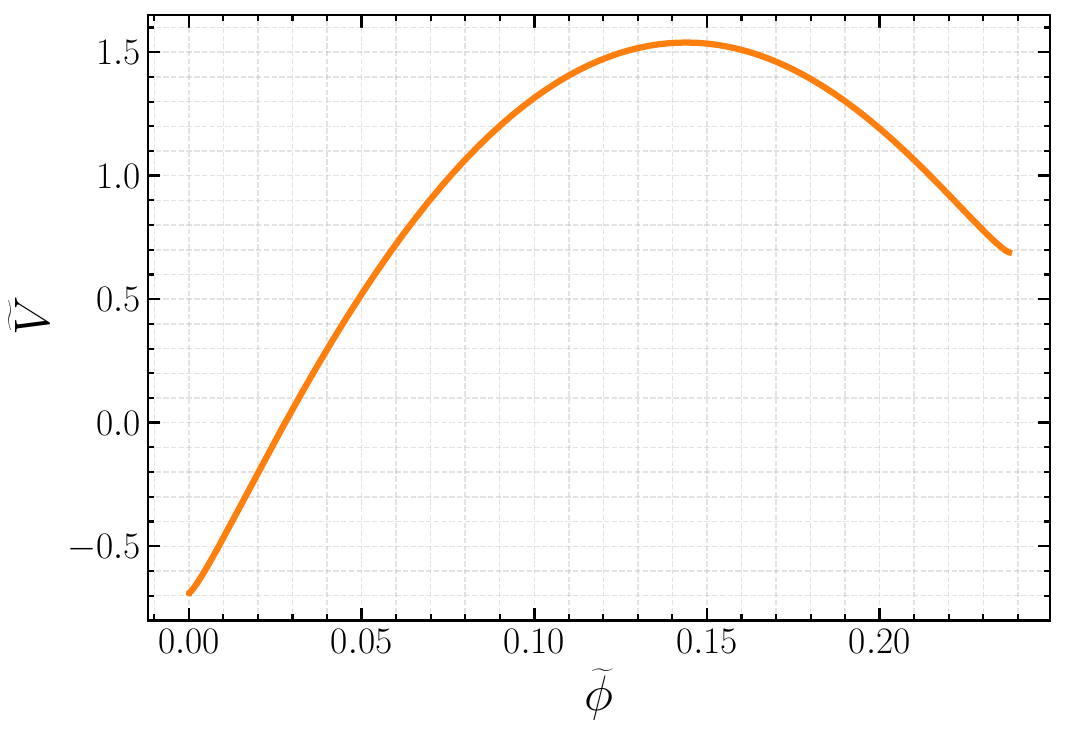}
    \includegraphics[width=0.32\linewidth]{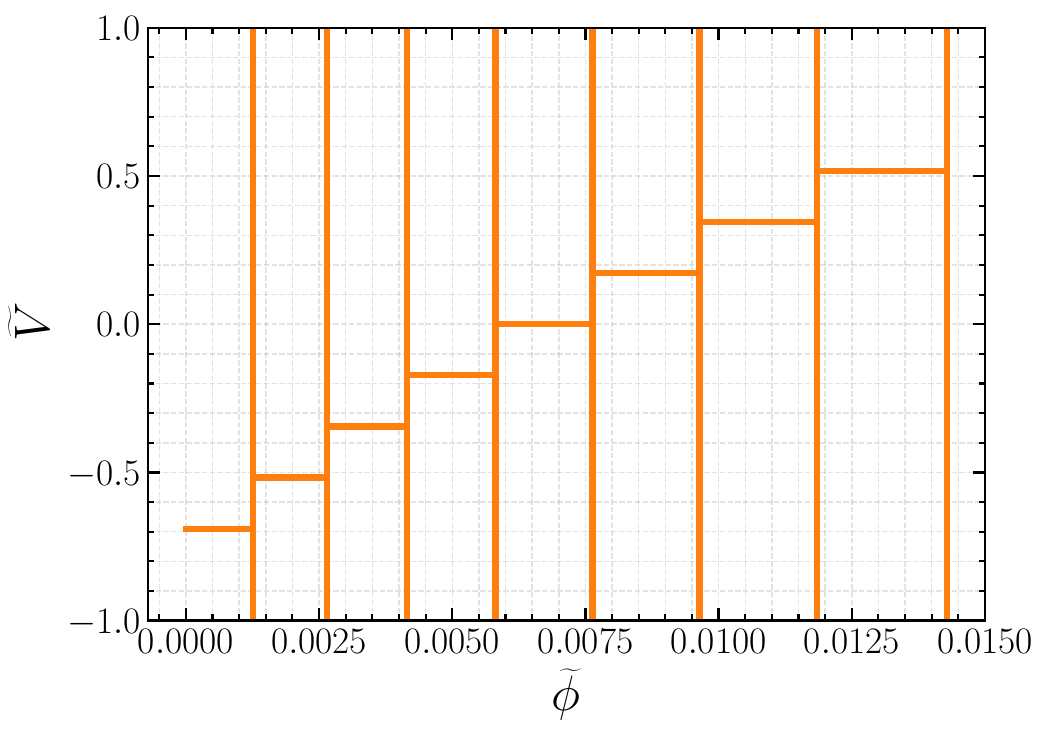}
    \caption{On-trajectory field-space reconstructions of $\widetilde V= V/\rho_{\rm c0}$ as a function of $\widetilde\phi=\phi/M_{\rm Pl}$. The left and middle panels show the ECDM and SSCDM targets, respectively. The ECDM curve is regular over the sampled interval, whereas the exact compact SSCDM target is $C^1$ but not $C^2$ at its transition endpoints. The right panel is only a finite-resolution, regulator-dependent diagnostic of the distributional L$\Lambda$CDM history: the strict Heaviside ladder does not define an ordinary regular single-field potential $V(\phi)$. Its vertical features are clipped at the displayed limits.}
    \label{fig:Vphi_background}
\end{figure*}

Figure~\ref{fig:p_background} presents the reconstructed DE pressure. The ECDM history produces a broad smooth minimum, whereas SSCDM produces a narrower feature confined to its compact transition interval and satisfies $p_{\rm de}=-\rho_{\rm de}$ on either plateau. For the ladder history, Eq.~\eqref{eq:pressure} gives the stepwise regular contribution $-\rho_{\rm de}$ together with negative Dirac impulses at the jump redshifts. The vertical features in the plotted ladder curve are therefore only a finite-resolution representation of a distribution-valued pressure rather than finite physical spikes.

Figure~\ref{fig:Vphi_background} shows the corresponding field-space targets. The horizontal origin and orientation of each regular curve depend on the field translation and reflection convention discussed below Eq.~\eqref{eq:phi_branches}; the descriptions here refer to the branch shown. For ECDM (left panel), the regular on-trajectory potential has a smooth bump-like profile, rising from negative values through zero to a broad maximum before approaching its late-time positive regime. For SSCDM (middle panel), the retained field interval likewise contains a pronounced maximum, but the exact compact reconstruction is not globally analytic: as shown in Eq.~\eqref{eq:sscdm_endpoint_scaling}, it is $C^1$ but not $C^2$ at the transition endpoints. Smooth ans\"atze can therefore approximate this target over the retained field interval but cannot reproduce its exact compact endpoint behavior. No analogous regular statement applies to the strict L$\Lambda$CDM history. Its displayed field-space curve is generated only after a finite-resolution prescription and is regulator dependent; in the Heaviside limit, no ordinary locally square-integrable field profile can have a squared velocity equal to a Dirac measure. The numerical parametric-potential comparisons below therefore apply only to the regular ECDM curve and to the retained transition-side interval of SSCDM; no such comparison is assigned to the exact Ladder.

\begin{figure}
    \centering
    \includegraphics[width=\columnwidth]{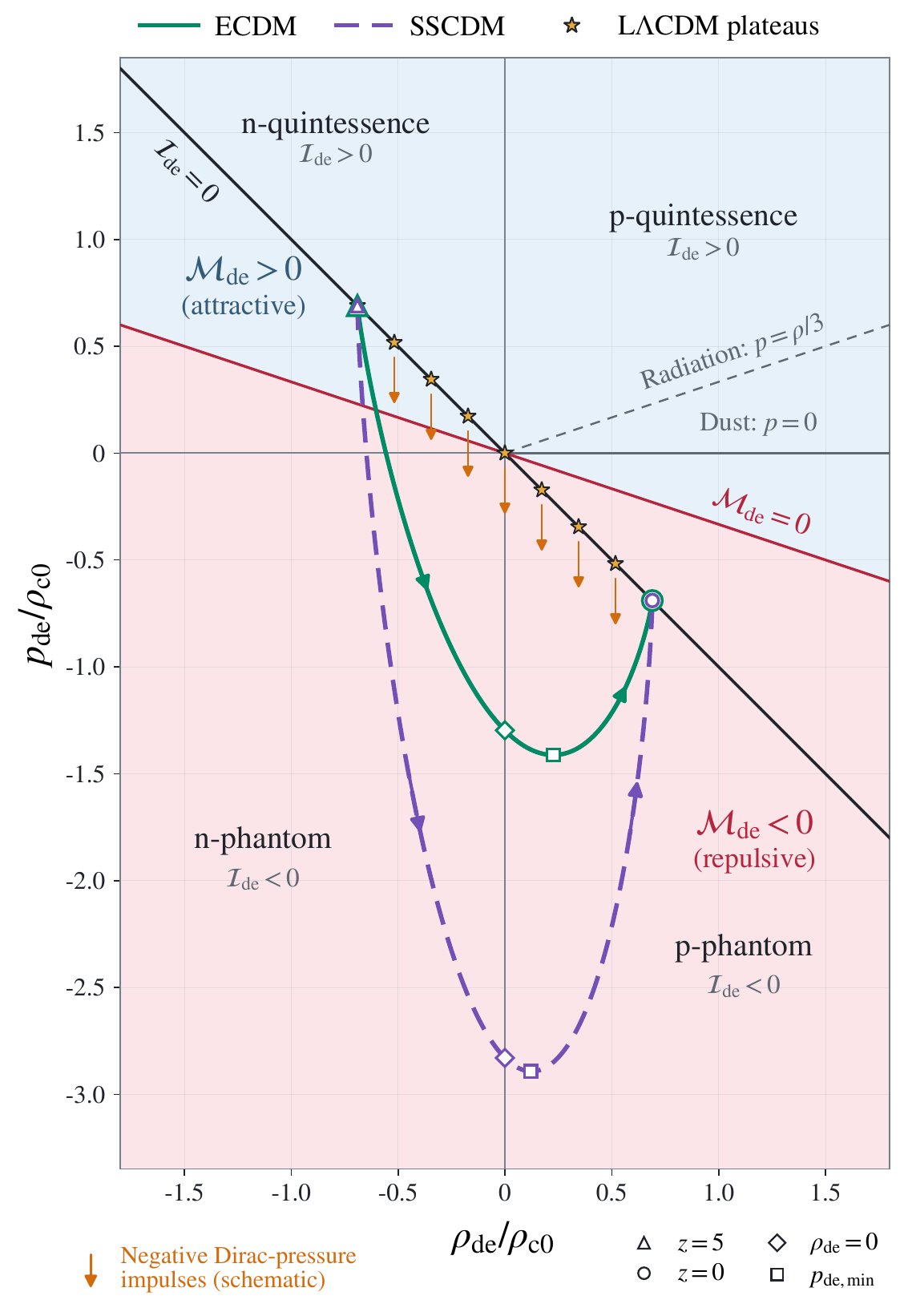}
    \caption{Pressure--density portrait for the ECDM $(z_\dagger,\eta)=(1.8,5.0)$, SSCDM $(z_\dagger,\Delta x)=(1.8,0.4)$, and L$\Lambda$CDM benchmarks. Both axes use the same scale, with ticks separated by $0.5$. Green solid and purple dashed curves show ECDM and SSCDM over $0\leq z\leq5$; colored arrowheads indicate forward evolution from triangles ($z=5$) to circles ($z=0$). Nearly coincident endpoint symbols are nested without displacing their coordinates. Diamonds mark density zeros; squares mark sampled pressure minima. The black line is the null-energy boundary $\mathcal I_{\rm de}\equiv\rho_{\rm de}+p_{\rm de}=0$, and the red line is $\mathcal M_{\rm de}\equiv\rho_{\rm de}+3p_{\rm de}=0$. Blue and pink shading indicate $\mathcal M_{\rm de}>0$ (attractive) and $\mathcal M_{\rm de}<0$ (repulsive) contributions of the DE sector, not the sign of the total cosmic acceleration. The n/p labels denote negative/positive density, while quintessence/phantom labels denote positive/negative $\mathcal I_{\rm de}$. Dust and radiation are shown only as positive-density reference equations of state, not additional DE trajectories. Stars mark Ladder plateaus; orange downward arrows schematically indicate negative Dirac-pressure impulses. Their lengths and placement are illustrative and do not define finite connecting trajectories.}
    \label{fig:nec}
\end{figure}

Figure~\ref{fig:nec} uses the same $N_z=30000$ ECDM and SSCDM density arrays as Fig.~\ref{fig:p_background}. The arrays span $100\geq z\geq0$ with $|\delta z|=100/29999$. We evaluate $p_{\rm de}/\rho_{\rm c0}=-\widetilde\Omega_{\rm de}+(1+z)\widetilde\Omega'_{\rm de}/3$ on the full grid with a second-order edge treatment and then restrict the portrait to $0\leq z\leq5$. This gives $(z_{\rm min},p_{\rm de,min}/\rho_{\rm c0})\simeq(1.637,-1.4125)$ for ECDM and $(1.760,-2.893)$ for SSCDM; at $\rho_{\rm de}=0$, the corresponding pressures are approximately $-1.2975$ and $-2.8296$. These values provide a direct grid-level cross-check of Fig.~\ref{fig:p_background}.

Figure~\ref{fig:nec} recasts the prescribed histories in the normalized pressure--density plane. Irrespective of the finite plotting resolution, the defining equations imply that, for ECDM and SSCDM, ${\rm d}\rho_{\rm de}/{\rm d}z<0$ wherever the density evolves and hence $\mathcal I_{\rm de}=(1+z)\rho'_{\rm de}/3<0$. Read forward in cosmic time, the curves leave or approach the negative-density NEC boundary, enter the n-phantom region, cross $\mathcal M_{\rm de}=0$ while $\rho_{\rm de}<0$, cross $\rho_{\rm de}=0$ at finite negative pressure, continue through the positive-density phantom sector, and reach or approach the positive-density NEC boundary. SSCDM lies exactly on $\mathcal I_{\rm de}=0$ on its two constant plateaus, whereas ECDM approaches it as its derivative decays away from the transition. The strict L$\Lambda$CDM plateaus also lie on $\mathcal I_{\rm de}=0$, but its jumps carry Dirac impulses in $p_{\rm de}$ and do not define ordinary connecting curves in this plane. The orange downward arrows are schematic indicators of negative distributional pressure impulses; their drawn lengths and placement do not define finite connecting trajectories. The blue/pink shading tracks the sign of $\mathcal M_{\rm de}$, independently of the $\rho_{\rm de}$--$\mathcal I_{\rm de}$ sector labels. The positive-density loci $p=0$ and $p=\rho/3$ are included only as dust and radiation reference equations of state, not as additional DE trajectories. Crossing the red $\mathcal M_{\rm de}=0$ line changes the sign of the DE sector's contribution to Raychaudhuri focusing; the sign of the total acceleration depends on all cosmic components.

This hierarchy is the principal theoretical lesson of the reconstruction. The apparent smoothness or sharpness of $\rho_{\rm de}(z)$ is not by itself a criterion for scalar-field regularity: eliminating redshift can turn a finite-order compact interpolation into a non-Lipschitz field-space force, whereas a discontinuous fluid limit can leave the ordinary scalar configuration space altogether. A scalar interpretation should therefore be screened in the following order: use stress-tensor combinations that remain regular at $\rho_{\rm de}=0$, impose the fixed-sign condition in Eq.~\eqref{eq:single_field_condition}, and then test the invertibility and endpoint regularity of the reconstructed $V(\phi)$. Only after these steps is it meaningful to approximate the on-shell curve by a chosen parametric family and study its forward dynamics.

%%-----------------------------------------------------------------------------%%
\subsection{Conditional potential-space template comparison}
%%-----------------------------------------------------------------------------%%

%%-----------------------------------------------------------------------------%%
\subsubsection{ECDM target}
%%-----------------------------------------------------------------------------%%

\begin{figure*}
\centering

% ---------------- Row 1 ----------------
\begin{minipage}{0.32\linewidth}
    \centering
    \includegraphics[width=\linewidth]{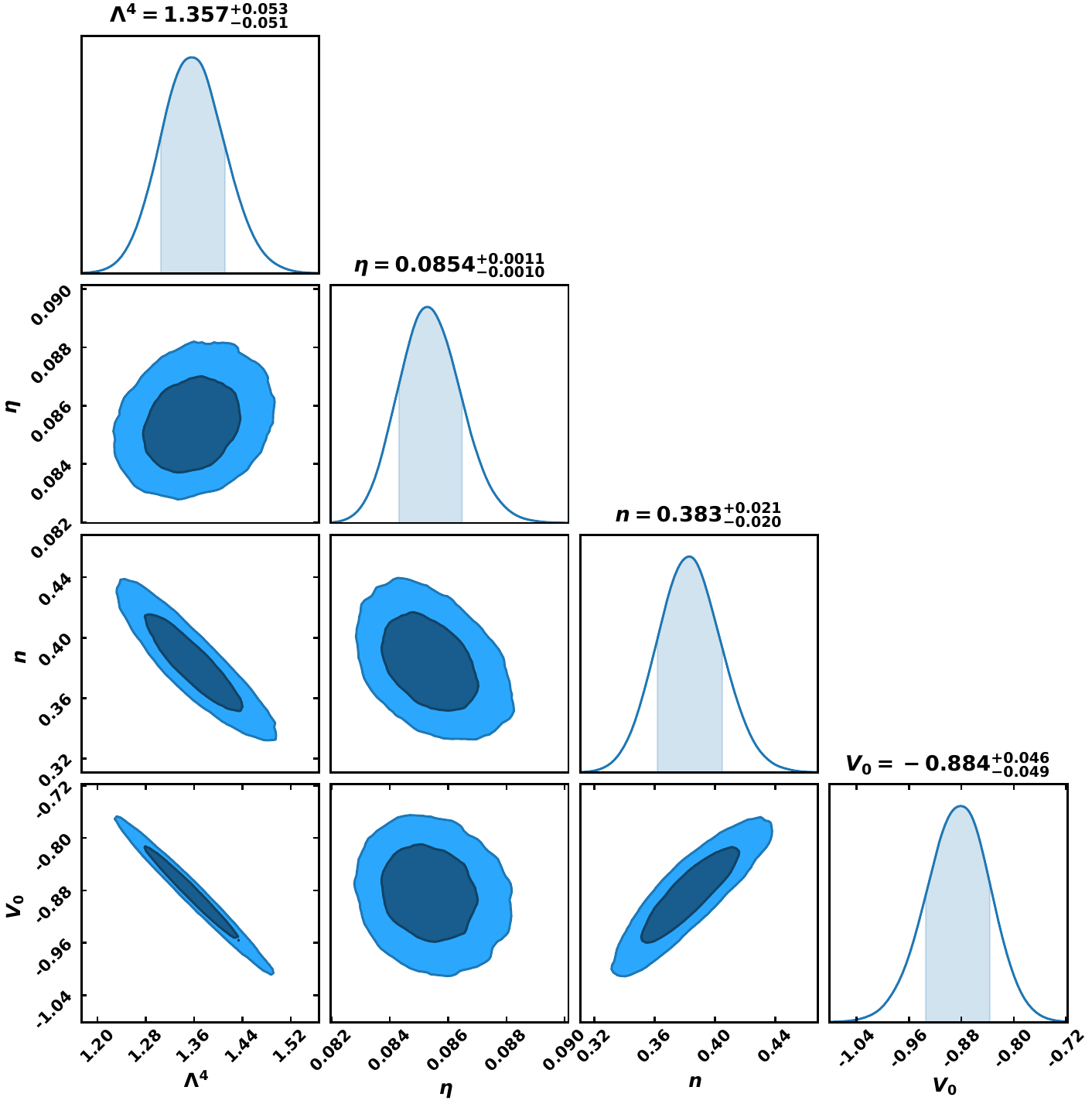}
    
    \vspace{2pt}
    (a) Generalized axion-like
\end{minipage}
\hfill
\begin{minipage}{0.32\linewidth}
    \centering
    \includegraphics[width=\linewidth]{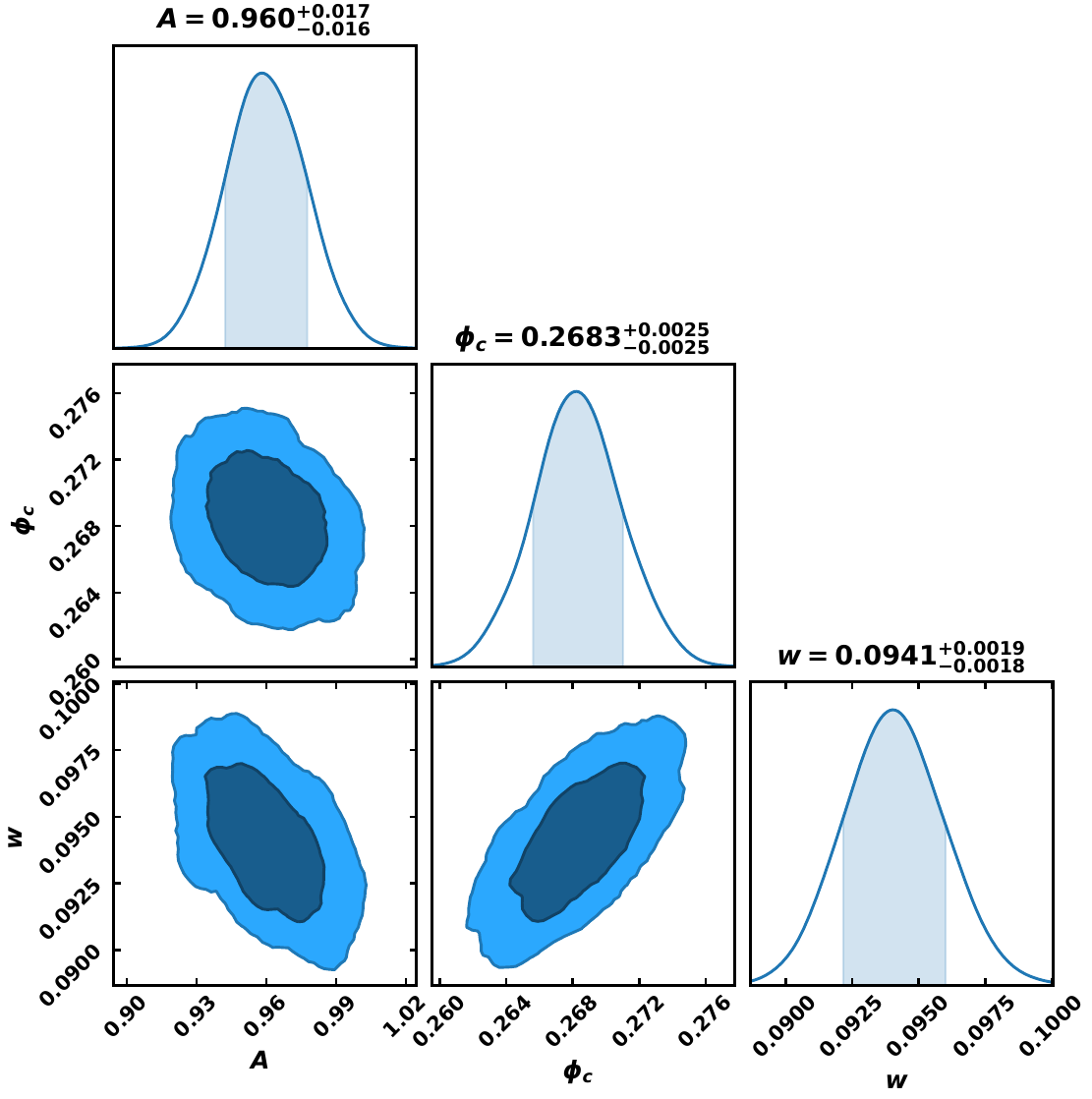}
    
    \vspace{2pt}
    (b) Gaussian feature
\end{minipage}
\hfill
\begin{minipage}{0.32\linewidth}
    \centering
    \includegraphics[width=\linewidth]{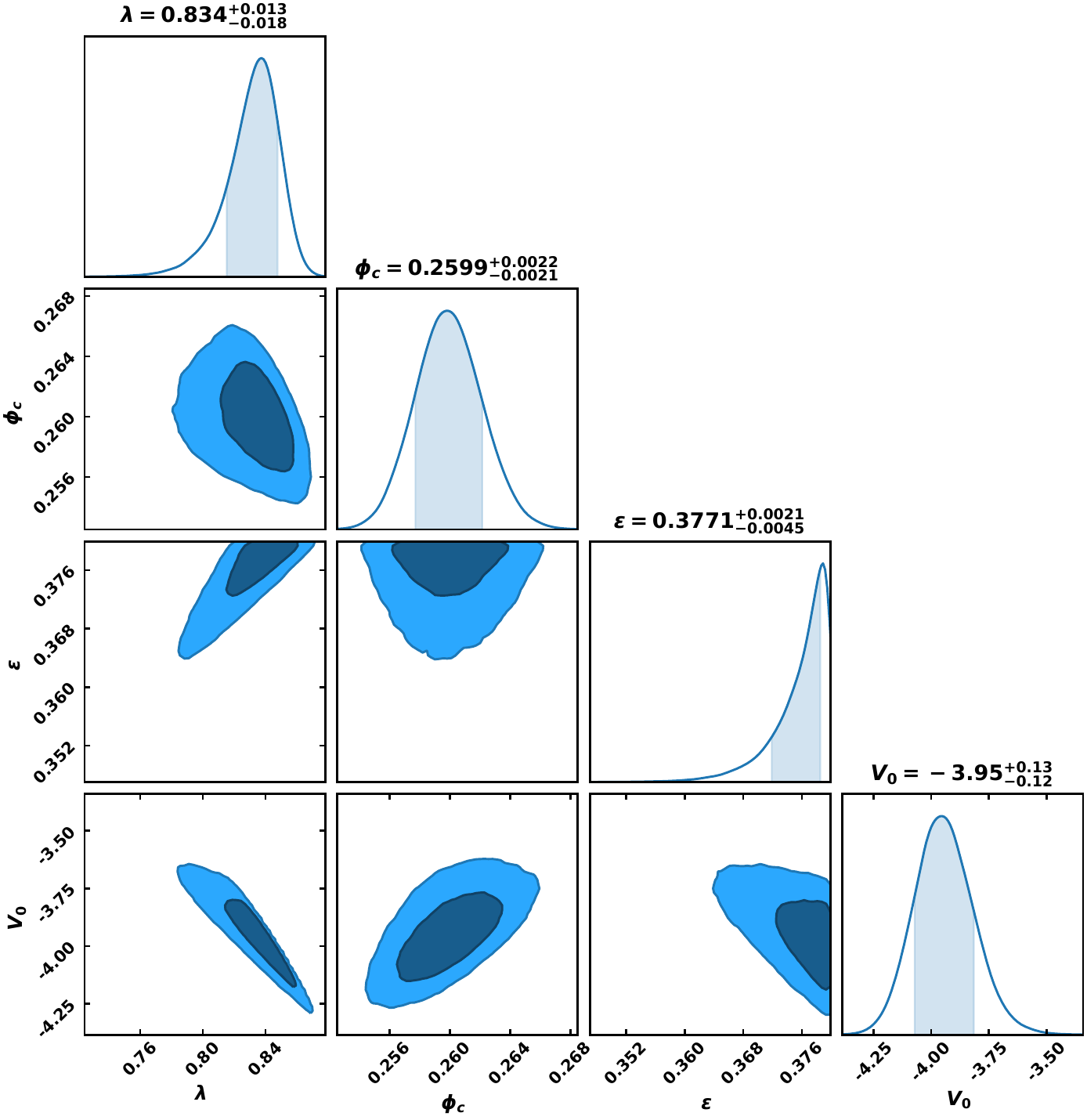}
    
    \vspace{2pt}
    (c) Regularized inverse-quadratic
\end{minipage}

\vspace{0.6cm}

% ---------------- Row 2 ----------------
\begin{minipage}{0.46\linewidth}
    \centering
    \includegraphics[width=\linewidth]{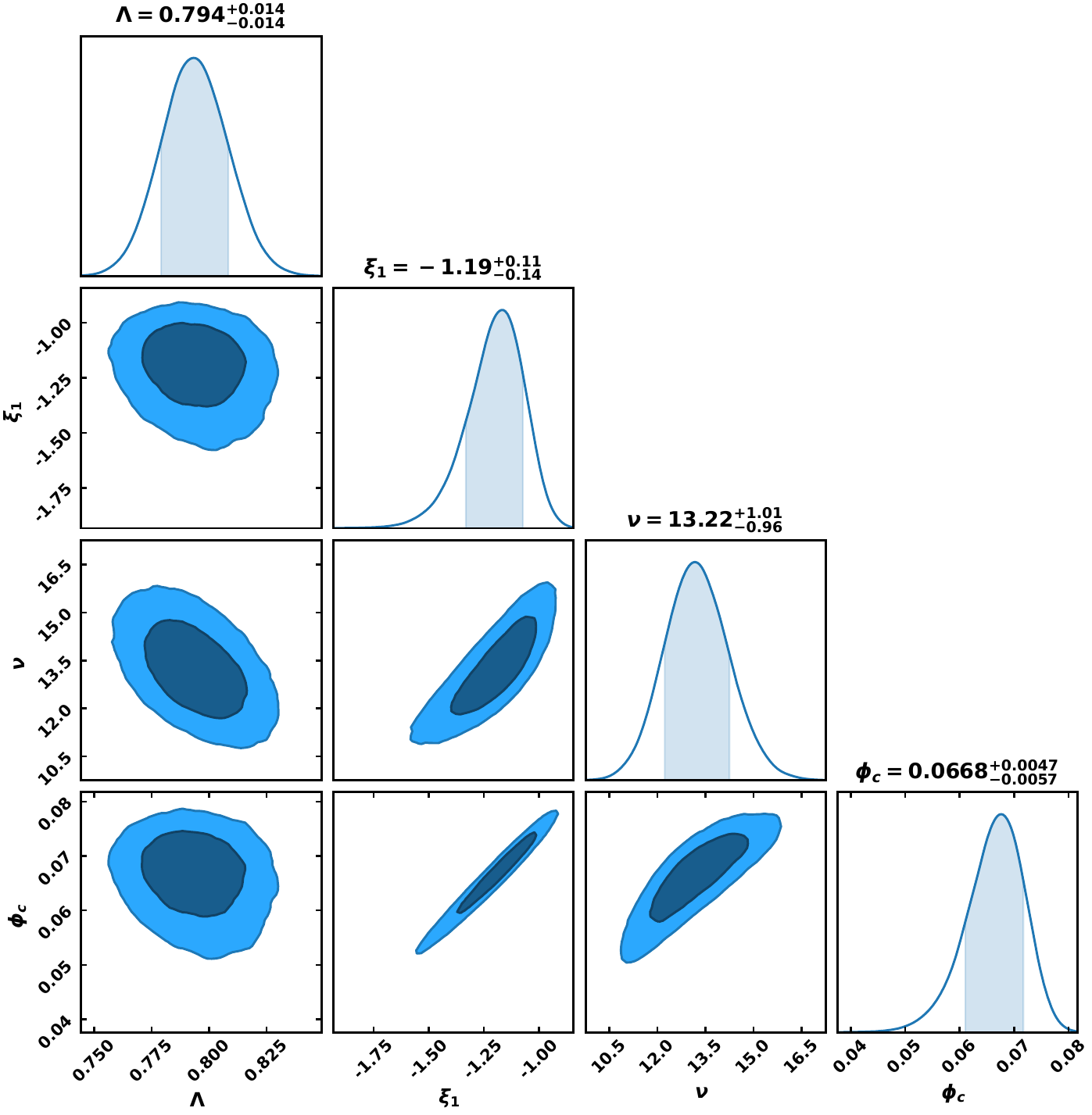}
    
    \vspace{2pt}
    (d) Shifted-$\tanh$
\end{minipage}
\hfill
\begin{minipage}{0.46\linewidth}
    \centering
    \includegraphics[width=\linewidth]{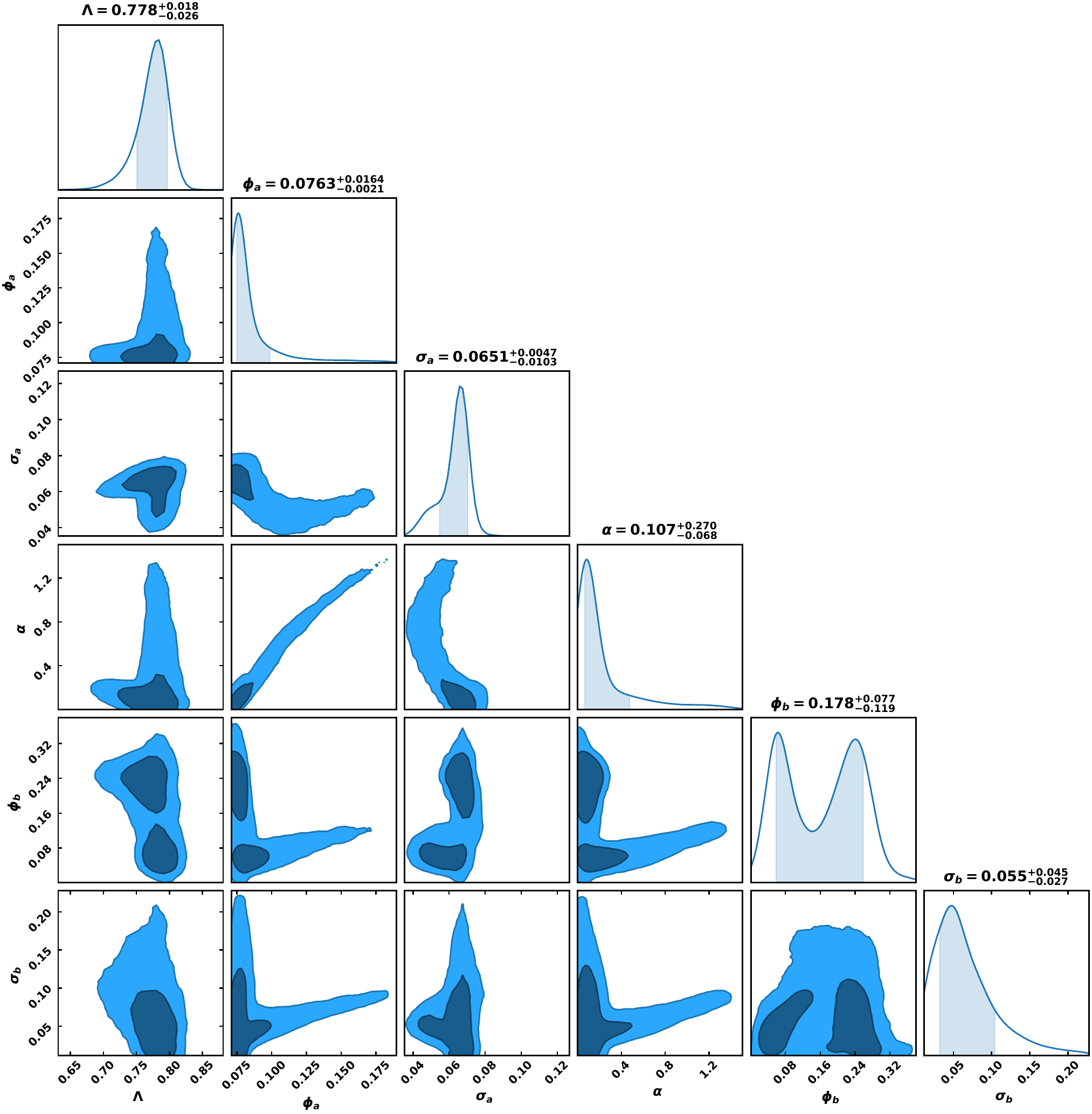}
    
    \vspace{2pt}
    (e) Sigmoid--Gaussian feature
\end{minipage}

\caption{Marginal posterior distributions for the five parametric potential families fitted to the fixed synthetic ECDM potential-space dataset. These are template-fit posteriors under the synthetic experiment of Sec.~\ref{subsec:bayes}, not observational constraints. The contours show the weighted, smoothed posterior densities and enclose the $68\%$ and $95\%$ marginal credible regions. The diagonal annotations report direct weighted medians and weighted equal-tail $16$th/$84$th percentiles from the same final posterior samples, with the saved weights applied once. Potential amplitudes and offsets are expressed in units of $\rho_{\rm c0}$, while field locations and widths refer to the dimensionless coordinate $\widetilde\phi=\phi/M_{\rm Pl}$.}
\label{fig:corner_ecdm_all}
\end{figure*}

We first report the potential-space fits to the ECDM target. The marginal posteriors are shown in Fig.~\ref{fig:corner_ecdm_all}. A narrow posterior does not by itself establish that a template adequately represents the target: a misspecified family can have a precisely determined best approximation while retaining large, structured residuals.

% ============================
%   ECDM: Corner plots
% ============================

The ECDM evidence entries, parameter summaries, corner panels, and posterior bands use one common final mock and configuration, together with the corresponding final model-specific run for each of the five potential families.

For the generalized axion-like template, the weighted marginal summaries are
\begin{equation*}
\begin{aligned}
\frac{\Lambda^4}{\rho_{\rm c0}}&=1.357^{+0.053}_{-0.051},&
\eta&=0.0854^{+0.0011}_{-0.0010},\\
n&=0.383^{+0.021}_{-0.020},&
\frac{V_0}{\rho_{\rm c0}}&=-0.884^{+0.046}_{-0.049}.
\end{aligned}
\end{equation*}
Figure~\ref{fig:corner_ecdm_all}(a) shows the posterior correlations among the amplitude, field-space scale, power, and offset. These relations reflect the dependence of the extrema on the combination $V_0+2^n\Lambda^4$, while $\eta$ primarily controls their positions in field space. Here the $n<1/2$ regularity problem is not confined to an unvisited part of the global periodic potential. The zero-phase convention and Eq.~\eqref{eq:field_orientation_fit} place the high-redshift fitting endpoint at $\widetilde\phi=0$, and the interpolation grid includes that endpoint. It is a periodic minimum of Eq.~\eqref{eq:axion_template}. With the fitted median $n=0.383$, the force behaves as
\begin{equation}
\frac{{\rm d}\widetilde V}{{\rm d}\widetilde\phi}\propto{\rm sgn}(\widetilde\phi)|\widetilde\phi|^{2n-1}\simeq{\rm sgn}(\widetilde\phi)|\widetilde\phi|^{-0.234},
\end{equation}
and therefore diverges at the included high-redshift endpoint. The function-value likelihood remains mathematically defined, but this fitted member is not an ordinary differentiable Klein--Gordon potential on the closed fitting interval and cannot reproduce the regular ECDM endpoint force. Its score measures only the representation of the sampled potential values. A regular dynamical comparison would require a consistently treated phase or field translation, exclusion of the singular endpoint, or a regularity prior on $n$. The $n=1$ axion trajectories evolved in Sec.~\ref{subsec:append_axion} are distinct regular members of the same broad family, not an evolution of this fitted potential. Here $\eta$ is the axion field-space scale and should not be confused with the ECDM transition-sharpness parameter in Eq.~\eqref{eq:pheno_model}.

% ============================
%   ECDM: Predictive bands
% ============================
\begin{figure*}
\centering

% ---------------- Row 1 ----------------
\begin{minipage}{0.43\linewidth}
    \centering
    \includegraphics[width=\linewidth]{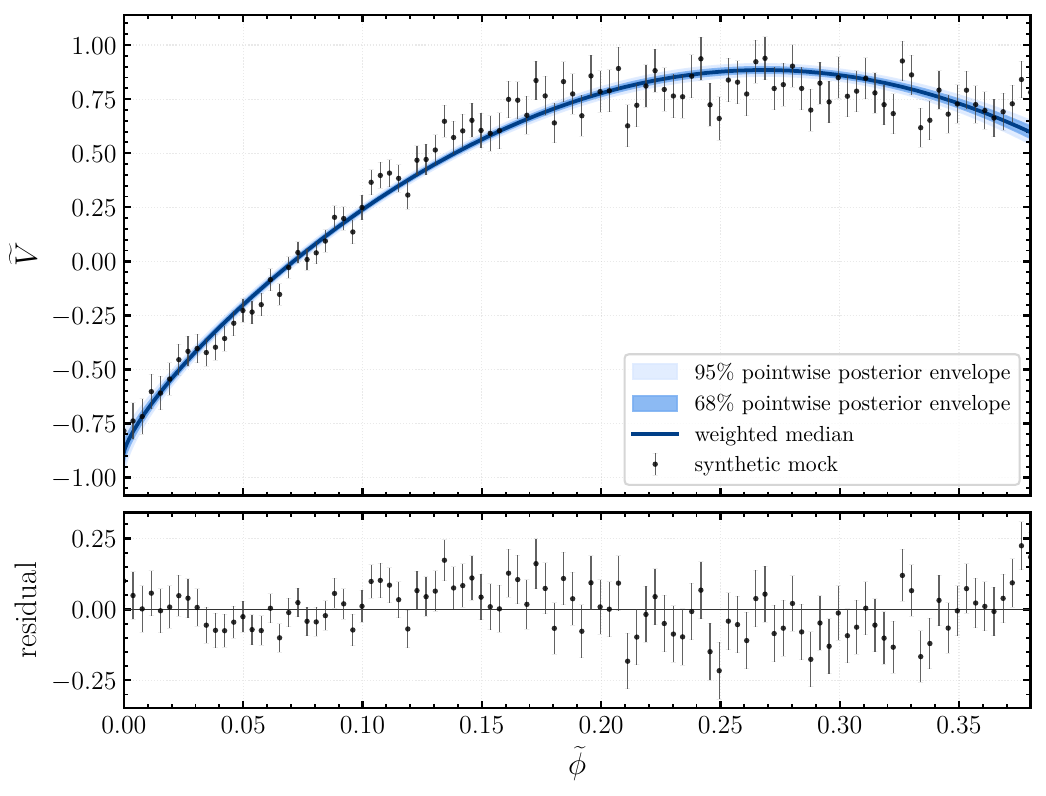}
    
    \vspace{2pt}
    (a) Generalized axion-like
\end{minipage}
\hfill
\begin{minipage}{0.43\linewidth}
    \centering
    \includegraphics[width=\linewidth]{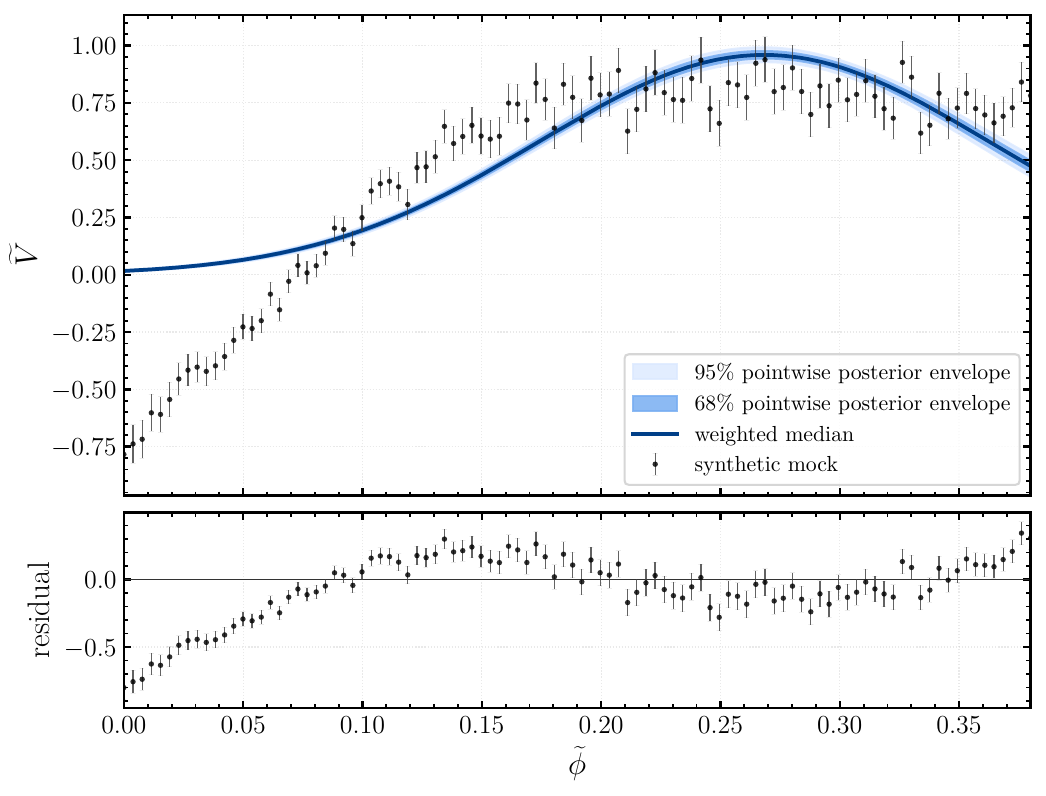}
    
    \vspace{2pt}
    (b) Gaussian feature
\end{minipage}

\vspace{0.2cm}

% ---------------- Row 2 ----------------
\begin{minipage}{0.43\linewidth}
    \centering
    \includegraphics[width=\linewidth]{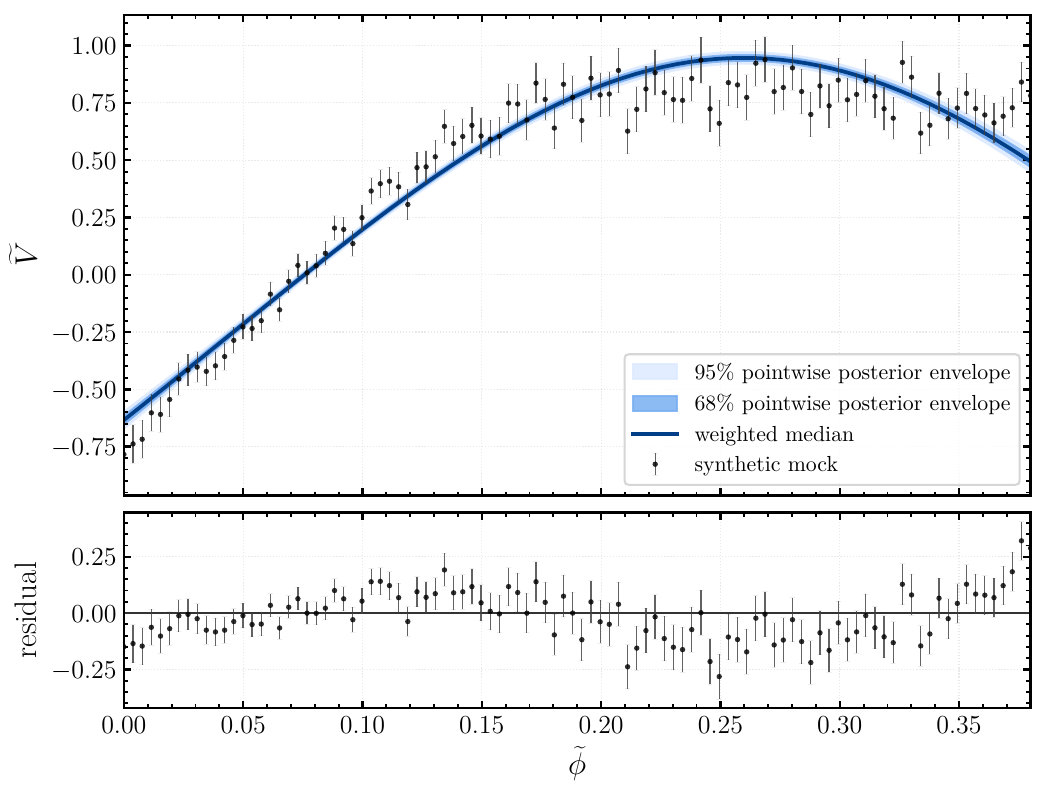}
    
    \vspace{2pt}
    (c) Regularized inverse-quadratic
\end{minipage}
\hfill
\begin{minipage}{0.43\linewidth}
    \centering
    \includegraphics[width=\linewidth]{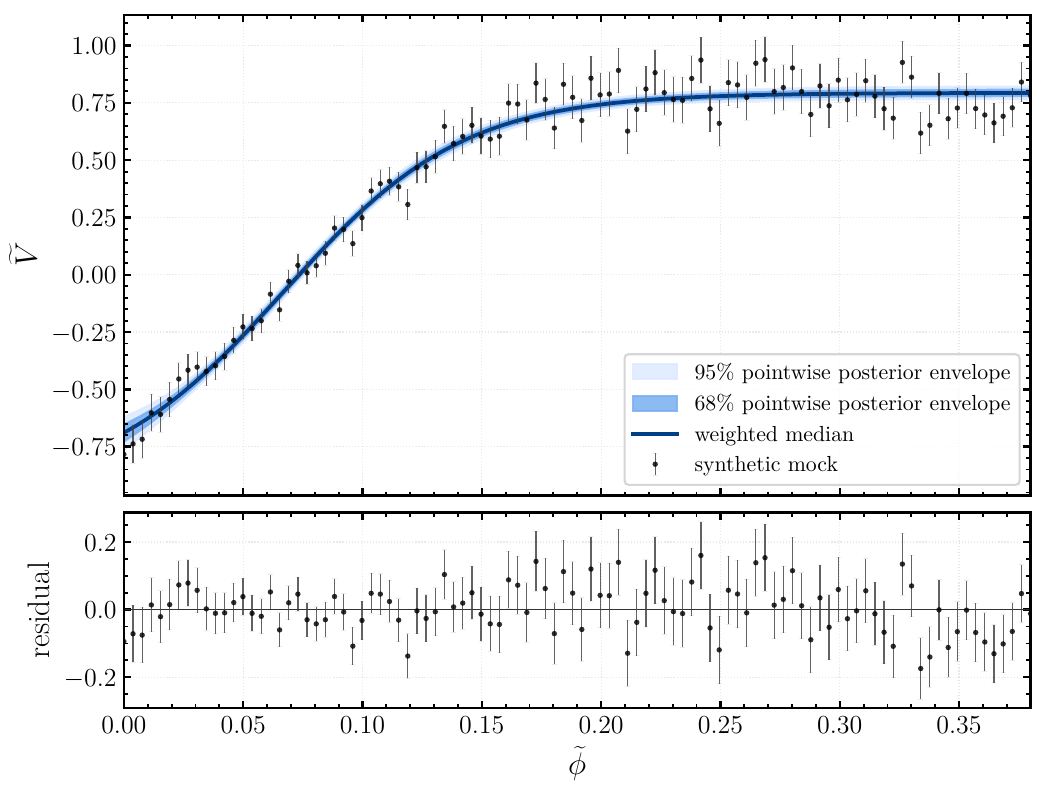}
    
    \vspace{2pt}
    (d) Shifted-$\tanh$
\end{minipage}

\vspace{0.2cm}

% ---------------- Row 3 ----------------
\begin{minipage}{0.43\linewidth}
    \centering
    \includegraphics[width=\linewidth]{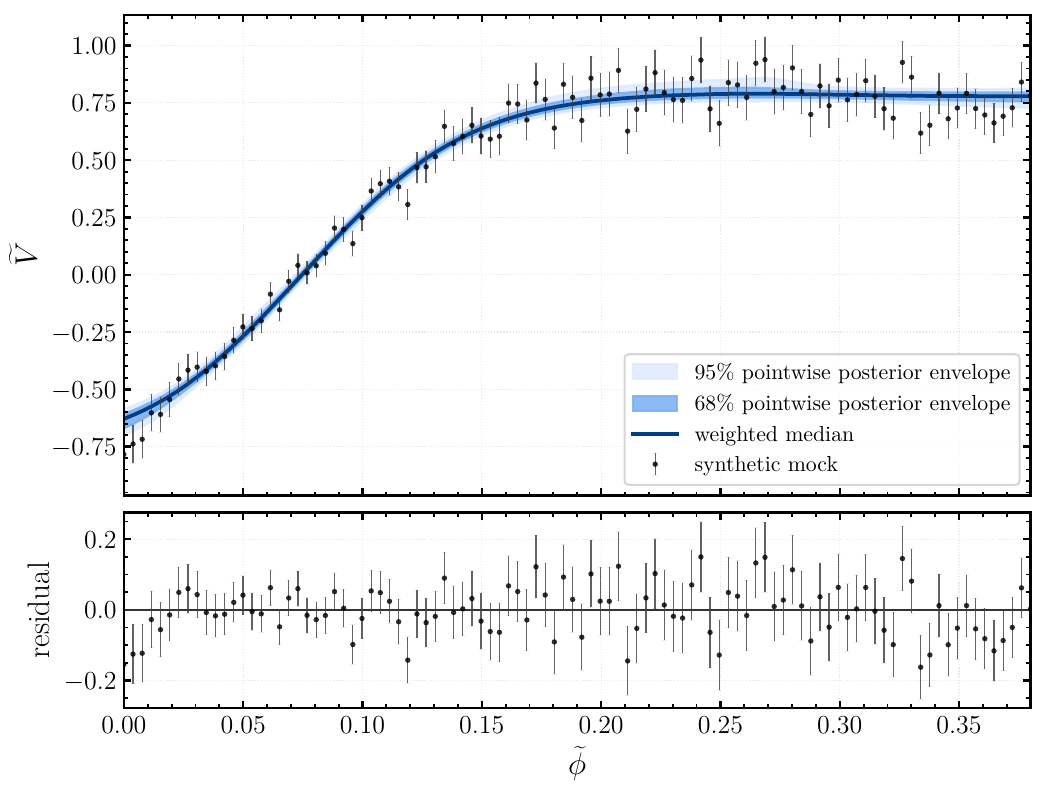}
    
    \vspace{2pt}
    (e) Sigmoid--Gaussian feature
\end{minipage}

\caption{Pointwise ECDM curve summaries for the five potential templates fitted to the fixed synthetic potential-space experiment. Black markers show the synthetic mock values $\widetilde V_i^{\rm(mock)}$, with design uncertainties $\sigma_i$ defined by Eq.~\eqref{eq:mock_variance}. Solid curves show the direct pointwise weighted medians of the deterministic templates, and the dark- and light-shaded regions show the corresponding direct weighted equal-tail $68\%$ and $95\%$ posterior envelopes. The saved posterior weights are applied exactly once; no posterior resampling or display smoothing is used. The envelopes describe parameter-induced spread within the synthetic experiment, not uncertainty on the target history. The lower panels show the synthetic-mock-minus-median residuals.}
\label{fig:band_ecdm}
\end{figure*}

For the Gaussian-feature template, we obtain
\begin{equation*}
\begin{aligned}
\frac{A}{\rho_{\rm c0}}&=0.960^{+0.017}_{-0.016},&
\widetilde\phi_{\rm c}&=(268.3^{+2.5}_{-2.5})\times10^{-3},&\\
w&=(94.1^{+1.9}_{-1.8})\times10^{-3}.
\end{aligned}
\end{equation*}
The likelihood localizes the center and width of the best Gaussian approximation precisely. The posterior correlations reflect compensating adjustments of the feature height, position, and curvature. These narrow marginal intervals do not imply an adequate global fit: because this template has no offset and cannot take both signs, Fig.~\ref{fig:band_ecdm} exhibits large systematic residuals over the negative-potential part of the target.

For the regularized inverse-quadratic template, we find
\begin{equation*}
\begin{aligned}
\frac{\lambda}{\sqrt{\rho_{\rm c0}}}&=0.834^{+0.013}_{-0.018},&
\widetilde\phi_{\rm c}&=(259.9^{+2.2}_{-2.1})\times10^{-3},\\
\epsilon&=(377.1^{+2.1}_{-4.5})\times10^{-3},&
\frac{V_0}{\rho_{\rm c0}}&=-3.95^{+0.13}_{-0.12}.
\end{aligned}
\end{equation*}
The correlation between $\lambda$ and $\epsilon$ reflects the dependence of the peak height on $\lambda^2/\epsilon^2$, while the offset compensates for changes in the overall potential level. The fitted $\epsilon\simeq0.377$ lies close to its imposed upper limit $\epsilon_{\max}=\widetilde\phi_{\rm sc}=0.38$, and its marginal is correspondingly prior truncated. The inverse-quadratic evidence is therefore conditional on this bound.

The shifted-$\tanh$ fit yields
\begin{equation*}
\begin{aligned}
\frac{\Lambda}{\rho_{\rm c0}}&=0.794^{+0.014}_{-0.014},&
\xi_1&=-1.19^{+0.11}_{-0.14},\\
\nu&=13.22^{+1.01}_{-0.96},&
\widetilde\phi_{\rm c}&=(66.8^{+4.7}_{-5.7})\times10^{-3}.
\end{aligned}
\end{equation*}
The posterior correlations couple the plateau ratio, transition sharpness, and transition position, as expected because the central slope is proportional to $\Lambda\nu(1-\xi_1)/2$. The template captures the principal monotonic transition but cannot reproduce the subsequent broad decline of the ECDM target exactly.

For the sigmoid--Gaussian template, we obtain
\begin{equation*}
\begin{aligned}
\frac{\Lambda}{\rho_{\rm c0}}&=0.778^{+0.018}_{-0.026},&
\widetilde\phi_{\rm a}&=(76.3^{+16.4}_{-2.1})\times10^{-3},\\
\sigma_{\rm a}&=(65.1^{+4.7}_{-10.3})\times10^{-3},&
\alpha&=0.107^{+0.270}_{-0.068},\\
\widetilde\phi_{\rm b}&=(178^{+77}_{-119})\times10^{-3},&
\sigma_{\rm b}&=(55^{+45}_{-27})\times10^{-3}.
\end{aligned}
\end{equation*}
The posterior is broad, strongly non-Gaussian, and substantially degenerate. Although $\Lambda/\rho_{\rm c0}$ remains comparatively localized, the transition parameters $(\widetilde\phi_{\rm a},\sigma_{\rm a})$ possess asymmetric tails and are correlated with the localized-feature parameters. The posterior for $\widetilde\phi_{\rm b}$ is visibly multimodal, while $\alpha$ and $\sigma_{\rm b}$ have broad, skewed distributions extending over a substantial fraction of their allowed ranges. The curved joint contours show that different combinations of feature amplitude, position, and width can reproduce similar potential values over the sampled field interval. The ECDM target nevertheless gives this family the highest evidence among the five templates: it represents the target curve most efficiently without pinning down its individual feature parameters.

The posterior widths and shapes in Fig.~\ref{fig:corner_ecdm_all} should be interpreted together with the residuals in Fig.~\ref{fig:band_ecdm}. The sigmoid--Gaussian template provides the most efficient overall representation of the synthetic target, although its localized-feature parameters remain broad, non-Gaussian, and substantially degenerate. The shifted-$\tanh$ template reproduces the principal transition and gives the second-highest evidence, but it retains structured residuals around the maximum and at large $\widetilde\phi$, where the target descends toward its plateau from above while the template saturates from below [cf. Eq.~\eqref{eq:endpoint_sign}]. The generalized axion-like template also follows the main target structure but is penalized relative to the first two families and has a singular field derivative at the included zero-phase endpoint. The regularized inverse-quadratic form exhibits systematic endpoint deviations, while the zero-offset Gaussian fails over the negative-potential portion of the target; its comparatively narrow posterior only identifies the best member of a structurally inadequate three-parameter family.

Table~\ref{tab:model_comparison_combined} gives the corresponding coordinate-fixed representational scores. The reported ECDM experiment ranks the sigmoid--Gaussian family first, followed by the shifted-$\tanh$ family with $\Delta\log\mathcal Z=-4.754\pm0.150$. The generalized axion-like, regularized inverse-quadratic, and Gaussian families give $\Delta\log\mathcal Z=-18.250\pm0.166$, $-58.043\pm0.164$, and $-539.253\pm0.259$, respectively. These differences are resolved relative to the sampler-reported numerical errors, although their robustness under changes of mock realization, field convention, point grid, weighting, and prior volume has not been tested. The axion score concerns sampled function values despite the singular gradient at the included zero-phase endpoint, and the poor Gaussian score reflects the inability of a zero-offset Gaussian to represent both signs of the target rather than evidence against localized features as a class.

% ============================
%   SSCDM: Corner plots
% ============================
\begin{figure*}
\centering

% ---------------- Row 1 ----------------
\begin{minipage}{0.32\linewidth}
    \centering
    \includegraphics[width=\linewidth]{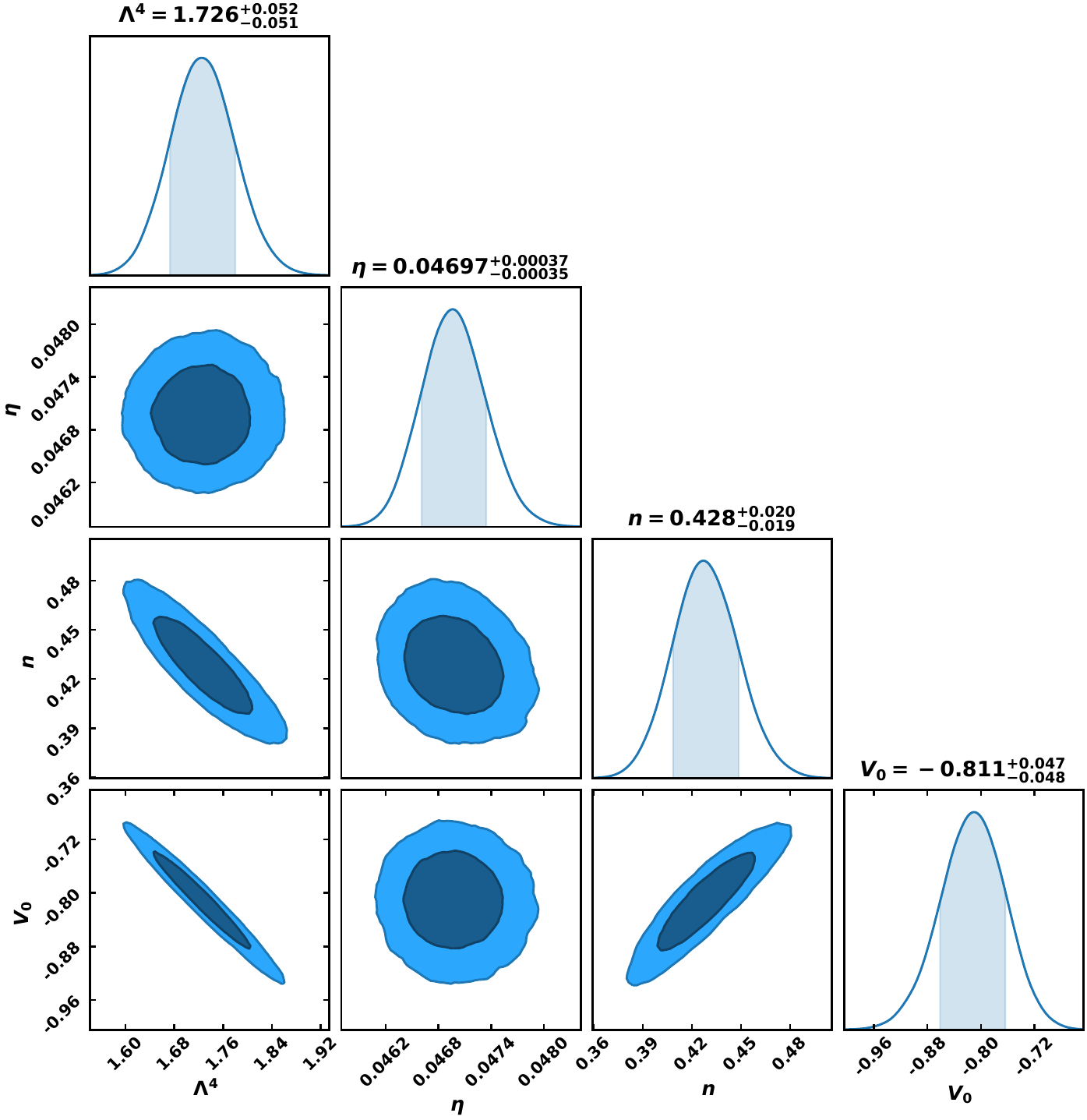}
    
    \vspace{2pt}
    (a) Generalized axion-like
\end{minipage}
\hfill
\begin{minipage}{0.32\linewidth}
    \centering
    \includegraphics[width=\linewidth]{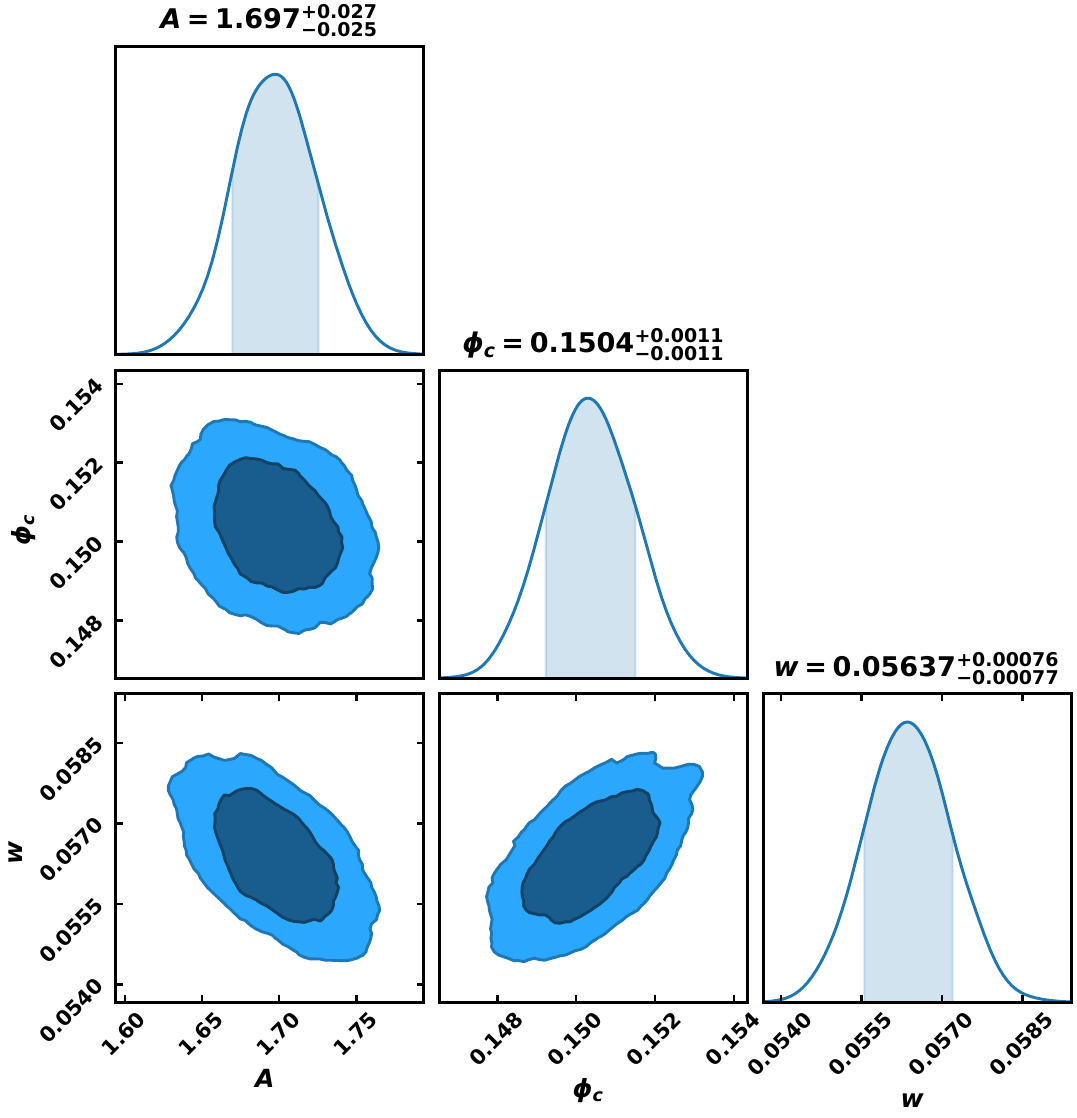}
    
    \vspace{2pt}
    (b) Gaussian feature
\end{minipage}
\hfill
\begin{minipage}{0.32\linewidth}
    \centering
    \includegraphics[width=\linewidth]{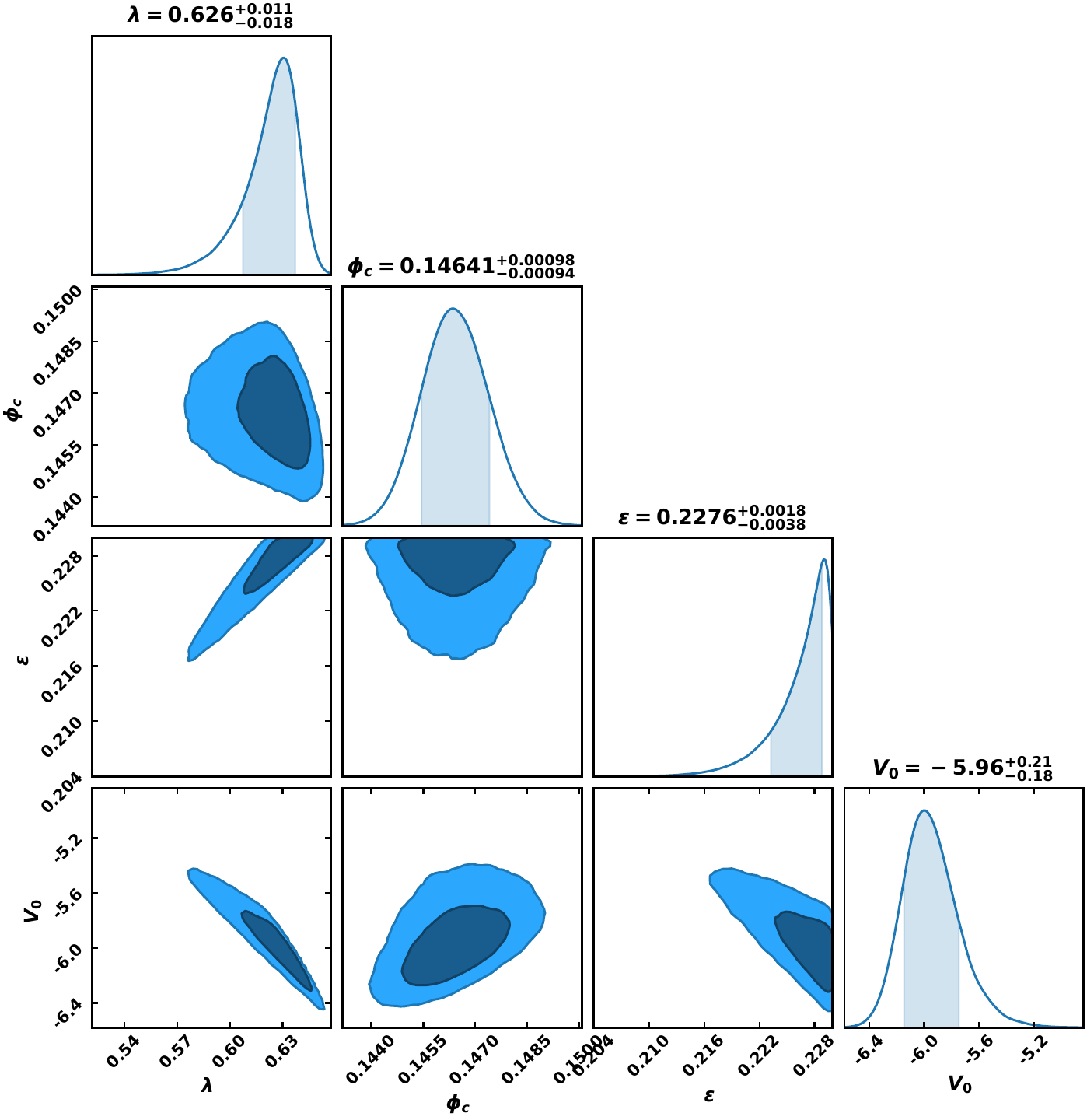}
    
    \vspace{2pt}
    (c) Regularized inverse-quadratic
\end{minipage}

\vspace{0.6cm}

% ---------------- Row 2 ----------------
\begin{minipage}{0.46\linewidth}
    \centering
    \includegraphics[width=\linewidth]{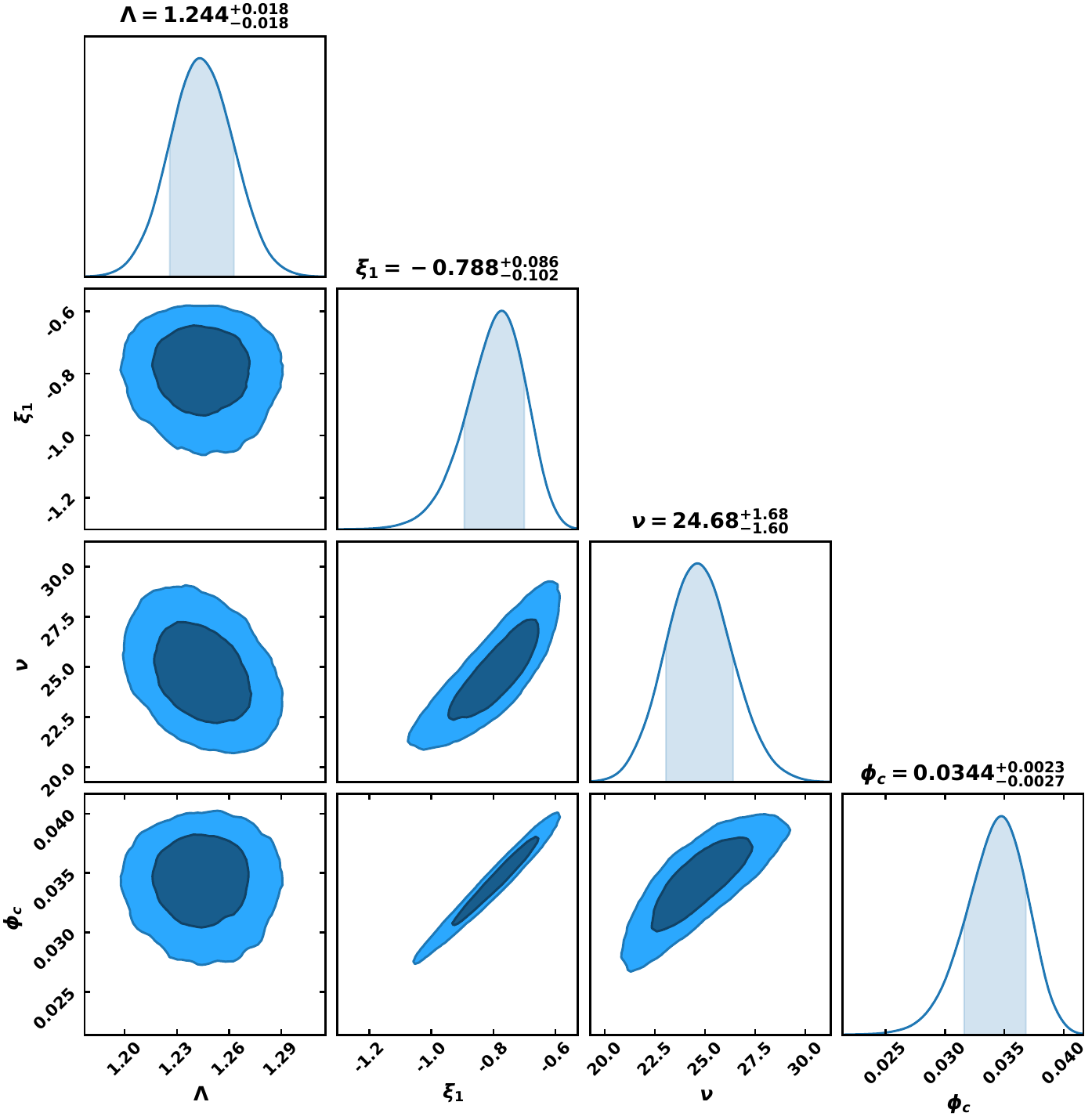}
    
    \vspace{2pt}
    (d) Shifted-$\tanh$
\end{minipage}
\hfill
\begin{minipage}{0.46\linewidth}
    \centering
    \includegraphics[width=\linewidth]{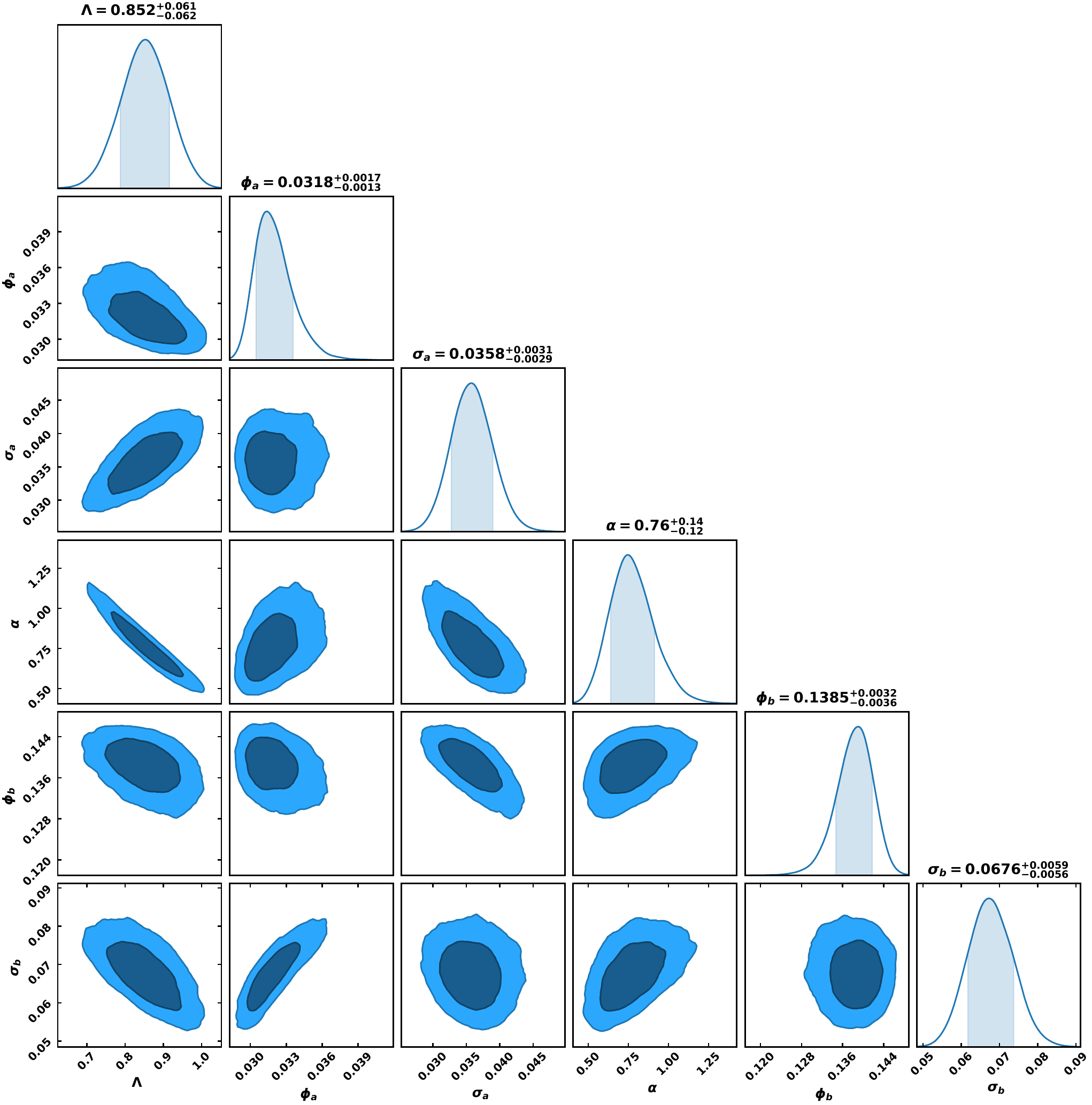}
    
    \vspace{2pt}
    (e) Sigmoid--Gaussian feature
\end{minipage}

\caption{Marginal posterior distributions for the five potential templates fitted to the SSCDM target under the adopted synthetic potential-space likelihood. Contours enclose the $68\%$ and $95\%$ posterior credible regions. The diagonal annotations report direct weighted medians and weighted equal-tail $16$th/$84$th percentiles from the same final posterior samples, with the saved weights applied once. The plotted field coordinate is the dimensionless ratio $\widetilde\phi=\phi/M_{\rm Pl}$, and potential amplitudes are in units of $\rho_{\rm c0}$. These are conditional template-fit posteriors, not observational cosmological constraints. In the sigmoid--Gaussian panel the amplitude parameter is interior to its imposed range, with $\alpha=0.76^{+0.14}_{-0.12}$, rather than prior-boundary limited.}
\label{fig:corner_sscdm_all}
\end{figure*}

% ============================
%   SSCDM: Predictive bands
% ============================
\begin{figure*}
\centering

% ---------------- Row 1 ----------------
\begin{minipage}{0.43\linewidth}
    \centering
    \includegraphics[width=\linewidth]{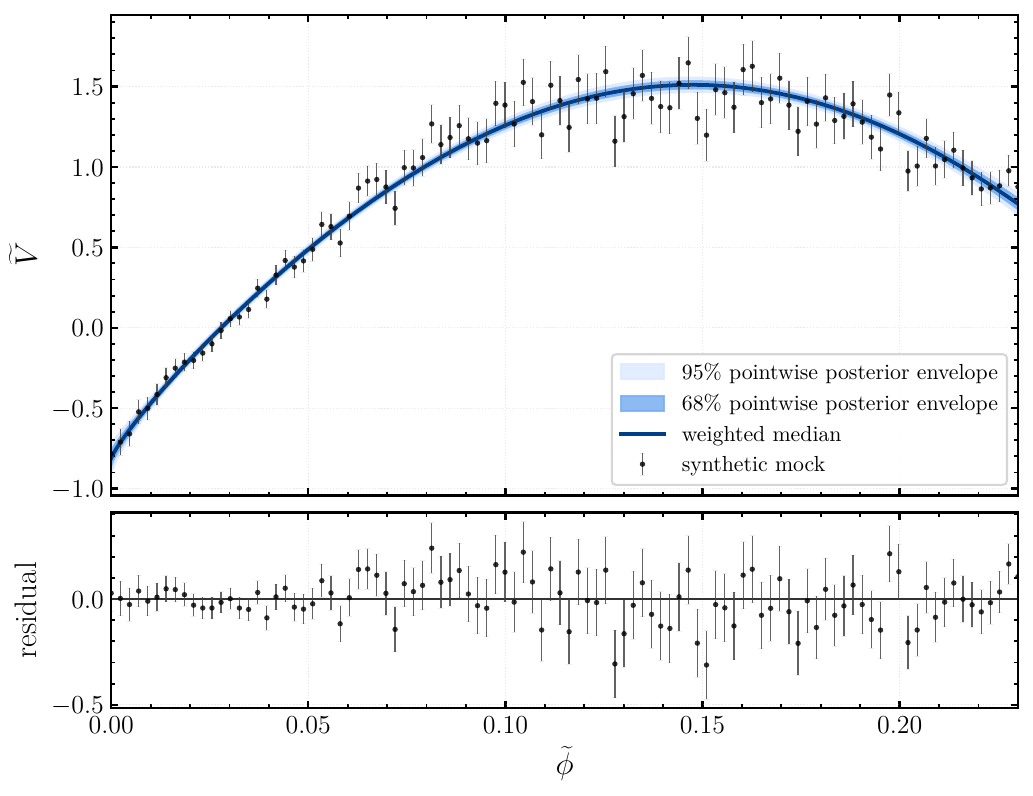}
    
    \vspace{2pt}
    (a) Generalized axion-like
\end{minipage}
\hfill
\begin{minipage}{0.43\linewidth}
    \centering
    \includegraphics[width=\linewidth]{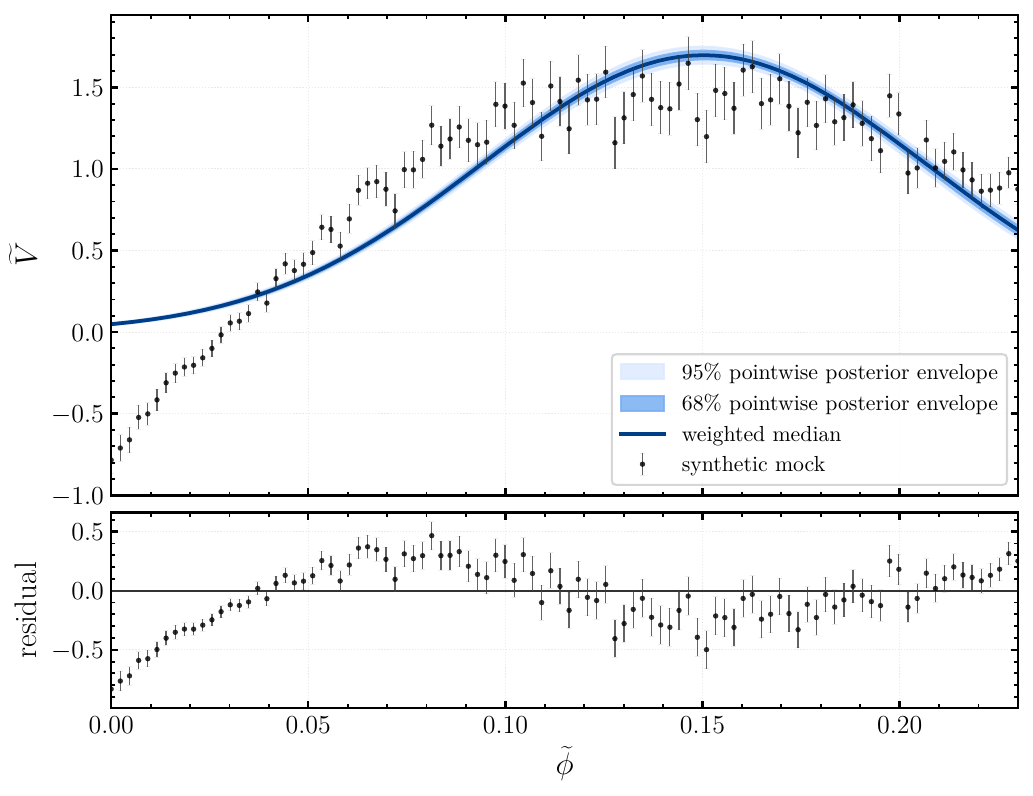}
    
    \vspace{2pt}
    (b) Gaussian feature
\end{minipage}

\vspace{0.15cm}

% ---------------- Row 2 ----------------
\begin{minipage}{0.43\linewidth}
    \centering
    \includegraphics[width=\linewidth]{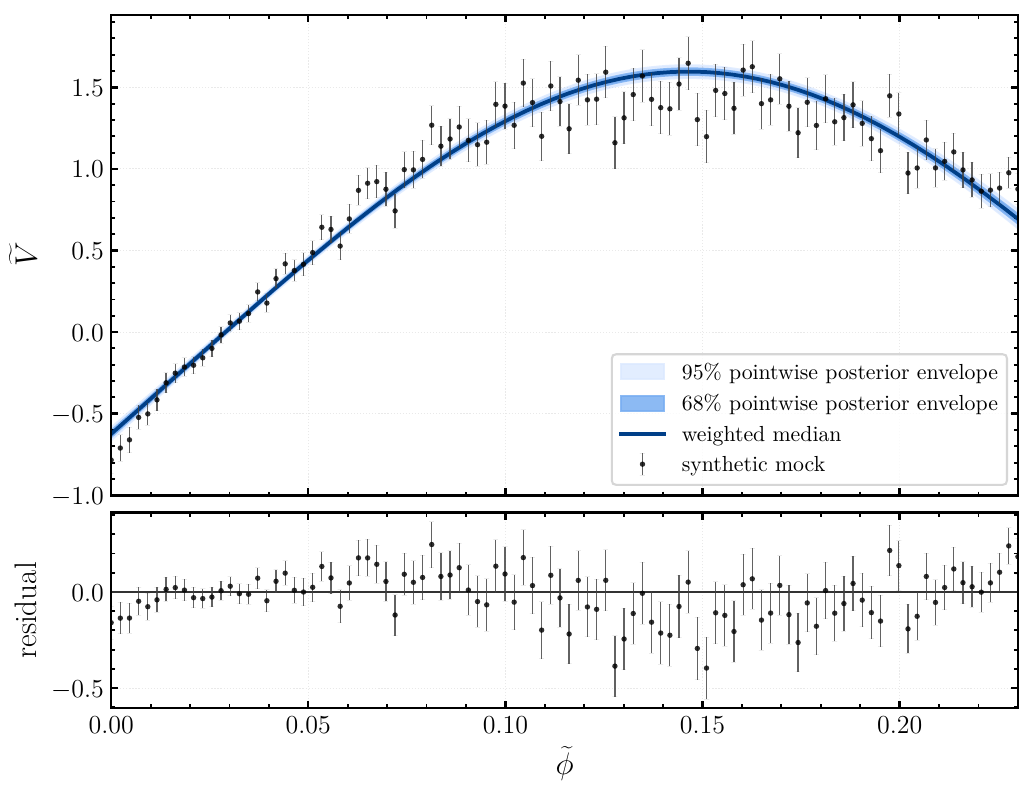}
    
    \vspace{2pt}
    (c) Regularized inverse-quadratic
\end{minipage}
\hfill
\begin{minipage}{0.43\linewidth}
    \centering
    \includegraphics[width=\linewidth]{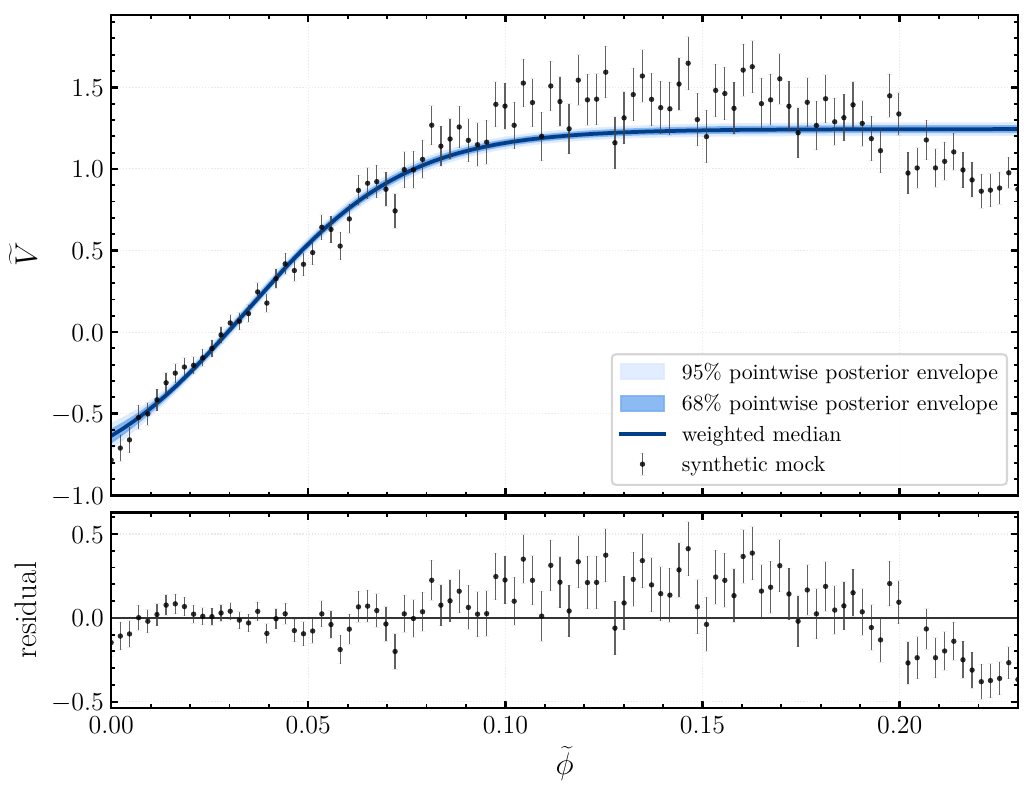}
    
    \vspace{2pt}
    (d) Shifted-$\tanh$
\end{minipage}

\vspace{0.15cm}

% ---------------- Row 3 ----------------
\begin{minipage}{0.43\linewidth}
    \centering
    \includegraphics[width=\linewidth]{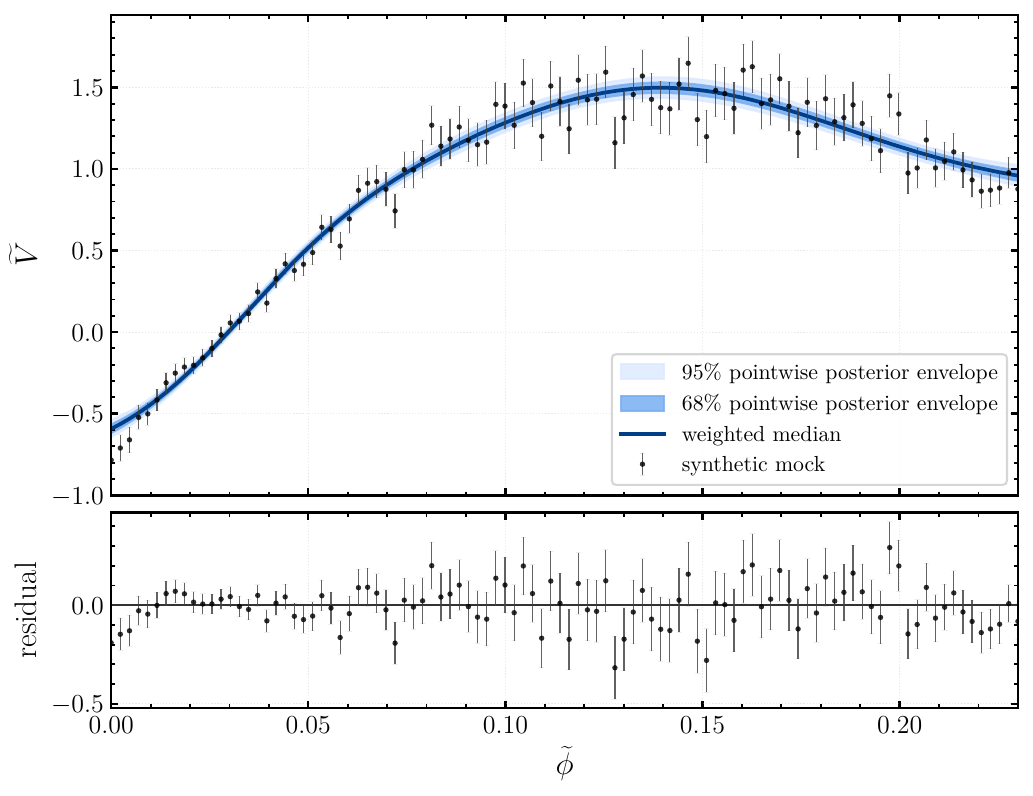}
    
    \vspace{2pt}
    (e) Sigmoid--Gaussian feature
\end{minipage}

\caption{Pointwise SSCDM curve summaries for the five potential templates fitted on the retained field interval $0\leq\widetilde\phi\leq0.23$. The panels show (a) the generalized axion-like potential, (b) the Gaussian feature, (c) the regularized inverse-quadratic potential, (d) the shifted-$\tanh$ potential, and (e) the sigmoid--Gaussian feature. Black points with error bars denote the single synthetic mock realization used in the fits. Solid dark-blue curves show the direct pointwise weighted medians, while the dark- and light-blue regions show the corresponding direct weighted equal-tail $68\%$ and $95\%$ posterior envelopes. The saved posterior weights are applied exactly once; no posterior resampling or display smoothing is used. The lower panels show the mock residuals relative to the pointwise weighted median. For all five reconstructions, the $68\%$ envelope remains contained within the $95\%$ envelope. These are parameter-induced bands for each template under the synthetic experiment, not observational posterior-predictive intervals.}
\label{fig:band_sscdm}
\end{figure*}

%%-----------------------------------------------------------------------------%%
\subsubsection{SSCDM target}
%%-----------------------------------------------------------------------------%%

We next examine the five potential families for the SSCDM target with $(z_\dagger,\Delta x)=(1.8,0.4)$; the marginal posteriors are shown in Fig.~\ref{fig:corner_sscdm_all}. The reconstructed on-shell potential is a broad, approximately symmetric profile over the retained field interval. All intervals quoted below are weighted marginal posterior summaries under the adopted synthetic likelihood; their widths are not directly comparable with the ECDM ones because the target curve, field interval, mock, scale, and priors differ.

Throughout this subsection, field locations and widths refer to the dimensionless coordinate $\widetilde\phi=\phi/M_{\rm Pl}$. Potential amplitudes are reported in units of $\rho_{\rm c0}$: the fitted quantities are $\Lambda^4/\rho_{\rm c0}$ for the axion, $\Lambda/\rho_{\rm c0}$ for the shifted-$\tanh$ and sigmoid--Gaussian templates, $A/\rho_{\rm c0}$ for the Gaussian template, $\lambda/\sqrt{\rho_{\rm c0}}$ for the regularized inverse-quadratic template, and $V_0/\rho_{\rm c0}$ for the offsets.

For the generalized axion-like template, we obtain
\begin{equation*}
\begin{aligned}
 \frac{\Lambda^4}{\rho_{\rm c0}}=&1.726^{+0.052}_{-0.051}, &
 \eta=&(469.7^{+3.7}_{-3.5})\times10^{-4},\\
 n=&0.428^{+0.020}_{-0.019}, & \frac{V_0}{\rho_{\rm c0}}=&-0.811^{+0.047}_{-0.048}.
\end{aligned}
\end{equation*}
The corner plot shows strong anticorrelations of $\Lambda^4/\rho_{\rm c0}$ with both $n$ and $V_0/\rho_{\rm c0}$, along with a strong positive correlation between $n$ and $V_0/\rho_{\rm c0}$. The scale $\eta$ is comparatively weakly correlated with the other parameters.

The fitted interval includes $\widetilde\phi=0$, the high-redshift endpoint of the compact transition, and the zero-phase axion places a periodic minimum there. The posterior favors $n<1/2$, rather than merely overlapping that threshold, so $V_{,\widetilde\phi}\propto\widetilde\phi^{\,2n-1}$ diverges as $\widetilde\phi\to0^+$. The leading axion score therefore quantifies function-value representation but does not define a differentiable Klein--Gordon potential on the closed fitting interval.

For the shifted-$\tanh$ template, we obtain
\begin{equation*}
\begin{aligned}
 \frac{\Lambda}{\rho_{\rm c0}}&=1.244^{+0.018}_{-0.018}, &
 \xi_1&=-0.788^{+0.086}_{-0.102}, \\
 \nu&=24.68^{+1.68}_{-1.60}, &
 \widetilde\phi_{\rm c}&=(34.4^{+2.3}_{-2.7})\times10^{-3}.
\end{aligned}
\end{equation*}

The strongest contour structure couples $\xi_1$, $\nu$, and $\widetilde\phi_{\rm c}$. $\xi_1$ is positively correlated with both $\nu$ and $\widetilde\phi_{\rm c}$, and $\nu$ is positively correlated with $\widetilde\phi_{\rm c}$. The amplitude is more weakly correlated with the remaining parameters. The inferred $\nu$ describes a relatively sharp transition within this template family.

For the Gaussian-feature template, we find
\begin{equation*}
\begin{aligned}
 \frac{A}{\rho_{\rm c0}}&=1.697^{+0.027}_{-0.025}, &
 \widetilde\phi_{\rm c}&=(150.4^{+1.1}_{-1.1})\times10^{-3},\\
 w&=(563.7^{+7.6}_{-7.7})\times10^{-4}.
\end{aligned}
\end{equation*}
The width is anticorrelated with the amplitude and positively correlated with the center, while the amplitude and center show a weaker anticorrelation. The narrow marginal contours identify the preferred member of this restrictive family; they do not establish agreement with the target.

\begin{table}[t]
\centering
\footnotesize
\renewcommand{\arraystretch}{1.10}
\setlength{\tabcolsep}{3pt}

\textit{Target: ECDM}\\[3pt]
\textbf{(a) Evidence values}\\[2pt]
\begin{tabular}{@{}lrrr@{}}
\toprule
Potential template & $N_{\rm par}$ & $\log\mathcal Z$ & $\sigma_{\log\mathcal Z}$\\
\midrule
Sigmoid--Gaussian feature & 6 & $\mathbf{107.718}$ & 0.1404\\
Shifted-$\tanh$ & 4 & 102.964 & 0.0520\\
Generalized axion-like & 4 & 89.467 & 0.0883\\
Regularized inverse-quadratic & 4 & 49.675 & 0.0843\\
Gaussian feature & 3 & $-431.535$ & 0.2181\\
\bottomrule
\end{tabular}

\medskip
\textbf{(b) Relative scores and sampling cost}\\[2pt]
\begin{tabular}{@{}lrr@{}}
\toprule
Potential template & $\Delta\log\mathcal Z$ & $N_{\rm like}$\\
\midrule
Sigmoid--Gaussian feature & $0$ & $53\,389\,099$\\
Shifted-$\tanh$ & $-4.754\pm0.150$ & $2\,518\,550$\\
Generalized axion-like & $-18.250\pm0.166$ & $655\,144$\\
Regularized inverse-quadratic & $-58.043\pm0.164$ & $3\,672\,431$\\
Gaussian feature & $-539.253\pm0.259$ & $40\,138$\\
\bottomrule
\end{tabular}

\medskip
\textit{Target: SSCDM}\\[3pt]
\textbf{(c) Evidence values}\\[2pt]
\begin{tabular}{@{}lrrr@{}}
\toprule
Potential template & $N_{\rm par}$ & $\log\mathcal Z$
& $\sigma_{\log\mathcal Z}$\\
\midrule
Generalized axion-like
& 4 & $\mathbf{72.700}$ & 0.0494\\
Sigmoid--Gaussian feature
& 6 & 66.721 & 0.1060\\
Regularized inverse-quadratic
& 4 & 49.309 & 0.0953\\
Shifted-$\tanh$
& 4 & $-3.160$ & 0.0857\\
Gaussian feature
& 3 & $-365.831$ & 0.1625\\
\bottomrule
\end{tabular}

\medskip
\textbf{(d) Relative scores and sampling cost}\\[2pt]
\begin{tabular}{@{}lrr@{}}
\toprule
Potential template & $\Delta\log\mathcal Z$ & $N_{\rm like}$\\
\midrule
Generalized axion-like
& $0$ & $785\,597$\\
Sigmoid--Gaussian feature
& $-5.979\pm0.117$ & $7\,773\,842$\\
Regularized inverse-quadratic
& $-23.391\pm0.107$ & $3\,697\,800$\\
Shifted-$\tanh$
& $-75.860\pm0.099$ & $2\,611\,495$\\
Gaussian feature
& $-438.531\pm0.170$ & $45\,693$\\
\bottomrule
\end{tabular}
\caption{Coordinate-fixed conditional marginal-likelihood scores for the five potential families fitted to the synthetic ECDM and SSCDM potential-space experiments. Panels (a) and (c) report the absolute evidence values, while panels (b) and (d) report the corresponding within-target differences and sampling costs. The highest evidence within each target is shown in bold. The scores are internal to each synthetic experiment (Sec.~\ref{subsec:bayes}); they are not observational Bayes factors, and absolute values cannot be compared across the two targets. Evidence differences are calculated from the unrounded absolute values. For a nonzero $\Delta\log\mathcal Z$, the displayed uncertainty is the quadrature combination of the two sampler-reported numerical errors. Here $N_{\rm par}$ is the number of sampled parameters and $N_{\rm like}$ is the number of likelihood evaluations, a computational cost rather than a goodness-of-fit statistic.}
\label{tab:model_comparison_combined}
\end{table}

For the regularized inverse-quadratic template, the marginal summaries are
\begin{equation*}
\begin{aligned}
 \frac{\lambda}{\sqrt{\rho_{\rm c0}}}&=0.626^{+0.011}_{-0.018}, &
 \widetilde\phi_{\rm c}&=(146.41^{+0.98}_{-0.94})\times10^{-3},\\
 \frac{V_0}{\rho_{\rm c0}}&=-5.96^{+0.21}_{-0.18}, &
 \epsilon&=(227.6^{+1.8}_{-3.8})\times10^{-3}.
\end{aligned}
\end{equation*}
The dominant correlations couple $\lambda$, $\epsilon$, and $V_0$: $\lambda$ and $\epsilon$ are positively correlated, whereas $V_0$ is anticorrelated with both. The center is less strongly constrained by these combinations. The fitted $\epsilon\simeq0.228$ lies close to its imposed upper bound $\epsilon_{\max}=\widetilde\phi_{\rm sc}=0.23$, and the marginal distribution is visibly prior truncated; the inverse-quadratic evidence is therefore conditional on that bound. Within the adopted range, the preferred feature is broad rather than a narrow regularized pole, while the negative offset compensates for the large contribution $\lambda^2/\epsilon^2$.

For the sigmoid--Gaussian template, we obtain
\begin{equation*}
\begin{aligned}
 \frac{\Lambda}{\rho_{\rm c0}}&=0.852^{+0.061}_{-0.062}, &
 \widetilde\phi_{\rm a}&=(31.8^{+1.7}_{-1.3})\times10^{-3},\\
 \sigma_{\rm a}&=(35.8^{+3.1}_{-2.9})\times10^{-3}, &
 \alpha&=0.76^{+0.14}_{-0.12},\\
 \widetilde\phi_{\rm b}&=(138.5^{+3.2}_{-3.6})\times10^{-3}, &
 \sigma_{\rm b}&=(67.6^{+5.9}_{-5.6})\times10^{-3}.
\end{aligned}
\end{equation*}
The posterior for $\alpha$ is well within the adopted interval $0\leq\alpha\leq3$ and is not boundary limited. The most prominent correlations include an anticorrelation between $\Lambda$ and $\alpha$, a positive correlation between $\widetilde\phi_{\rm a}$ and $\sigma_{\rm b}$, an anticorrelation between $\sigma_{\rm a}$ and $\alpha$, and a positive correlation between $\alpha$ and $\widetilde\phi_{\rm b}$. The remaining parameter combinations show weaker or broader contour structures.

The posterior-band reconstructions are shown in Fig.~\ref{fig:band_sscdm}. The generalized axion-like potential gives the closest representation of the SSCDM target: its median follows the rise, broad maximum, and subsequent decline, with residuals distributed close to zero and without a prominent coherent trend. The sigmoid--Gaussian feature also reproduces the principal target structure, although modest structured residuals remain near the ends of the fitted interval. The regularized inverse-quadratic model captures the broad maximum but leaves an alternating residual pattern across the interval. The shifted-$\tanh$ model approaches an approximately constant plateau and consequently misses both the height and the descending side of the target maximum. The Gaussian feature gives the poorest reconstruction, with large coherent discrepancies on the low-field side and across the central feature. Thus, the visual reconstruction quality is consistent with the marginal-likelihood ordering: generalized axion-like, sigmoid--Gaussian feature, regularized inverse-quadratic, shifted-$\tanh$, and Gaussian feature.

Table~\ref{tab:model_comparison_combined} gives the corresponding conditional evidences. Taking the generalized axion-like template as the within-SSCDM reference, the sigmoid--Gaussian, regularized inverse-quadratic, shifted-$\tanh$, and Gaussian-feature templates give $\Delta\log\mathcal Z=-5.979\pm0.117,\quad -23.391\pm0.107,\quad -75.860\pm0.099,\quad -438.531\pm0.170,$ respectively. The displayed uncertainties are quadrature combinations of the two sampler-reported numerical errors.

The generalized axion-like template therefore has the largest SSCDM evidence, followed by the sigmoid--Gaussian, regularized inverse-quadratic, shifted-$\tanh$, and Gaussian-feature families; the leading axion result carries the $n<1/2$ regularity qualification described above.

The different within-target rankings reflect different field-space geometries under the adopted coordinate and prior conventions. The sampled ECDM curve combines a sign-changing trend with a broad localized excess, which the sigmoid--Gaussian family can represent explicitly. The retained SSCDM field interval is instead dominated by an approximately symmetric broad maximum, which the generalized axion-like family represents particularly efficiently.

%%-----------------------------------------------------------------------------%%
\section{Representative Klein--Gordon--Friedmann evolutions}
\label{sec:dynamics}
%%-----------------------------------------------------------------------------%%

The two potential-space comparisons of Sec.~\ref{sec:results} identify useful parametric families for representing reconstructed target curves. We now turn to the distinct dynamical question and use representative smooth members of two such families to illustrate homogeneous sign-switching trajectories of the coupled Klein--Gordon--Friedmann system. The parameter combinations evolved below are not the posterior-summary vectors of Sec.~\ref{sec:results}, and the tabulated shooting scans report numerical diagnostics relative to compressed background targets rather than fits to data. The plotted sigmoid--Gaussian trajectories are the accepted $H_0=73\,\mathrm{km\,s^{-1}\,Mpc^{-1}}$ rows of the three scan blocks; the offset-axion trajectories are the $H_0=75\,\mathrm{km\,s^{-1}\,Mpc^{-1}}$, $H_0=73\,\mathrm{km\,s^{-1}\,Mpc^{-1}}$, and $H_0=71\,\mathrm{km\,s^{-1}\,Mpc^{-1}}$ continuation rows, with their numerical classifications stated in Table~\ref{tab:axion_V0_scan}.

For the recombination-to-present shooting calculations of Secs.~\ref{subsec:append_sig_bump} and \ref{subsec:append_axion}, following Ref.~\cite{Akarsu:2025gwi}, we integrate from the reference redshift $z_*=1090$ to the present and restrict attention to expanding solutions with $E(z)>0$, so that redshift remains a valid evolution variable. We define $\varphi\equiv\phi/M_{\rm Pl}$, identical to the $\widetilde\phi$ of the preceding sections, $E\equiv H/H_0$, and $\widetilde\Omega_i\equiv\rho_i/\rho_{\rm c0}$, with $\rho_{\rm c0}=3M_{\rm Pl}^2H_0^2$. Both scalar contributions are normalized by $\rho_{\rm c0}$.
\begin{equation}
\begin{aligned}
 \widetilde\Omega_V(\varphi)&\equiv
 \frac{V(M_{\rm Pl}\varphi)}{\rho_{\rm c0}},\\
 \widetilde\Omega_K&\equiv
 \frac{\xi\dot\phi^{\,2}}{2\rho_{\rm c0}}=\frac{\xi}{6}(1+z)^2E^2
 \left(\frac{{\rm d}\varphi}{{\rm d}z}\right)^2.
\end{aligned}
\label{parameters}
\end{equation}
We then write $\widetilde\Omega_\phi=\widetilde\Omega_K+\widetilde\Omega_V$. A consistent recombination-to-present formulation distinguishes cold matter plus baryons, effectively massless radiation, and the single massive-neutrino species used in the Planck baseline. We define
\begin{equation}
 \widetilde\Omega_{cb}=\Omega_{cb0}(1+z)^3,
 \qquad
 \widetilde\Omega_{\rm r}=\Omega_{\rm r0}(1+z)^4,
\end{equation}
and write $\widetilde P_\nu\equiv P_\nu/\rho_{\rm c0}$. The quantities $\widetilde\Omega_\nu(z)$ and $\widetilde P_\nu(z)$ follow from the thermal Fermi--Dirac background and interpolate continuously between the relativistic and nonrelativistic limits. The background equations are then

\begin{align}
 E^2&=\widetilde\Omega_{cb}+\widetilde\Omega_{\rm r}
      +\widetilde\Omega_\nu+\widetilde\Omega_\phi,
\label{friedmann_1_dless}
\\
 2(1+z)E\frac{{\rm d}E}{{\rm d}z}
 &=3\widetilde\Omega_{cb}+4\widetilde\Omega_{\rm r}
   +3\bigl(\widetilde\Omega_\nu+\widetilde P_\nu\bigr)
   +6\widetilde\Omega_K.
\label{friedmann_2_dless}
\end{align}
The kinematical quantities plotted below are evaluated directly from the same background solution,
\begin{equation}
\begin{aligned}
 q(z)=&-1+(1+z)\frac{E'}{E},
 \\
 w_{\rm tot}(z)=&-1+\frac{2(1+z)}{3}\frac{E'}{E},
 \label{eq:kinematical_quantities}
\end{aligned} 
\end{equation}
where a prime denotes ${\rm d}/{\rm d}z$ in this section. Together with these equations, the field obeys
\begin{equation}
\begin{aligned}
 0={}&(1+z)^2E^2\frac{{\rm d}^2\varphi}{{\rm d}z^2}\\
 &+(1+z)^2E\frac{{\rm d}E}{{\rm d}z}
 \frac{{\rm d}\varphi}{{\rm d}z}\\
 &-2(1+z)E^2\frac{{\rm d}\varphi}{{\rm d}z}
 +3\xi\frac{{\rm d}\widetilde\Omega_V}{{\rm d}\varphi}.
\end{aligned}
\label{eq_phi_dless}
\end{equation}

For each integration, we specify
\begin{equation}
 \varphi_{\rm in}\equiv\varphi(z_*),
 \qquad
 \varphi'_{\rm in}\equiv
 \left.\frac{{\rm d}\varphi}{{\rm d}z}\right|_{z_*}.
\end{equation}
Because $\widetilde\Omega_K$ itself contains $E^2$, define at an arbitrary redshift
\begin{equation}
 B(z)\equiv\frac{\xi}{6}(1+z)^2[\varphi'(z)]^2,
 \qquad
 \widetilde\Omega_K(z)=B(z)E^2(z).
 \label{eq:B_definition}
\end{equation}
Solving the Friedmann constraint explicitly gives
\begin{equation}
 E^2(z)=
 \frac{\widetilde\Omega_{cb}(z)+\widetilde\Omega_{\rm r}(z)
 +\widetilde\Omega_\nu(z)+\widetilde\Omega_V[\varphi(z)]}
 {1-B(z)}.
 \label{eq:E_algebraic}
\end{equation}
This algebraic relation is imposed at every right-hand-side evaluation. The denominator is required to remain positive; on the phantom branch it is $1-B(z)=1+(1+z)^2[\varphi'(z)]^2/6>0$. At the initial point, $B_{\rm in}\equiv B(z_*)$ and the constraint specializes to
\begin{equation}
 E_{\rm in}^2=
 \frac{\widetilde\Omega_{cb}(z_*)+
       \widetilde\Omega_{\rm r}(z_*)+
       \widetilde\Omega_\nu(z_*)+
       \widetilde\Omega_V(\varphi_{\rm in})}
      {1-B_{\rm in}}.
 \label{eq:Ein_correct}
\end{equation}
All recombination-to-present shooting integrations in Secs.~\ref{subsec:append_sig_bump} and \ref{subsec:append_axion} use $\varphi'_{\rm in}=0$, for which $B_{\rm in}=0$; the correction in Eq.~\eqref{eq:Ein_correct} therefore does not change the reported numerical solutions. The closure test of Sec.~\ref{subsec:closure} instead uses the target-specific initial data stated there.

The physical-density calibration follows the Planck 2018 baseline TT,TE,EE+lowE+lensing best fit~\cite{Planck:2018vyg}. The present physical densities satisfy
\begin{equation}
\begin{aligned}
 \omega_{cb}&=\omega_{\rm b}+\omega_{\rm cdm}
             =0.022383+0.12011=0.142493,\\
 \omega_{\nu0}&\simeq\frac{0.06}{93.14}=0.000644,\\
 \omega_{\rm m}&=\omega_{cb}+\omega_{\nu0}\simeq0.143137,
\end{aligned}
\label{eq:Omega_M0}
\end{equation}
consistent with the rounded best-fit value $\omega_{\rm m}=0.14314$. Equations~\eqref{friedmann_1_dless} and \eqref{friedmann_2_dless} keep the cold-matter--baryon and massive-neutrino sectors distinct, thereby avoiding the simultaneous inclusion of the neutrino rest-mass density in both an $a^{-3}$ term and the evolved thermal background~\cite{Lesgourgues:2006nd}.

For the recombination-to-present forward integrations, we adopt the Planck 2018 one-massive-neutrino prescription translated into CLASS variables. The background contains one thermal non-cold species and a massless ultrarelativistic sector specified by
\begin{equation}
\begin{aligned}
 m_\nu&=0.06\,{\rm eV},&
 \frac{T_{\rm ncdm}}{T_\gamma}&=0.71611,\\
 N_{\rm ur}&=2.0328.&&
\end{aligned}
\end{equation}
In the ultrarelativistic limit, this prescription corresponds to $N_{\rm eff}\simeq3.046$. The present physical density of the massive species is normalized as
\begin{equation}
 \omega_{\nu0}=\frac{0.06}{93.14},
\end{equation}
consistent with the adopted Planck matter-density convention.

Defining
\begin{equation}
\begin{aligned}
 T_{\nu0}&\equiv0.71611\,T_{\rm CMB},&
 \mathcal T_{\nu0}&\equiv k_{\rm B}T_{\nu0},\\
 y(z)&\equiv \frac{m_\nu}{\mathcal T_{\nu0}(1+z)}.&&
\end{aligned}
\end{equation}
The Fermi--Dirac energy-density and pressure integrals are
\begin{align}
 I_\rho(y)&=\int_0^\infty \frac{q^2\sqrt{q^2+y^2}}{e^q+1}\,{\rm d}q,\\
 I_p(y)&=\int_0^\infty \frac{q^4}{3\sqrt{q^2+y^2}(e^q+1)}\,{\rm d}q.
\end{align}
The corresponding massive-neutrino contributions are
\begin{align}
 \widetilde\Omega_\nu(z)&= \Omega_{\nu0}(1+z)^4 \frac{I_\rho[y(z)]}{I_\rho[y(0)]},\\
 \widetilde P_\nu(z)&= \Omega_{\nu0}(1+z)^4 \frac{I_p[y(z)]}{I_\rho[y(0)]}.
\end{align}
The integrals are evaluated over $0\leq q\leq40$ with absolute and relative quadrature tolerances $10^{-10}$ and $10^{-9}$, respectively. They are tabulated at 700 points uniformly spaced in $\ln(1+z)$ over $0\leq z\leq1090$.

The photon-plus-massless contribution entering Eqs.~\eqref{friedmann_1_dless} and
\eqref{friedmann_2_dless} is
\begin{align}
 \omega_{\rm r}&=\omega_\gamma
 \left[1+\frac{7}{8}\left(\frac{4}{11}\right)^{4/3}N_{\rm ur}\right],\notag\\
 \Omega_{{\rm r}0}&=\frac{\omega_{\rm r}}{h^2},
 \qquad N_{\rm ur}=2.0328.
 \label{eq:radiation_dynamics}
\end{align}
The separately evolved massive species is not included in $\Omega_{{\rm r}0}$, thereby avoiding double counting between the massless-radiation and massive-neutrino sectors.

For a selected $H_0$, $\Omega_{cb0}=\omega_{cb}/h^2$ and $\Omega_{\nu0}=\omega_{\nu0}/h^2$ follow; for example, $H_0=73.04\,{\rm km\,s^{-1}\,Mpc^{-1}}$ gives $\Omega_{\rm m0}\equiv\Omega_{cb0}+\Omega_{\nu0}=0.26831$ \cite{Riess:2021jrx}. Thus $H_0$ is an input, not a quantity inferred by the calculations below.

The second numerical target is the comoving distance
\begin{equation}
\begin{aligned}
 D_{\rm M}^{\rm(cal)}(z_*)&=13869.57\,{\rm Mpc},\\
 D_{\rm M}(z_*)&=\int_0^{z_*}\frac{c\,{\rm d}z}{H(z)}.
\end{aligned}
\label{eq:D_M}
\end{equation}
Here the speed of light $c$ has been restored. The adopted value follows from the Planck 2018 best-fit quantities $r_*=144.394\,{\rm Mpc}$ and $100\theta_*=1.041085$ through $D_{\rm M}(z_*)=r_*/\theta_*$; it is a derived calibration quantity rather than a directly tabulated observable \cite{Planck:2018vyg}. We use it only as a compressed numerical target, not as a replacement for the Planck likelihood or its parameter covariance. In particular, the $25\,{\rm Mpc}$ scale entering the objective below is an optimizer weight, not a Planck standard deviation; combining separately quoted one-dimensional errors as if they were independent would neglect their covariance.

We monitor the Friedmann-constraint residual at the end of each integration,
\begin{equation}
 \mathcal O_{\rm F}\equiv
 \Omega_{cb0}+\Omega_{{\rm r}0}+\Omega_{\nu0}
 +\widetilde\Omega_V(\varphi_0)
 +\frac{\xi}{6}E_0^2(\varphi'_0)^2-E_0^2.
 \label{eq:closure_residual}
\end{equation}
For $\xi=-1$ and $E_0=1$, this reduces to the expression used in the numerical analysis. Near-machine precision values of $\mathcal O_{\rm F}$ are internal constraint-consistency checks and do not independently establish the accuracy of the numerical integration. For the three-term calibration criterion considered here, we define the least-squares residual vector
\begin{equation}
\begin{aligned}
 \boldsymbol R_{\rm ax}={}&\Bigg(
 \frac{E_0-1}{10^{-4}},\,
 \frac{D_{\rm M}(z_*)-D_{\rm M}^{\rm(cal)}(z_*)}
 {25\,{\rm Mpc}},\,
 \frac{z_\dagger-z_{\dagger,{\rm target}}}{0.03}
 \Bigg),\\
 \mathcal C_{\rm LS}^{({\rm ax})}={}&
 \frac{1}{2}\boldsymbol R_{\rm ax}^{\mathsf T}\boldsymbol R_{\rm ax}.
\end{aligned}
\label{eq:numerical_merit}
\end{equation}
The axion scan uses $z_{\dagger,{\rm target}}=1.8$ and minimizes Eq.~\eqref{eq:numerical_merit} with the massive-neutrino background described above. The denominators are numerical weights, not observational standard deviations. In this objective, $z_\dagger$ is prescribed, whereas $z_{\rm t}$ is derived after integration; consequently, a relation between the selected $H_0$ and $z_\dagger$ is not an independent prediction of the potential.

For the numerical shooting integrations reported in Secs.~\ref{subsec:append_sig_bump} and \ref{subsec:append_axion}, the coupled background and field equations are integrated in redshift from $z_*=1090$ to $z=0$ using the implicit Radau method in \texttt{scipy.integrate.solve\_ivp} \cite{Virtanen:2019joe}. The evolved ODE state is
\begin{equation}
 {\boldsymbol y}(z)=\left(\varphi,\frac{{\rm d}\varphi}{{\rm d}z},\chi\right),
\end{equation}
where $\chi$ is the accumulated line-of-sight comoving distance. The Hubble quantity is not evolved as an independent ODE variable. Instead, $E^2$ is imposed algebraically at every right-hand-side evaluation through Eq.~\eqref{eq:E_algebraic}, and its derivative is evaluated from Eq.~\eqref{friedmann_2_dless}.

The integration is initialized at $z=z_*$ with
\begin{equation}
 \chi(z_*)=0,\qquad\frac{{\rm d}\chi}{{\rm d}z}=-\frac{c}{H_0E(z)}.
\end{equation}
Consequently,
\begin{equation}
 \chi(0)=\frac{c}{H_0}\int_0^{z_*}\frac{{\rm d}z}{E(z)}=D_{\rm M}(z_*).
\end{equation}
The potential and scalar-density zero crossings are located from the adaptively integrated solution using their corresponding event functions.

The final integrations use $\texttt{rtol}=10^{-9}$, $\texttt{atol}_{\varphi,\varphi'}=10^{-11}$, $\texttt{atol}_{D_{\rm M}}=10^{-7}$, and $\texttt{max\_step}=0.2$. The tolerance-refinement calculation uses $\texttt{rtol}=3\times10^{-10}$, $\texttt{atol}_{\varphi,\varphi'}=3\times10^{-12}$, $\texttt{atol}_{D_{\rm M}}=3\times10^{-8}$, and $\texttt{max\_step}=0.1$. The near-machine-precision Friedmann closure residual is an internal algebraic-constraint check rather than an independent test of integration accuracy. Numerical accuracy is assessed from the changes in $E_0$, $D_{\rm M}$, $z_{\rm t}$, and $z_\dagger$ under tolerance refinement.

As an end-to-end background check, we compare the dynamical implementation with CLASS version 3.3.4 \cite{Blas:2011rf}. For this comparison, the CLASS input is specified by
\begin{equation}
\begin{aligned}
H_0&=67.32\,{\rm km\,s^{-1}\,Mpc^{-1}}, & T_{\rm CMB}&=2.7255\,{\rm K},\\
\omega_b&=0.022383, & \omega_{\rm cdm}&=0.12011,\\
\Omega_{k0}&=0, & N_{\rm ncdm}&=1,\\
m_{\rm ncdm}&=0.06\,{\rm eV}, & \omega_{\rm ncdm}&=\frac{0.06}{93.14},\\
\frac{T_{\rm ncdm}}{T_\gamma}&=0.71611, & N_{\rm ur}&=2.0328,
\end{aligned}
\end{equation}
with \texttt{deg\_ncdm=1}. The cosmological constant is determined by the flat background budget, while additional fluid and scalar-field dark energy components are disabled. Because both $m_{\rm ncdm}$ and $\omega_{\rm ncdm}$ are supplied, CLASS renormalizes the ncdm phase-space distribution to satisfy the specified mass and present density. It returns
\begin{align}
\omega_{\rm ncdm}^{\rm CLASS}&=6.441915396178\times10^{-4},\\
\omega_{\rm m}^{\rm CLASS}&=0.1431371261,
\end{align}
confirming the intended matter and massive-neutrino mapping.

The CLASS background is sorted in increasing redshift and interpolated onto a common grid containing 5001 linearly spaced points over $0\leq z\leq10$ and 5000 points logarithmically spaced in $1+z$ over $10\leq z\leq1090$. We define
\begin{equation}
E_{\rm CLASS}(z)=\frac{H_{\rm CLASS}(z)}{H_{\rm CLASS}(0)}
\end{equation}
and compare it with the independently evaluated dynamical-code background. Both implementations use the photon density associated with $T_{\rm CMB}=2.7255\,{\rm K}$,
\begin{equation}
\omega_\gamma=2.4729792808613565\times10^{-5},
\end{equation}
together with identical photon, massless-neutrino, massive-neutrino, and matter inputs. The same-input comparison gives
\begin{equation}
\max_{0\leq z\leq1090}\left|\frac{E_{\rm code}(z)-E_{\rm CLASS}(z)}{E_{\rm CLASS}(z)}\right|=1.51426\times10^{-7},
\end{equation}
with the maximum occurring near $z=1088$. The corresponding distances are
\begin{align}
D_{\rm M}^{\rm CLASS}(1090)&=13869.63897898\,{\rm Mpc},\\
D_{\rm M}^{\rm code}(1090)&=13869.63875033\,{\rm Mpc},
\end{align}
and hence
\begin{equation}
D_{\rm M}^{\rm code}(1090)-D_{\rm M}^{\rm CLASS}(1090)=-2.2865\times10^{-4}\,{\rm Mpc}.
\end{equation}
This agreement validates the background implementation under identical input conventions. The comparison is an end-to-end numerical code check rather than an observational likelihood test. Its distance difference is distinct from the separate offset between the code reference distance and the fixed compressed target $D_{\rm M}^{({\rm cal})}=13869.57\,{\rm Mpc}$.

Both shooting scans below use SciPy's bounded trust-region-reflective \texttt{least\_squares} implementation with \texttt{x\_scale='jac'}, $\texttt{xtol}=\texttt{ftol}=\texttt{gtol}=10^{-8}$, and $\texttt{max\_nfev}=100$. For the offset-axion scan, the algorithm adjusts $(\ln A,\widetilde V_0,\varphi_{\rm in})$ at fixed $(\eta,n,\varphi'_{\rm in},\xi)=(0.2,1,0,-1)$. The bounds are $10^{-4}\leq A\leq50$, $-2\leq\widetilde V_0\leq-0.01$, and $10^{-5}\leq\varphi_{\rm in}\leq\pi\eta-10^{-5}$. The nominal start $(A,\widetilde V_0,\varphi_{\rm in})=(1,-0.75,0.14)$ is supplemented by $(2,-0.95,0.20)$ and $(0.5,-0.30,0.05)$. When available, the preceding $H_0$ solution is tested first as a continuation start. If its cost is below $10^{-3}$, the remaining starts are skipped; otherwise, all starts are evaluated and the lowest-cost result is retained. A large penalty is assigned to an integration failure, an invalid $E^2$, or the absence of a finite density crossing.

%%-----------------------------------------------------------------------------%%
\subsection{Sigmoid--Gaussian potential}
\label{subsec:append_sig_bump}
%%-----------------------------------------------------------------------------%%

We first consider representative homogeneous evolutions for the sigmoid--Gaussian family of Eq.~\eqref{eq:sigmoid_template}. The three parameter combinations displayed below differ from the ECDM posterior summary in Sec.~\ref{sec:results}; the comparison is therefore at the level of the potential family rather than of a posterior-summary reconstructed function. The equations depend only on field displacements, such as $(\varphi-\varphi_{\rm a})/\sigma_{\rm a}$, and are invariant under a common translation of the field and both potential centers. We remove this redundant freedom by setting
\begin{equation}
(\varphi_{\rm in},\varphi'_{\rm in})=(0,0).
\end{equation}
At fixed $H_0$ and fixed shape parameters $(\sigma_{\rm a},\alpha,\varphi_{\rm b},\sigma_{\rm b})$, the shooting procedure adjusts $(\ln\widetilde\Lambda,\varphi_{\rm a})$. The amplitude is therefore determined by the numerical shooting calculation rather than by an approximate closure relation.

\begin{figure*}
\centering
\includegraphics[width=\linewidth]{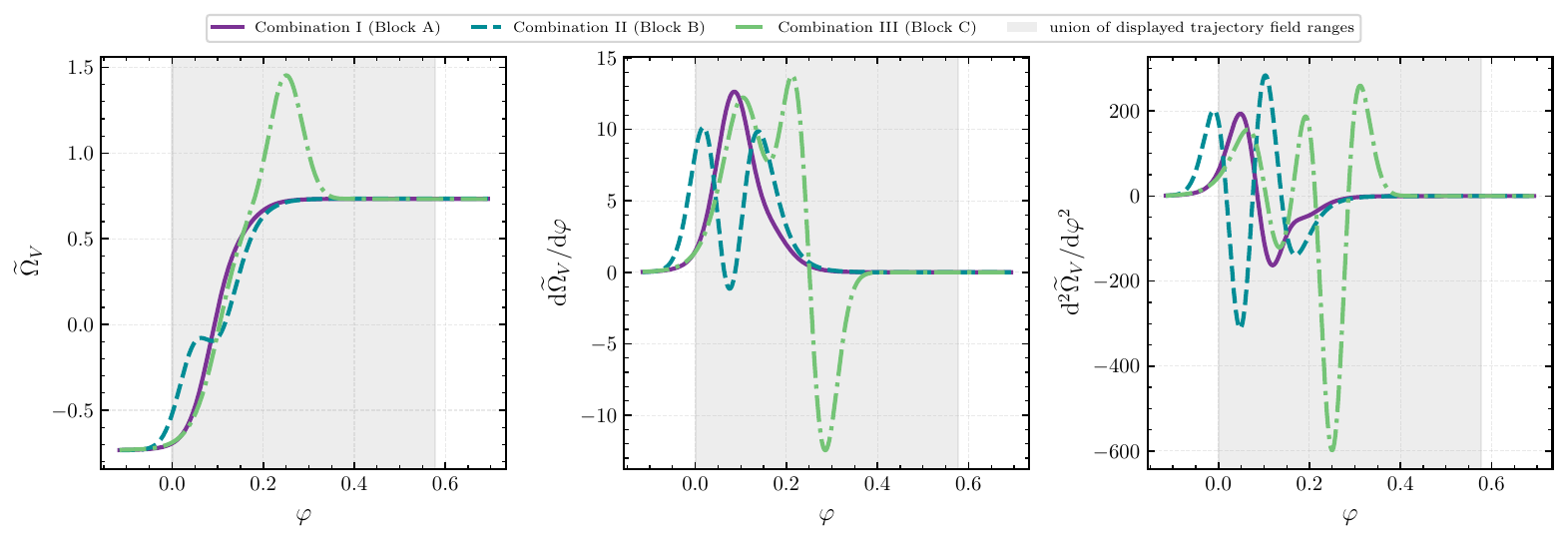}
\caption{Potential geometry for the three accepted sigmoid--Gaussian trajectories: $\widetilde\Omega_V$ (left), ${\rm d}\widetilde\Omega_V/{\rm d}\varphi$ (center), and ${\rm d}^2\widetilde\Omega_V/{\rm d}\varphi^2$ (right). The colored curves correspond to Combinations I--III at $H_0=73\,\mathrm{km\,s^{-1}\,Mpc^{-1}}$, as defined in Fig.~\ref{fig:density_parameters_sigmoid}. The gray interval $0\leq\varphi\leq0.5777$ is the union of the field ranges sampled by the three Block A--C trajectories over the full integration interval $0\leq z\leq1090$. The shading identifies the dynamically sampled region of field space and is not an uncertainty, confidence, or stability band. The homogeneous force is proportional to $-\xi\,{\rm d}\widetilde\Omega_V/{\rm d}\varphi$; for the phantom choice $\xi=-1$, it has the same sign as the displayed gradient and the opposite sign from the canonical case. Correspondingly, homogeneous linear stability near an extremum selects the opposite curvature sign from that of a canonical scalar.}
\label{fig:potential_vs_field_sigmoid}
\end{figure*}

\begin{table}
\centering
\footnotesize
\renewcommand{\arraystretch}{1.18}
\setlength{\tabcolsep}{2.2pt}
\begin{tabular}{ccccccc}
\toprule
Block & $H_0$ & $\Omega_{\rm m0}$ & $\varphi_{\rm a}$ & $z_{\rm t}$ & $z_\dagger$ & $\Delta z_{{\rm t}\dagger}$\\
\midrule
\multirow{6}{*}{A} & 75 & 0.2545 & 0.122882 & 1.907738 & 1.574453 & 0.333285\\
 & \textbf{73} & \textbf{0.2686} & \textbf{0.110750} & \textbf{2.123768} & \textbf{1.765987} & \textbf{0.357780} \\
 & 71 & 0.2839 & 0.095248 & 2.479368 & 2.094887 & 0.384481\\
 & 69 & 0.3006 & 0.070313 & 3.272284 & 2.814609 & 0.457675\\
 & 68 & 0.3096 & 0.044930 & 4.439269 & 3.819041 & 0.620228\\
 & 67 & 0.3189 & -- & -- & -- & --\\
\midrule
\multirow{6}{*}{B} & 75 & 0.2545 & 0.124552 & 1.556874 & 1.385973 & 0.170900\\
 & \textbf{73} & \textbf{0.2686} & \textbf{0.122896} & \textbf{1.814198} & \textbf{1.611151} & \textbf{0.203046}\\
 & 71 & 0.2839 & 0.116962 & 2.364344 & 2.070369 & 0.293976\\
 & 69 & 0.3006 & 0.093974 & 4.641862 & 3.786711 & 0.855150\\
 & 68 & 0.3096 & 0.053107 & 6.926558 & 5.956700 & 0.969857\\
 & 67 & 0.3189 & -- & -- & -- & --\\
\midrule
\multirow{6}{*}{C} & 75 & 0.2545 & 0.111921 & 1.745315 & 1.484474 & 0.260841\\
 & \textbf{73} & \textbf{0.2686} & \textbf{0.104241} & \textbf{1.902305} & \textbf{1.607630} & \textbf{0.294676}\\
 & 71 & 0.2839 & -- & -- & -- & --\\
 & 69 & 0.3006 & -- & -- & -- & --\\
 & 68 & 0.3096 & -- & -- & -- & --\\
 & 67 & 0.3189 & -- & -- & -- & --\\
\bottomrule
\end{tabular}
\caption{Sigmoid--Gaussian shooting solutions obtained using the Planck 2018 one-massive-neutrino prescription translated into CLASS variables, $(T_{\rm ncdm}/T_\gamma,N_{\rm ur})=(0.71611,2.0328)$, corresponding to $N_{\rm eff}\simeq3.046$. The selected values of $H_0$ are in $\mathrm{km\,s^{-1}\,Mpc^{-1}}$, with $\Omega_{\rm m0}=\omega_{\rm m}/h^2$ and $(\varphi_{\rm in},\varphi'_{\rm in})=(0,0)$. For each row, $(\widetilde\Lambda,\varphi_{\rm a})$ are adjusted using Eq.~\eqref{eq:sigmoid_merit}. The fixed shape parameters $(\sigma_{\rm a},\alpha,\varphi_{\rm b},\sigma_{\rm b})$ are $(0.06,0.30,0.10,0.05)$ for A, $(0.06,0.70,0.05,0.05)$ for B, and $(0.06,1.00,0.25,0.05)$ for C. The tabulated solutions and Eq.~\eqref{eq:sigmoid_template} use the $1/e$ half-width convention $\exp[-(\varphi-\varphi_{\rm b})^2/\sigma_{\rm b}^2]$; no width conversion is required. The event-located separation $\Delta z_{{\rm t}\dagger}\equiv z_{\rm t}-z_\dagger$ is evaluated from the unrounded adaptive-event locations. A row is retained only when the optimizer terminates successfully with $\mathcal C_{\rm LS}^{({\rm sig})}<10^{-3}$, the integration remains finite with positive $E^2$ over $1090\geq z\geq0$, both required crossings are finite, and neither adjusted parameter is pinned to a bound. A dash indicates that no trajectory passed these restricted criteria and is not a non-existence statement. Bold rows correspond to the trajectories shown in Figs.~\ref{fig:potential_vs_field_sigmoid}--\ref{fig:cosmo_metrics_field_sigmoid}.}
\label{tab:tanh_gaussian_simple}
\end{table}

The numerical implementation uses precisely the Gaussian convention of Eq.~\eqref{eq:sigmoid_template},
\begin{equation}
G(\varphi)=\exp\left[-\frac{(\varphi-\varphi_{\rm b})^2}{\sigma_{\rm b}^2}\right],
\end{equation}
so that $\sigma_{\rm b}$ is the $1/e$ half-width. The Gaussian contribution and its first two field derivatives are
\begin{align}
G_{,\varphi}&=-\frac{2(\varphi-\varphi_{\rm b})}{\sigma_{\rm b}^2}G,\\
G_{,\varphi\varphi}&=\left[\frac{4(\varphi-\varphi_{\rm b})^2}{\sigma_{\rm b}^4}-\frac{2}{\sigma_{\rm b}^2}\right]G.
\end{align}

For each fixed shape-parameter block and selected $H_0$, we define
\begin{align}
R_E^{({\rm sig})}&=\frac{E_0-1}{10^{-4}}, & R_D^{({\rm sig})}&=\frac{D_{\rm M}-D_{\rm M}^{({\rm cal})}}{25\,{\rm Mpc}},\\
\boldsymbol R_{\rm sig}&=\bigl(R_E^{({\rm sig})},R_D^{({\rm sig})}\bigr), & \mathcal C_{\rm LS}^{({\rm sig})}&=\frac{1}{2}\boldsymbol R_{\rm sig}^{\mathsf T}\boldsymbol R_{\rm sig}.
\label{eq:sigmoid_merit}
\end{align}
The bounded trust-region-reflective optimizer uses
\begin{equation}
10^{-4}\leq\widetilde\Lambda\leq50,\qquad 0.01\leq\varphi_{\rm a}\leq0.50.
\end{equation}
Its ordered multistart list is
\begin{equation}
\begin{aligned}
(\widetilde\Lambda,\varphi_{\rm a})={}&(1,\varphi_{\rm a}^{(0)}),(0.5,\varphi_{\rm a}^{(0)}),(2,\varphi_{\rm a}^{(0)}),\\
&(1,0.08),(1,0.16),
\end{aligned}
\end{equation}
where $\varphi_{\rm a}^{(0)}$ is a previously obtained estimate or the preceding continuation solution. The starts are evaluated in the stated order, and the sequence stops at the first successful solution with $\mathcal C_{\rm LS}^{({\rm sig})}<10^{-3}$. If no start reaches that threshold, all five are evaluated and the lowest-cost result is retained only for diagnostic purposes. A row enters the shooting table only if the optimizer terminates successfully, $\mathcal C_{\rm LS}^{({\rm sig})}<10^{-3}$, the integration remains finite, and neither fitted parameter is pinned to a bound.

\begin{figure}
\centering
\includegraphics[width=0.98\linewidth]{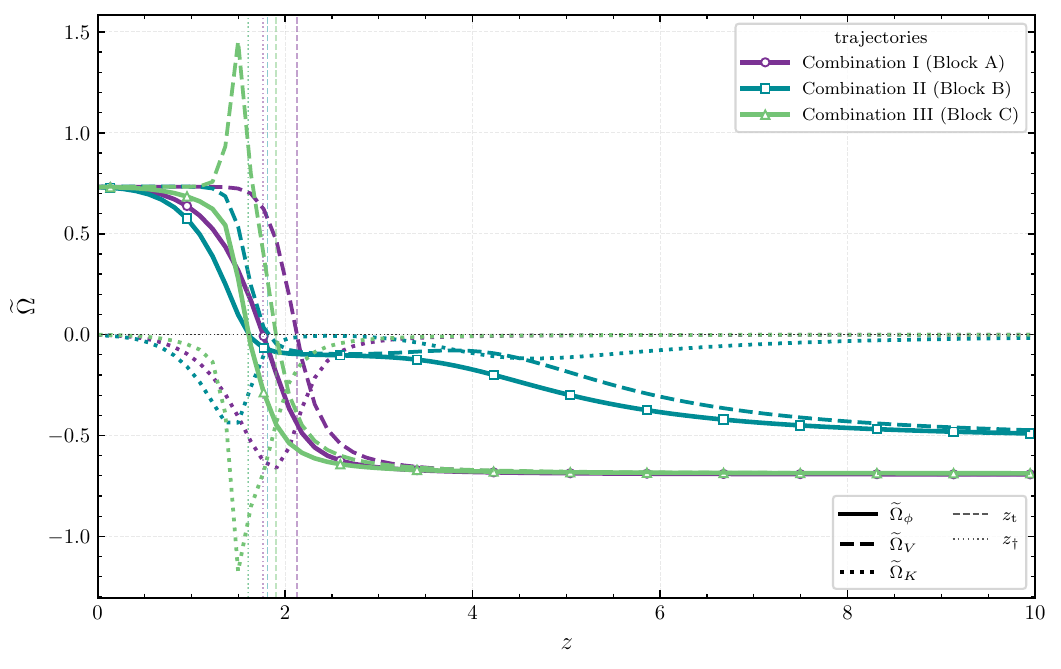}
\caption{Evolution of $\widetilde\Omega_\phi=\rho_\phi/\rho_{\rm c0}$ (solid), $\widetilde\Omega_V=V/\rho_{\rm c0}$ (dashed), and the signed phantom kinetic contribution $\widetilde\Omega_K=K/\rho_{\rm c0}$ (dotted) for the three accepted sigmoid--Gaussian trajectories at the selected input $H_0=73\,\mathrm{km\,s^{-1}\,Mpc^{-1}}$. The figure displays $0\leq z\leq10$; the integrations and acceptance checks cover the full interval $0\leq z\leq1090$. Combinations I--III (purple, turquoise, and light green) are the bold Block A--C rows of Table~\ref{tab:tanh_gaussian_simple}; their full parameter vectors and numerical diagnostics are given there and in Table~\ref{tab:tanh_gaussian_diagnostics}. Both the integrations and Eq.~\eqref{eq:sigmoid_template} use $\exp[-(\varphi-\varphi_{\rm b})^2/\sigma_{\rm b}^2]$, so $\sigma_{\rm b}$ is the $1/e$ half-width. Vertical dashed and dotted lines mark, respectively, the potential zero $z_{\rm t}$ and scalar-density zero $z_\dagger$.}
\label{fig:density_parameters_sigmoid}
\end{figure}

All four sigmoid--Gaussian figures and their derived quantities were exported from the unrounded diagnostics rows and parameter-validated trajectory caches corresponding to the accepted $H_0=73\,\mathrm{km\,s^{-1}\,Mpc^{-1}}$ rows of Table~\ref{tab:tanh_gaussian_simple}. Before rounding in Tables~\ref{tab:tanh_gaussian_simple} and \ref{tab:tanh_gaussian_diagnostics}, the adjusted pairs are
\begin{equation}
\begin{aligned}
 (\widetilde\Lambda,\varphi_{\rm a})_{\rm A}
   &=(0.73305158,0.11075025),\\
 (\widetilde\Lambda,\varphi_{\rm a})_{\rm B}
   &=(0.73419977,0.12289609),\\
 (\widetilde\Lambda,\varphi_{\rm a})_{\rm C}
   &=(0.73219468,0.10424053).
\end{aligned}
\end{equation}

\begin{figure*}
\centering
\includegraphics[width=0.75\linewidth]{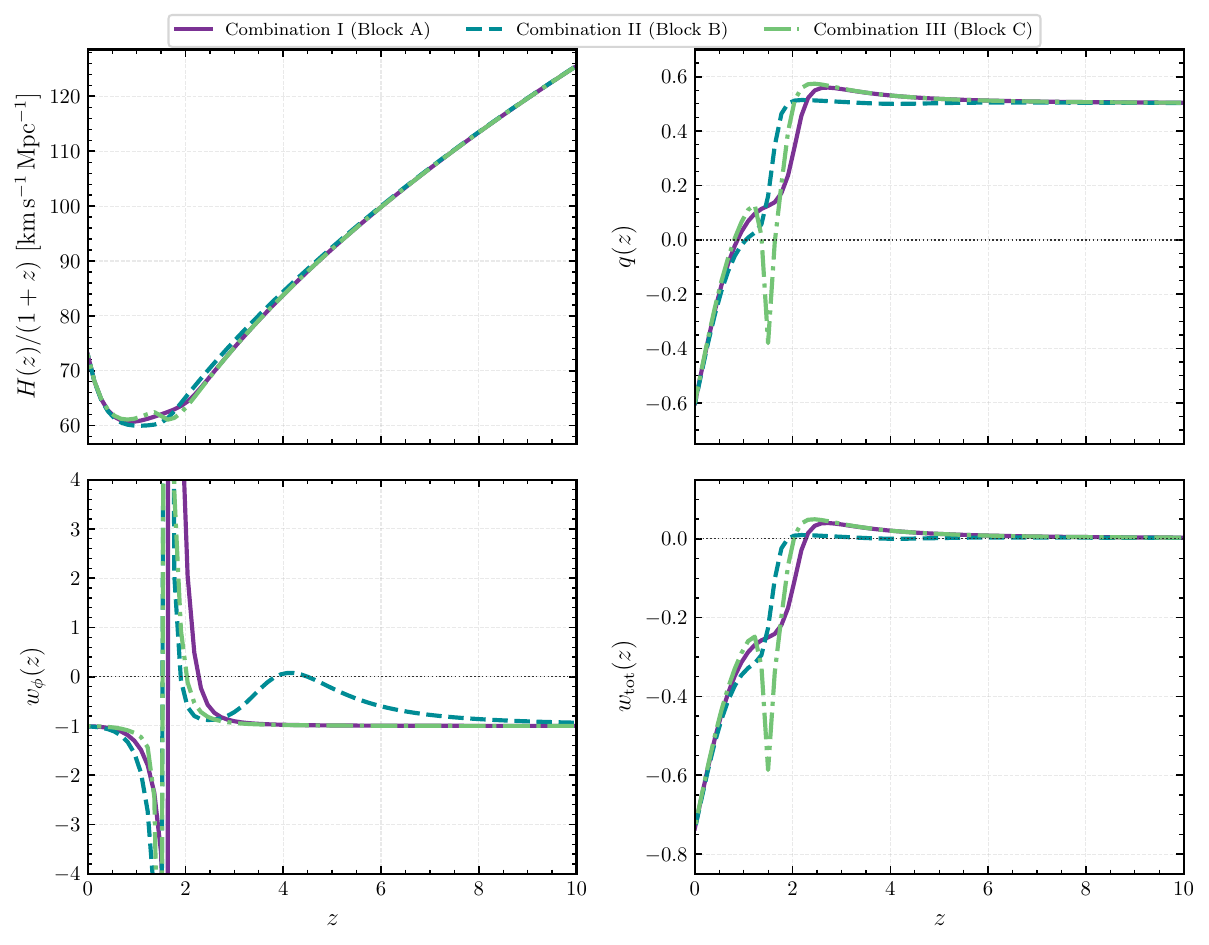}
\caption{Background quantities over the displayed interval $0\leq z\leq10$ for the three accepted sigmoid--Gaussian trajectories defined in Fig.~\ref{fig:density_parameters_sigmoid}: $H(z)/(1+z)$ in $\mathrm{km\,s^{-1}\,Mpc^{-1}}$ (top left), the deceleration parameter $q(z)$ (top right), the scalar equation-of-state ratio $w_\phi(z)=p_\phi/\rho_\phi$ (bottom left), and the total effective ratio $w_{\rm tot}(z)=p_{\rm tot}/\rho_{\rm tot}$ (bottom right). The integrations and acceptance checks cover $0\leq z\leq1090$. The $w_\phi$ panel is restricted to the displayed vertical range because this ratio diverges at $\widetilde\Omega_\phi=0$, although $\rho_\phi$ and $p_\phi$ remain finite. Combination I is purple, Combination II is turquoise, and Combination III is light green; their parameters and table-row mappings are given in Fig.~\ref{fig:density_parameters_sigmoid}.}
\label{fig:cosmo_metrics_sigmoid}
\end{figure*}

The numerical implementation and Eq.~\eqref{eq:sigmoid_template} both use $\exp[-(\varphi-\varphi_{\rm b})^2/\sigma_{\rm b}^2]$, so the quoted $\sigma_{\rm b}=0.05$ is the $1/e$ half-width and no $\sqrt{2}$ conversion is applied. The union of the complete sampled field ranges of the three trajectories is $0\leq\varphi\leq0.57774786$, with the upper endpoint supplied by Block B.

Two distinct zero crossings occur. We define $z_{\rm t}$ by $\widetilde\Omega_V(z_{\rm t})=0$ and $z_\dagger$ by $\widetilde\Omega_\phi(z_\dagger)=0$. On the phantom branch,
\begin{equation}
 \widetilde\Omega_\phi=\widetilde\Omega_V+\widetilde\Omega_K,
 \quad
 \widetilde\Omega_K=-\frac{1}{6}(1+z)^2E^2(\varphi')^2\leq0.
\end{equation}
At the density crossing one therefore has $\widetilde\Omega_V(z_\dagger)=-\widetilde\Omega_K(z_\dagger)\geq0$. When the potential rises from negative to positive values as cosmic time increases, its zero occurs first, and all tabulated cases consequently satisfy $z_{\rm t}>z_\dagger$. Neither zero is generically an inflection point. Accordingly, the zero of the potential must not be identified with the background DE transition unless the signed kinetic contribution is negligible; in a scalar implementation, $z_\dagger$, not $z_{\rm t}$, is the redshift at which the effective scalar density actually changes sign.

\begin{figure*}
\centering
\includegraphics[width=0.75\linewidth]{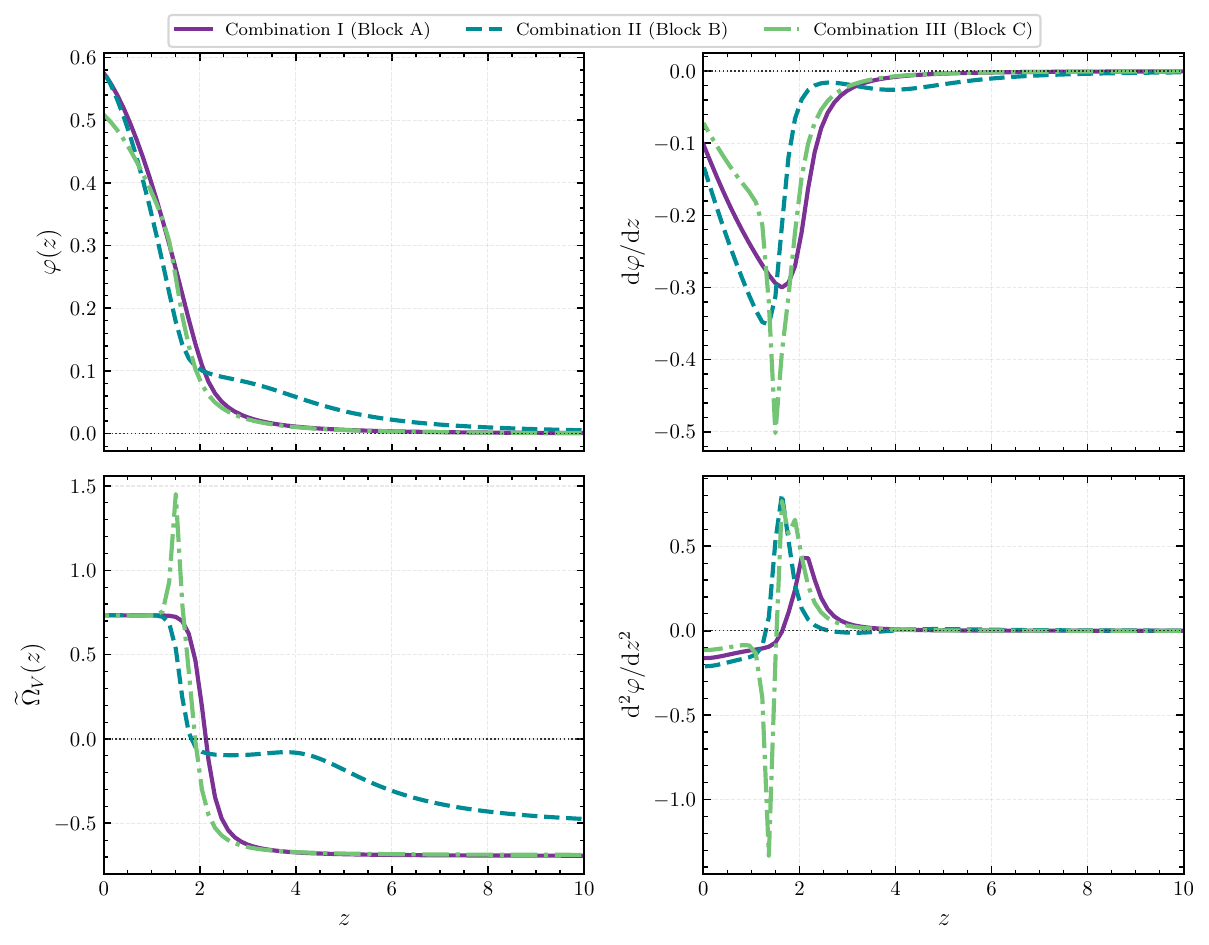}
\caption{Field evolution over the displayed interval $0\leq z\leq10$ for the three accepted sigmoid--Gaussian trajectories at $H_0=73\,\mathrm{km\,s^{-1}\,Mpc^{-1}}$ defined in Fig.~\ref{fig:density_parameters_sigmoid}: the dimensionless field $\varphi(z)$ (top left), its redshift derivative ${\rm d}\varphi/{\rm d}z$ (top right), the dimensionless potential $\widetilde\Omega_V(z)$ (bottom left), and the numerical redshift-coordinate acceleration ${\rm d}^2\varphi/{\rm d}z^2$ (bottom right). The integrations and acceptance checks cover $0\leq z\leq1090$. Combination I (Block A) is purple, Combination II (Block B) is turquoise, and Combination III (Block C) is light green. The curves correspond to the accepted rows of Table~\ref{tab:tanh_gaussian_simple}. The lower-right quantity is a coordinate-dependent trajectory diagnostic; it is neither the field-space curvature ${\rm d}^2\widetilde\Omega_V/{\rm d}\varphi^2$ nor a perturbative-stability measure.}
\label{fig:cosmo_metrics_field_sigmoid}
\end{figure*}

Figure~\ref{fig:potential_vs_field_sigmoid} displays the potential geometry sampled by the representative $H_0=73\,\mathrm{km\,s^{-1}\,Mpc^{-1}}$ solutions, whereas their detailed dynamical evolution as functions of redshift is tracked in Figs.~\ref{fig:density_parameters_sigmoid}--\ref{fig:cosmo_metrics_field_sigmoid}. Across these trajectories, the energy components remain finite through both zero crossings. The pole in $w_\phi=p_\phi/\rho_\phi$ is solely a ratio singularity at $\rho_\phi=0$, while $w_{\rm tot}$ and $q$ remain finite. The displayed Combination III curve also contains an additional transient interval with $q<0$ near the transition. This interval was identified from the zeros of $q(z)$ along the Block-C trajectory; it is a feature of this selected trajectory, not a generic prediction of the family, and no quantitative inference is attached to it. Overall, these quantities illustrate regular homogeneous background evolution for the selected cases, rather than perturbative or quantum stability.

The ranges below describe only the restricted numerical exploration used to construct the representative examples. They are neither observational priors nor confidence intervals, and failure of the shooting algorithm outside them does not prove that the corresponding solutions do not exist. Because $\varphi_{\rm in}=0$ fixes an otherwise arbitrary field origin, only relative locations such as $\varphi_{\rm b}-\varphi_{\rm a}$ are invariant.

\begin{itemize}
\item \textbf{Gaussian width $\sigma_{\rm b}$.} The scan uses $\sigma_{\rm b}\in[0.03,0.10]$, where $\sigma_{\rm b}$ is the $1/e$ half-width in $\exp[-(\varphi-\varphi_{\rm b})^2/\sigma_{\rm b}^2]$. A narrow feature may be traversed over only a small field interval or missed by the trajectory, whereas a broader feature modifies a larger portion of the evolution and can interfere with the assumed late-time plateau. For some wider choices, the restricted shooting search returned no trajectory passing the acceptance criteria.

\item \textbf{Relative feature amplitude $\alpha$.} The dynamical scan was restricted to $\alpha\in[0.30,1.25]$ with $\alpha>0$. The limit $\alpha\to0$ reduces the family to the pure sigmoid. The restricted search returned no accepted trajectories at larger amplitudes.

\item \textbf{Sigmoid width $\sigma_{\rm a}$.} The scan used $\sigma_{\rm a}\in[0.03,0.10]$. This parameter controls the transition width in field space, but the transition rate in redshift also depends on the field velocity and the Gaussian parameters.

\item \textbf{Feature location $\varphi_{\rm b}$.} The explored interval was $\varphi_{\rm b}\in[0.05,0.60]$ in the convention $\varphi_{\rm in}=0$. Within the explored grid, larger locations generally required narrower features for the trajectory to reach the late-time plateau.

\item \textbf{Sigmoid location $\varphi_{\rm a}$.} The search allowed $\varphi_{\rm a}\in[0.01,0.50]$. The tabulated cases in Table~\ref{tab:tanh_gaussian_simple} span approximately $0.0449$--$0.1246$.
\end{itemize}

The three displayed sigmoid--Gaussian trajectories provide a family-level illustration of regular homogeneous negative-to-positive scalar-density crossings; because they correspond to the accepted $H_0=73\,\mathrm{km\,s^{-1}\,Mpc^{-1}}$ rows, the residuals in Table~\ref{tab:tanh_gaussian_diagnostics} document their numerical convergence to the compressed targets.

%%-----------------------------------------------------------------------------%%
\subsection{Representative offset-axion solutions}
\label{subsec:append_axion}
%%-----------------------------------------------------------------------------%%

We next study homogeneous solutions in the generalized axion-like family. The evolutions shown here deliberately use the regular cosine choice $\eta=0.2$ and $n=1$. They are not the parameter vectors obtained in either the ECDM or SSCDM potential-space comparison. This subsection presents representative homogeneous realizations within smooth members of the family; it does not evolve or validate a posterior-summary potential.

\begin{figure*}
\centering
\includegraphics[width=0.98\linewidth]{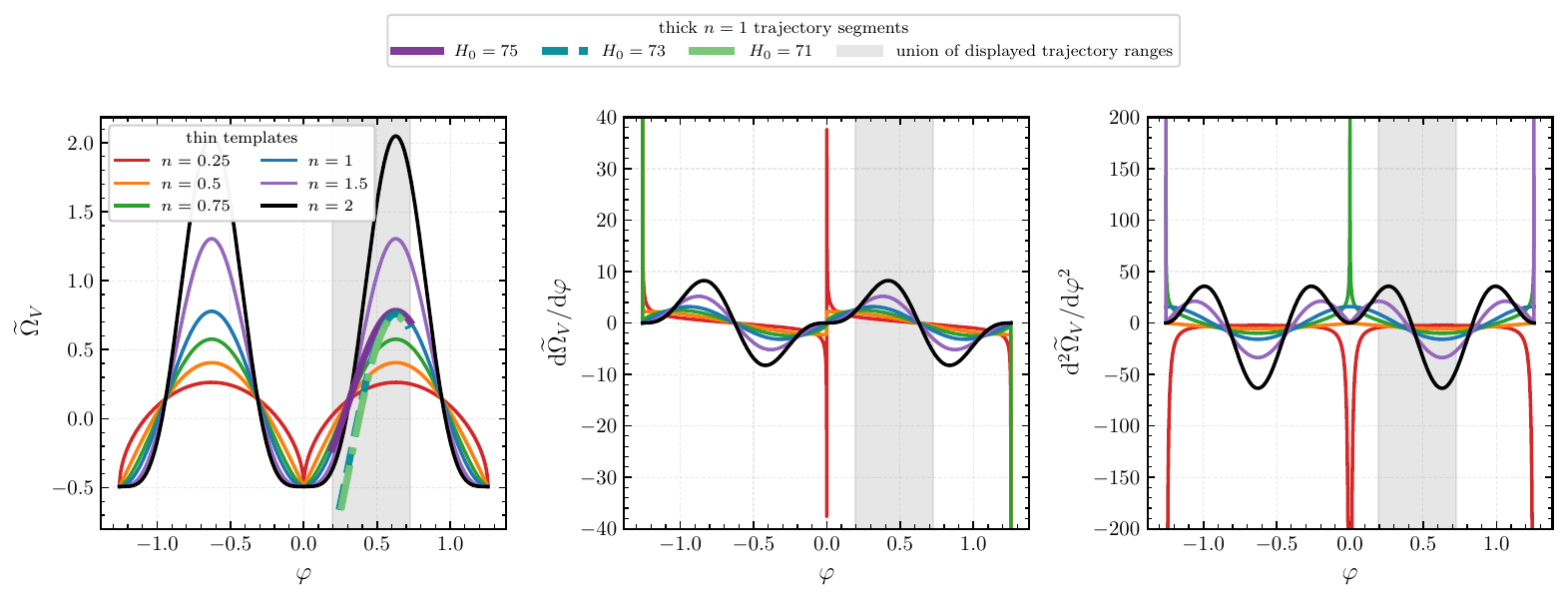}
\caption{The dimensionless cosine-power potential $\widetilde\Omega_V$ (left), its first derivative ${\rm d}\widetilde\Omega_V/{\rm d}\varphi$ (center), and its second derivative ${\rm d}^2\widetilde\Omega_V/{\rm d}\varphi^2$ (right) for $n=\{0.25,0.5,0.75,1,1.5,2\}$. All six thin template curves use the parameters of the accepted $H_0=75\,\mathrm{km\,s^{-1}\,Mpc^{-1}}$ continuation row, $(A,\eta,\widetilde V_0)=(0.635408,0.2,-0.492263)$; only $n$ is varied. For $n<1/2$ the slope diverges at a periodic minimum; at $n=1/2$ it has a finite jump; for $1/2<n<1$ the slope vanishes there while the curvature diverges; $n=1$ is the smooth cosine; and $n>1$ produces progressively flatter minima. Divergent derivative curves are clipped at the displayed vertical limits. The thick colored segments in the left panel show the portions of the regular $n=1$ potentials sampled by the three continuation trajectories plotted in Figs.~\ref{fig:append_Omega_axion}--\ref{fig:append_phase_axion}. Purple, turquoise, and light green correspond, respectively, to the $H_0=75\,\mathrm{km\,s^{-1}\,Mpc^{-1}}$, $H_0=73\,\mathrm{km\,s^{-1}\,Mpc^{-1}}$, and $H_0=71\,\mathrm{km\,s^{-1}\,Mpc^{-1}}$ rows of Table~\ref{tab:axion_V0_scan}, with $(A,\eta,\widetilde V_0)=(0.635408,0.2,-0.492263)$, $(1.049063,0.2,-1.338754)$, and $(1.083161,0.2,-1.423469)$. Only the $H_0=75\,\mathrm{km\,s^{-1}\,Mpc^{-1}}$ row satisfies the full acceptance criterion; the other two are the $\dagger$-marked diagnostic continuation rows of Table~\ref{tab:axion_V0_scan}. The gray interval $0.1936\leq\varphi\leq0.7276$ is the union of the field ranges sampled by these three trajectories over the full integration interval $0\leq z\leq1090$. The shading identifies the dynamically sampled field region and is not an uncertainty or stability band. Only the regular $n=1$ case is dynamically evolved; the remaining thin curves diagnose the mathematical regularity of the broader template family.}
\label{fig:append_potential_axion}
\end{figure*}

In the notation of this section, the dimensionless potential is
\begin{equation}\label{eq:axion_potential}
\begin{aligned}
 \widetilde\Omega_V(\varphi)&=
 A\left(1-\cos\frac{\varphi}{\eta}\right)^n+\widetilde V_0,\\
 A&\equiv\frac{\Lambda^4}{\rho_{\rm c0}}, \qquad \widetilde V_0\equiv\frac{V_0}{\rho_{\rm c0}}.
\end{aligned}
\end{equation}
For $A>0$, the periodic range is $\widetilde V_0\leq\widetilde\Omega_V \leq\widetilde V_0+2^nA$. Thus the potential itself takes both signs iff $-2^nA<\widetilde V_0<0$; unlike the sigmoid family, it has no field-space asymptotic plateau. For the regular $n=1$ potential used in the dynamical examples, the initial conditions
\begin{equation}
 \varphi_{\rm in}=0,
 \qquad
 \varphi'_{\rm in}=0,
\end{equation}
place the field exactly at a periodic minimum, where $\widetilde\Omega_{V,\varphi}=0$. With zero velocity, uniqueness of the smooth Klein--Gordon initial-value problem then leaves the field frozen. A nontrivial trajectory consequently requires a nonzero initial displacement. We retain a zero initial field derivative with respect to redshift,
\begin{equation}
 \varphi'_{\rm in}=0,
\end{equation}
and treat $\varphi_{\rm in}$ as a shooting parameter. For each selected $H_0$ in the tabulated continuation scan, the parameters $(A,\widetilde V_0,\varphi_{\rm in})$ were adjusted within the continuation procedure described above. Because the shooting prescription treats $z_\dagger$ as a target whereas $z_{\rm t}$ is derived after integration, the tabulated $z_\dagger$ values and their apparent dependence on selected $H_0$ are not independent predictions.

The first two field derivatives of Eq.~\eqref{eq:axion_potential} are

\begin{equation}\label{eq:axion_potential_derivative}
 \frac{{\rm d}\widetilde\Omega_V}{{\rm d}\varphi}
 =\frac{An}{\eta}
 \left(1-\cos\frac{\varphi}{\eta}\right)^{n-1}
 \sin\frac{\varphi}{\eta}.
\end{equation}

\begin{equation} \label{eq:axion_potential_sec_deriv}
\begin{aligned}
 \frac{{\rm d}^2\widetilde\Omega_V}{{\rm d}\varphi^2}
 ={}&\frac{An(n-1)}{\eta^2}
 \left(1-\cos\frac{\varphi}{\eta}\right)^{n-2}
 \sin^2\frac{\varphi}{\eta}\\
 &+\frac{An}{\eta^2}
 \left(1-\cos\frac{\varphi}{\eta}\right)^{n-1}
 \cos\frac{\varphi}{\eta}.
\end{aligned}
\end{equation}

\begin{table}
\centering
\footnotesize
\renewcommand{\arraystretch}{1.10}
\setlength{\tabcolsep}{2.2pt}
\textbf{(a) Parameter values}\\[2pt]
\begin{tabular}{@{}ccccc@{}}
\toprule
$H_0$ & $\Omega_{\rm m0}$ & $A$ & $\widetilde V_0$ & $\varphi_{\rm in}$\\
\midrule
$\mathbf{75}$ & 0.2545 & 0.635408 & $-0.492263$ & 0.193560\\
$73^\dagger$ & 0.2686 & 1.049063 & $-1.338754$ & 0.241600\\
$71^\dagger$ & 0.2839 & 1.083161 & $-1.423469$ & 0.254103\\
$69^\dagger$ & 0.3006 & 1.119116 & $-1.513231$ & 0.266738\\
$68^\dagger$ & 0.3096 & 1.137886 & $-1.560246$ & 0.273119\\
$67^\dagger$ & 0.3189 & 1.157216 & $-1.608781$ & 0.279550\\
\bottomrule
\end{tabular}

\medskip
\textbf{(b) Derived zero crossings}\\[2pt]
\begin{tabular}{@{}cccc@{}}
\toprule
$H_0$ & $z_{\rm t}$ & $z_\dagger$ & $\Delta z_{{\rm t}\dagger}$\\
\midrule
$\mathbf{75}$ & 2.193163 & 1.800000 & 0.393163\\
$73^\dagger$ & 2.222453 & 1.803568 & 0.418884\\
$71^\dagger$ & 2.243888 & 1.819894 & 0.423994\\
$69^\dagger$ & 2.262801 & 1.834134 & 0.428667\\
$68^\dagger$ & 2.271351 & 1.840503 & 0.430848\\
$67^\dagger$ & 2.279308 & 1.846379 & 0.432929\\
\bottomrule
\end{tabular}

\caption{Offset-axion continuation scan for $\eta=0.2$ and $n=1$, obtained using the Planck 2018 one-massive-neutrino prescription translated into CLASS variables, $(T_{\rm ncdm}/T_\gamma,N_{\rm ur})=(0.71611,2.0328)$, corresponding to $N_{\rm eff}\simeq3.046$, and expressed in the critical-density normalization of Eq.~\eqref{eq:axion_potential}. The tabulated values of $H_0$ are in $\mathrm{km\,s^{-1}\,Mpc^{-1}}$. The adjusted quantities are $(A,\widetilde V_0,\varphi_{\rm in})$, with $\varphi'_{\rm in}=0$, $z_{\dagger,{\rm target}}=1.8$, and $\Omega_{\rm m0}=\omega_{\rm m}/h^2$. Here $\Delta z_{{\rm t}\dagger}\equiv z_{\rm t}-z_\dagger$ is evaluated from the unrounded adaptive-event locations; the endpoint and optimization diagnostics of every row are collected in Table~\ref{tab:axion_diagnostics} of Appendix~\ref{app:diag}. The rows are obtained sequentially, using the preceding $H_0$ solution as a continuation start. The bold $H_0=75\,\mathrm{km\,s^{-1}\,Mpc^{-1}}$ row is the only row satisfying the complete operational acceptance rule: successful optimizer termination, $\mathcal C_{\rm LS}^{({\rm ax})}<10^{-3}$, a finite integration with positive $E^2$, finite $z_{\rm t}$ and $z_\dagger$, and no adjusted parameter pinned to a bound. Rows marked by $\dagger$ have valid integrations and crossings and are not bound-pinned, but fail the cost threshold; they are retained only as lowest-cost diagnostic continuation trajectories. A dagger is not a non-existence statement for the corresponding input $H_0$.}
\label{tab:axion_V0_scan}
\end{table}

Equations~\eqref{eq:axion_potential_derivative} and \eqref{eq:axion_potential_sec_deriv} are pointwise formulas away from the periodic minima $\varphi=2\pi k\eta$. At a minimum, a derivative is defined by its limiting value only when that limit exists.

The six thin curves in Fig.~\ref{fig:append_potential_axion} are evaluated on $-2\pi\eta\leq\varphi\leq2\pi\eta$.

\begin{figure}
\centering
\includegraphics[width=\columnwidth]{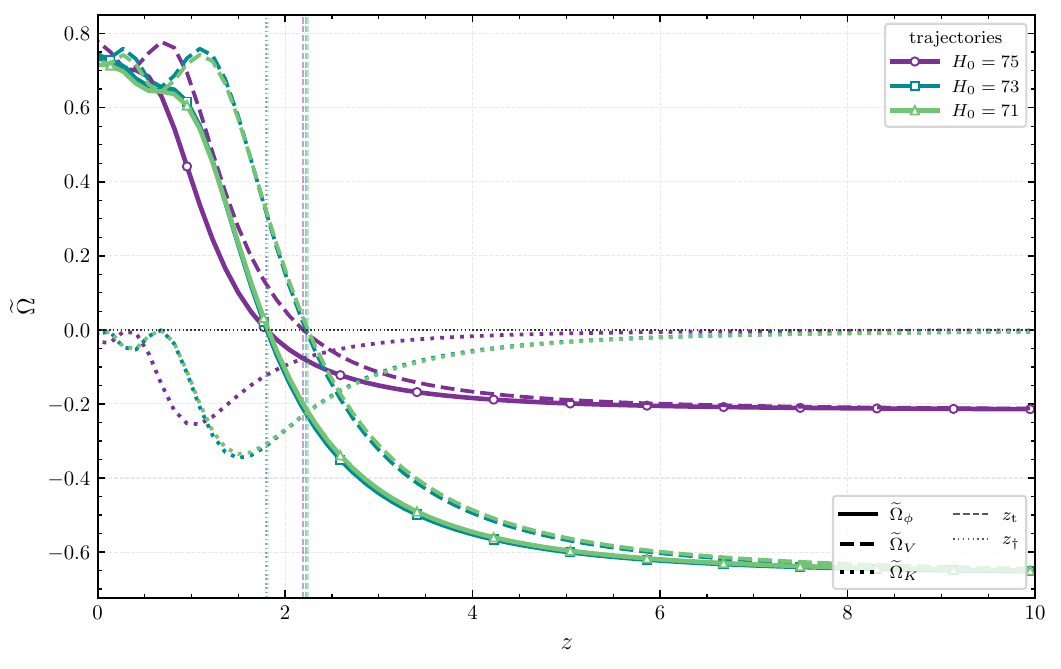}
\caption{Energy components over the displayed interval $0\leq z\leq10$ for three representative regular offset-axion continuation trajectories with $\eta=0.2$, $n=1$, $\xi=-1$, and $\varphi'_{\rm in}=0$. The integrations and acceptance checks cover $0\leq z\leq1090$. Solid, dashed, and dotted curves show $\widetilde\Omega_\phi=\rho_\phi/\rho_{\rm c0}$, $\widetilde\Omega_V=V/\rho_{\rm c0}$, and the signed phantom kinetic contribution $\widetilde\Omega_K=K/\rho_{\rm c0}\leq0$, respectively. Purple, turquoise, and light green correspond to the $H_0=75\,\mathrm{km\,s^{-1}\,Mpc^{-1}}$, $H_0=73\,\mathrm{km\,s^{-1}\,Mpc^{-1}}$, and $H_0=71\,\mathrm{km\,s^{-1}\,Mpc^{-1}}$ rows of Table~\ref{tab:axion_V0_scan}, with $(A,\widetilde V_0,\varphi_{\rm in})=(0.635408,-0.492263,0.193560)$, $(1.049063,-1.338754,0.241600)$, and $(1.083161,-1.423469,0.254103)$, respectively. Thin vertical dashed lines mark $z_{\rm t}$, defined by $\widetilde\Omega_V(z_{\rm t})=0$, while thin vertical dotted lines mark $z_\dagger$, defined by $\widetilde\Omega_\phi(z_\dagger)=0$. Only the $H_0=75\,\mathrm{km\,s^{-1}\,Mpc^{-1}}$ row satisfies the full acceptance criterion; the other two are the $\dagger$-marked diagnostic continuation rows of Table~\ref{tab:axion_V0_scan}. The curves correspond to the rows identified in Table~\ref{tab:axion_V0_scan}. They are representative smooth members of the regular offset-axion family and are not the ECDM or SSCDM posterior-summary potentials.}
\label{fig:append_Omega_axion}
\end{figure}

To make the regularity distinction explicit, expand about a periodic minimum $\varphi_k=2\pi k\eta$ and define $\Delta\varphi\equiv\varphi-\varphi_k$. For $|\Delta\varphi|/\eta\ll1$,
\begin{equation}
 1-\cos\frac{\varphi}{\eta}\simeq
 \frac{1}{2}\left(\frac{\Delta\varphi}{\eta}\right)^2,
 \qquad
 \sin\frac{\varphi}{\eta}\simeq\frac{\Delta\varphi}{\eta}.
\end{equation}
Hence
\begin{equation}
 \widetilde\Omega_V(\varphi)\simeq
 \widetilde V_0+A2^{-n}
 \left|\frac{\Delta\varphi}{\eta}\right|^{2n}.
\end{equation}
Using Eqs.~\eqref{eq:axion_potential_derivative} and \eqref{eq:axion_potential_sec_deriv}, the leading behavior of the first derivative is
\begin{equation}
\frac{{\rm d}\widetilde\Omega_V}{{\rm d}\varphi}
\simeq
A n\,2^{1-n}\,\eta^{-2n}\,
{\rm sgn}(\Delta\varphi)\,
|\Delta\varphi|^{2n-1},
\end{equation}
while the second derivative behaves as
\begin{equation}
\frac{{\rm d}^2\widetilde\Omega_V}{{\rm d}\varphi^2}
\simeq
A n(2n-1)\,2^{1-n}\,\eta^{-2n}\,
|\Delta\varphi|^{2n-2},
\end{equation}
for $\Delta\varphi\neq0$. At $n=1/2$ the second derivative is distributional at the cusp and is not represented by substituting $n=1/2$ directly into this pointwise expression. For $n<1/2$ the first derivative diverges; at $n=1/2$ the potential is locally proportional to $|\Delta\varphi|$ and its slope jumps; and for $1/2<n<1$ the first derivative vanishes but the second derivative diverges. The case $n=1$ is the usual smooth cosine with finite curvature, while $n>1$ gives a vanishing curvature at the minimum. These qualifications apply to the global template family. The trajectories below use $n=1$ and therefore do not dynamically test the noninteger members favored by either potential-space comparison.

\begin{figure*}
\centering
\includegraphics[width=0.75\linewidth]{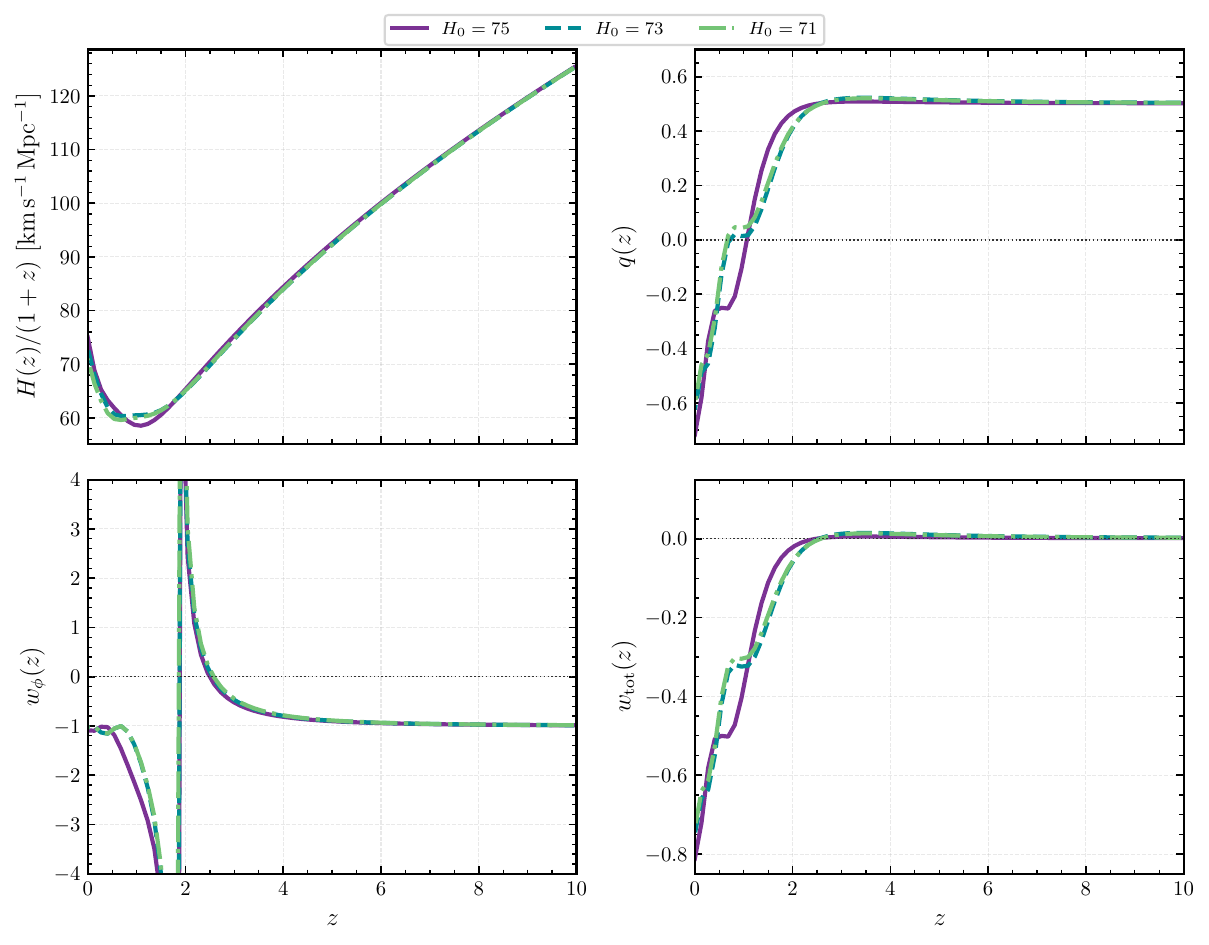}
\caption{Background quantities over the displayed interval $0\leq z\leq10$ for the three representative regular offset-axion continuation trajectories shown in Fig.~\ref{fig:append_Omega_axion}: $H(z)/(1+z)$ in $\mathrm{km\,s^{-1}\,Mpc^{-1}}$ (top left), the deceleration parameter $q(z)$ (top right), the scalar equation-of-state ratio $w_\phi(z)=p_\phi/\rho_\phi$ (bottom left), and the total effective ratio $w_{\rm tot}(z)=p_{\rm tot}/\rho_{\rm tot}$ (bottom right). The integrations and acceptance checks cover $0\leq z\leq1090$. Purple, turquoise, and light green denote the $H_0=75\,\mathrm{km\,s^{-1}\,Mpc^{-1}}$, $H_0=73\,\mathrm{km\,s^{-1}\,Mpc^{-1}}$, and $H_0=71\,\mathrm{km\,s^{-1}\,Mpc^{-1}}$ rows of Table~\ref{tab:axion_V0_scan}, respectively. Only the $H_0=75\,\mathrm{km\,s^{-1}\,Mpc^{-1}}$ row satisfies the full acceptance criterion; the other two are the $\dagger$-marked diagnostic continuation rows of Table~\ref{tab:axion_V0_scan}. The $w_\phi$ panel is restricted to the displayed vertical range because the ratio diverges at $\widetilde\Omega_\phi=0$, although $\rho_\phi$ and $p_\phi$ remain finite. All three curves correspond to the rows identified in Table~\ref{tab:axion_V0_scan}.}
\label{fig:append_cosmology_axion}
\end{figure*}

\begin{figure*}
\centering
\includegraphics[width=0.75\linewidth]{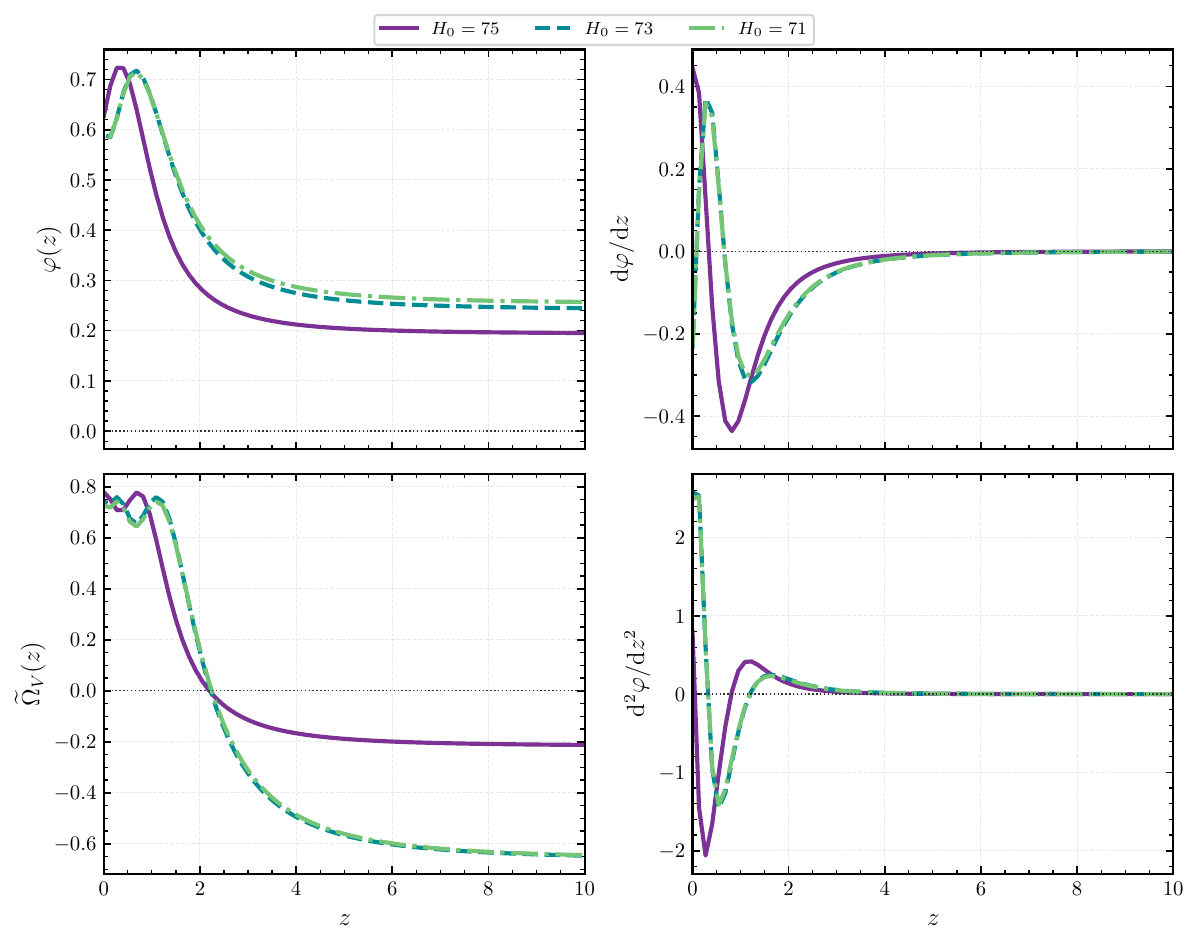}
\caption{Field evolution over the displayed interval $0\leq z\leq10$ for the three representative regular offset-axion continuation trajectories specified in Fig.~\ref{fig:append_Omega_axion}: the dimensionless field $\varphi(z)$ (top left), its redshift derivative ${\rm d}\varphi/{\rm d}z$ (top right), the dimensionless potential $\widetilde\Omega_V(z)$ (bottom left), and the numerical redshift-coordinate acceleration ${\rm d}^2\varphi/{\rm d}z^2$ (bottom right). The integrations and acceptance checks cover $0\leq z\leq1090$. Purple, turquoise, and light green denote the $H_0=75\,\mathrm{km\,s^{-1}\,Mpc^{-1}}$, $H_0=73\,\mathrm{km\,s^{-1}\,Mpc^{-1}}$, and $H_0=71\,\mathrm{km\,s^{-1}\,Mpc^{-1}}$ rows of Table~\ref{tab:axion_V0_scan}, respectively. Only the $H_0=75\,\mathrm{km\,s^{-1}\,Mpc^{-1}}$ row satisfies the full acceptance criterion; the other two are the $\dagger$-marked diagnostic continuation rows of Table~\ref{tab:axion_V0_scan}. Each field is initialized with $\varphi'_{\rm in}=0$ and subsequently evolves toward the vicinity of a potential maximum, as expected for a homogeneous phantom field. The lower-right quantity is a coordinate-dependent trajectory diagnostic; it is neither the field-space potential curvature ${\rm d}^2\widetilde\Omega_V/{\rm d}\varphi^2$, nor ${\rm d}^2\widetilde\Omega_V/{\rm d}z^2$, nor a perturbative-stability measure. All curves correspond to the rows identified in Table~\ref{tab:axion_V0_scan}.}
\label{fig:append_phase_axion}
\end{figure*}
Figure~\ref{fig:append_potential_axion} visualizes the regularity statements above across the broader cosine-power family. The extrema lie at $\varphi=2\pi k\eta$ and $(2k+1)\pi\eta$. For the regular $n=1$ solutions actually evolved here, the phantom climbs toward a potential maximum and the sampled field interval stays away from the periodic minima. Homogeneous linearization about an extremum gives $\delta\ddot\varphi+3H\delta\dot\varphi -3H_0^2\widetilde\Omega_{V,\varphi\varphi}(\varphi_*)\delta\varphi=0$ for $\xi=-1$, so the potential-curvature criterion is reversed relative to a canonical scalar; this is the reversal underlying the endpoint classification of Eq.~\eqref{eq:endpoint_sign}.

The continuation diagnostics define the parameter rows listed in Table~\ref{tab:axion_V0_scan}. The parameter triplets quoted in the corresponding figure captions are their six-decimal roundings. The plotted trajectories and zero crossings were exported from the same unrounded diagnostics rows and parameter-validated trajectory caches. For the shading and thick potential segments in Fig.~\ref{fig:append_potential_axion}, the field-range boundaries are evaluated from the integration endpoints and event-located field turning points, rather than from the extrema of a coarse plotting grid. The shading covers the combined field range $0.1936\leq\varphi\leq0.7276$ traversed over the full integration interval $0\leq z\leq1090$.

In the continuation scan, $H_0$ is an input rather than an inferred parameter. Moreover, because $z_\dagger$ is part of the shooting prescription, the scan does not predict a relation between $H_0$ and the density-crossing redshift.

Figures~\ref{fig:append_Omega_axion}--\ref{fig:append_phase_axion} show three regular $n=1$ examples. Their signed kinetic contributions are non-positive (and negative while the field evolves), the field climbs toward a potential maximum, and the total scalar density changes sign after the potential does, so $z_{\rm t}>z_\dagger$. The poles in $w_\phi$ occur only because the denominator $\rho_\phi$ passes through zero; the underlying density, pressure, field, and expansion variables remain finite in the plotted solutions.

Figure~\ref{fig:append_phase_axion} makes the time orientation of the phantom motion explicit. The field starts below the nearest potential maximum, climbs through its neighborhood, and in the displayed solutions overshoots before turning back toward it. This behavior is consistent with the reversed homogeneous curvature criterion for a phantom field.

The displayed axion integrations show that the smooth $n=1$ offset-cosine family contains homogeneous phantom trajectories with complete negative-to-positive scalar energy-density crossings. They are family-level illustrations; in particular, they do not evolve either fitted noninteger axion potential of Sec.~\ref{sec:results}.

%%-----------------------------------------------------------------------------%%
\subsection{A dynamical closure test}
\label{subsec:closure}
%%-----------------------------------------------------------------------------%%

\begin{figure*}[ht!]
\centering
\includegraphics[width=0.75\linewidth]{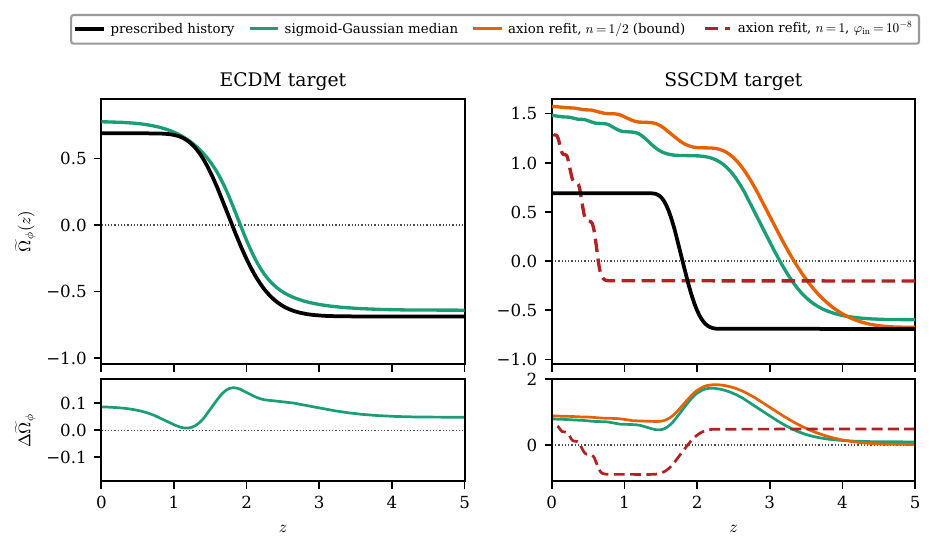}
\caption{Dynamical closure test. Forward Klein--Gordon--Friedmann evolution of fitted potentials over the reconstruction range $0\leq z\leq5$, on the same low-redshift background as the reconstruction arrays, compared with the prescribed histories (black). Left: the ECDM target and the evolution of the sigmoid--Gaussian marginal-median potential from the reconstructed initial data at $z=5$. Right: the SSCDM target and the evolutions of the sigmoid--Gaussian median, launched from the exactly frozen plateau state, and of two constrained axion refits---the cusp-bound $n=1/2$ diagnostic and the regular $n=1$ member---launched from the seeded state $(10^{-8},0)$ (see text). The lower panels show $\widetilde\Omega_\phi-\widetilde\Omega_{\rm de}$. These are pure initial-value integrations: no quantity is shot for and $E(0)=1$ is not imposed. The large right-panel residuals are therefore the intended closure diagnostic, not numerical failures: the sigmoid--Gaussian member departs immediately under its nonvanishing endpoint force, as does the seeded cusp under the positive-side prescription. The regular $n=1$ member is displaced from its unstable minimum by the stated seed, which sets the epoch at which its departure becomes appreciable; this is not a family-level non-existence statement.}
\label{fig:closure}
\end{figure*}

The three layers of the analysis have so far been kept separate: the reconstruction defines an on-shell target, the potential-space comparison scores parametric representations of it, and the preceding subsections evolve representative regular members. The remaining question is an initial-value closure test: once the fitted potential vectors and initial data are specified, how closely do their forward evolutions reproduce the prescribed histories without any dynamical retuning? For this test we integrate the homogeneous system over the reconstruction range $0\leq z\leq5$, with the same low-redshift background prescription as the reconstruction arrays (Sec.~\ref{subsec:dark_energy_profile}), so that the forward solution and its target are defined on identical backgrounds. The initial data at $z=5$ are $(\widetilde\phi,\widetilde\phi')=(0,-\mathcal Q(5))$ for ECDM, with $\mathcal Q(5)\simeq2.9\times10^{-5}$, and, for SSCDM, the exactly frozen reconstructed state $(0,0)$ in the sigmoid--Gaussian run and the explicitly seeded neighboring state $(10^{-8},0)$ in the two axion diagnostics. Throughout this test, $H_0$ and $\rho_{\rm c0}$ retain the target normalization; they are not redefined using the present-day expansion rate of the forward solution. These are pure initial-value integrations: in contrast with the shooting scans above, nothing is adjusted, $E(0)=1$ is not imposed, and its violation is part of the measured closure error. Figure~\ref{fig:closure} shows the outcome. The ECDM and sigmoid--Gaussian SSCDM vectors are the corresponding marginal-median points of Sec.~\ref{sec:results}, used as representative members rather than as posterior samples; the two axion vectors are new constrained least-squares refits to the noiseless SSCDM target, as described below.

For ECDM the loop closes approximately; its residual is ordinary smooth-template error rather than an endpoint obstruction. The top-ranked sigmoid--Gaussian median reproduces the target history with $\max_z|\widetilde\Omega_\phi-\widetilde\Omega_{\rm de}|=0.158$, displaces the density zero to $z_\dagger=1.913$ (target $1.800$), and yields $E(0)=1.043$ rather than the target normalization $E(0)=1$, a $4.3\%$ mismatch in $E(0)$. The residual at $z=0$, $+0.087$, is close to and largely accounted for by the asymptotic plateau offset $\Lambda/\rho_{\rm c0}-\Omega_{\rm de0}=+0.088$ of the fitted member: the trajectory approaches the template's own late-time plateau from below rather than the target's, in accordance with Eq.~\eqref{eq:endpoint_sign}. The late-time portion of this initial-value test also probes the continuation of the sigmoid--Gaussian ansatz beyond the fitted field interval $0\leq\widetilde\phi\leq0.38$.

For compact SSCDM the loop does not close: the large right-panel residuals are the intended structural diagnostic, not a numerical failure. The sigmoid--Gaussian median has a nonvanishing force at the frozen plateau state, so the field departs immediately at $z=5$: the transition occurs far too early and too broadly, with $\max_z|\widetilde\Omega_\phi-\widetilde\Omega_{\rm de}|=1.72$ and $z_\dagger=3.14$. Constrained least-squares refits of the axion family to the noiseless target, using the weights $\sigma_i$ of Eq.~\eqref{eq:mock_variance}, behave in the same way whenever the endpoint force does not vanish: excluding divergent-force exponents by imposing $n\geq1/2$ drives the fit to the boundary, where it still represents the sampled values well ($\chi^2/N=0.086$, against $0.033$ unconstrained) but again departs the plateau at once, giving $z_\dagger=3.33$. At the boundary the fitted potential has a cusp, $\widetilde V-\widetilde V(0)\propto|\widetilde\phi|$ near $\widetilde\phi=0$, so the endpoint force possesses only one-sided limits and this run is a cusp-bound diagnostic rather than a regular classical completion. The seed $\varphi_{\rm in}=10^{-8}$ selects the $\varphi>0$ branch. For the unregularized cusp, the positive-side force magnitude tends to $A/(\sqrt{2}\,\eta)\simeq24$ in the dimensionless units used here, so the trajectory generated with this one-sided prescription departs immediately. It neither defines classical evolution from the cusp itself nor establishes independence under a smooth regularization. The condition $n\geq1/2$ excludes a divergent one-sided force but includes the cusp boundary; a $C^1$ potential with continuous finite force requires $n>1/2$, while local Lipschitz regularity of the force and the standard uniqueness theorem apply for $n\geq1$. The regular $n=1$ potential admits the exactly frozen solution from $(0,0)$ by uniqueness. With the displayed seeded state $(10^{-8},0)$, the field instead evolves immediately; the seed controls the crossing timescale and yields $z_\dagger=0.64$. The continuation scans of Sec.~\ref{subsec:append_axion} show that jointly shooting the potential parameters and initial displacement can place the crossing at selected values within these families; such trajectories do not reproduce the exact finite-duration plateau from the same frozen reconstructed data. These refits are function-space fits under the synthetic weights; the evidence comparison is not repeated here. Relative to the unconstrained refit, the $n\geq1/2$ condition increases the noiseless-target best-fit cost by $\Delta\chi^2\simeq5.3$. This diagnostic is evaluated against the noiseless target, whereas the marginal likelihoods in Table~\ref{tab:model_comparison_combined} refer to the Gaussian mock and depend additionally on the prior and posterior volumes. It therefore carries no implication for the evidence ranking; a dedicated calculation under a precisely specified endpoint prior is deferred to future work.

The closure test therefore quantifies the separation of the three layers. For the smooth ECDM history, the evolved top-ranked representative tracks the target only approximately, with $\Delta z_\dagger\simeq0.11$ and a $4.3\%$ mismatch in $E(0)$. For compact SSCDM, no potential with a locally Lipschitz force can reproduce the exact compact history from its exactly frozen plateau data: a vanishing endpoint force leaves the solution frozen by uniqueness, whereas a nonzero endpoint force produces immediate departure. The fitted members tested here realize these alternatives, either breaking the frozen plateaus or turning their transition timescales into initial-data choices. Function-value scores and dynamical adequacy thus genuinely decouple, realizing the endpoint classification of Eq.~\eqref{eq:endpoint_sign} at the level of complete solutions.

%%-----------------------------------------------------------------------------%%
\section{Conclusions}
\label{sec:conclude}
%%-----------------------------------------------------------------------------%%

In this work, we studied the homogeneous inverse problem of representing prescribed sign-switching DE histories using a minimally coupled scalar field with a fixed kinetic sign. The useful reconstruction variable is $\rho_{\rm de}+p_{\rm de}=(1+z)\rho'_{\rm de}/3$, rather than the ratio $w_{\rm de}=p_{\rm de}/\rho_{\rm de}$. At a differentiable density zero, $w_{\rm de}$ has a ratio pole, while the density, pressure, signed kinetic contribution, potential, and homogeneous geometry can remain regular. For the continuous negative-to-positive histories considered here, the fixed-sign condition selects the phantom branch, $\xi=-1$, wherever the density evolves. This is a statement about their representation within the fixed-sign action adopted here, not an identification of their microscopic origin. The labels \emph{AdS-like} and \emph{dS-like} refer only to negative- and positive-vacuum-energy-like regimes of the DE sector, not to the global matter--radiation--DE spacetime.

The central result is that phenomenologically similar histories have different field-theoretic status. ECDM is smooth, and its field map is locally invertible at every finite redshift, including the density crossing, provided $E^2>0$. This finite-redshift result does not imply a unique global off-shell completion. Indeed, at the compactified infinite-future field endpoint, the force tends to zero while the curvature grows logarithmically, as shown in Eq.~\eqref{eq:ecdm_asymptotic_endpoint}; that endpoint is reached only asymptotically and produces no finite-time waiting ambiguity. Exact compact SSCDM also has a regular homogeneous trajectory, but its finite transition is encoded by a potential that is $C^1$ but not $C^2$ at the two endpoints. The local law $V-V_e\propto|\phi-\phi_e|^{4/3}$ gives a continuous non-Lipschitz force and divergent curvature. It is precisely this loss of Lipschitz regularity that permits the one-parameter family of delayed departures in Eq.~\eqref{eq:sscdm_waiting_solution}. This is a one-sided, on-shell statement: the inverse reconstruction fixes the traversed transition-side branch, not a unique off-shell continuation through either endpoint. Conventional analytic potentials may approximate the retained transition-side interval but cannot reproduce the exact compact plateaus globally.

These endpoint results are unified by the elementary criterion of Eq.~\eqref{eq:endpoint_sign}. On the phantom branch, a plateau is always departed from a one-sided minimum of the on-shell potential---an unstable equilibrium of the reversed-stability phantom dynamics---and is arrived at through a one-sided minimum precisely when the density settles faster than $a^{-6}$, the scale-factor dilution law of free kinetic energy. Both continuous targets satisfy this unstable-arrival condition, ECDM with a Gaussian tail and compact SSCDM at finite time, so in both reconstructions the late-time dS-like state is reached only along a fine-tuned approaching branch and is not an attractor of the scalar dynamics; for compact SSCDM, the nonunique classical solutions even include re-departures after arbitrary waiting times. The corresponding asymptotic plateaus of the shifted-$\tanh$ and sigmoid--Gaussian templates, together with the positive-amplitude maxima of the Gaussian and regularized inverse-quadratic templates and the maxima of regular axion members, have the opposite local phantom stability and can furnish attractors on the appropriate branch. This qualitative difference is not detected by agreement in sampled potential values over a finite interval. The best-fitting $n<1/2$ axion members instead possess a divergent-force minimum at the included high-redshift endpoint and do not furnish differentiable completions on the closed fitting interval.

The strict L$\Lambda$CDM Ladder lies one level further outside the regular one-field configuration space. Its derivative consists of Dirac measures, and the scalar mapping would require the square of an ordinary locally square-integrable field velocity to reproduce those singular measures. No such ordinary classical field exists. The associated pressure, kinetic contribution, and potential contain impulses, and a mollified step has $\Delta\phi=\mathcal O(\sqrt{\epsilon})$, while its pointwise kinetic and potential peaks grow as $\mathcal O(\epsilon^{-1})$. Although their integrated impulse weights are fixed, the finite-width shapes and field-space curve are regulator dependent. The plotted Ladder reconstruction is therefore a finite-resolution representation of a distributional fluid history rather than the potential of a regular scalar field.

This no-go statement is intentionally restricted to the minimally coupled, fixed-sign, single-real-field action used here. It neither invalidates the Ladder as a phenomenological benchmark nor excludes a smooth regularization, additional degrees of freedom \cite{Cai:2009zp}, noncanonical kinetic structures \cite{Creminelli:2006xe}, interacting dark sectors \cite{Escamilla:2023shf}, or modified-gravity and string-motivated realizations \cite{Perivolaropoulos:2005yv,Akarsu:2024qsi,Anchordoqui:2023woo}. More generally, an effective sign-changing DE density can depend on the chosen matter--DE split. The analysis determines whether a history admits this particular homogeneous embedding; it does not establish that the embedding is the unique or fundamental description of that history.

For the ECDM on-shell curve and the retained SSCDM field interval, we compared five closed-form potential families under fixed field conventions, synthetic mocks, and target-dependent pre-mock prior scales. For ECDM, the sigmoid--Gaussian template ranks first, with the shifted-$\tanh$ family second at $\Delta\log\mathcal Z=-4.75$; for SSCDM, the generalized axion-like family ranks first, with the sigmoid--Gaussian second at $\Delta\log\mathcal Z=-5.98$; the remaining families follow at substantially larger deficits (Table~\ref{tab:model_comparison_combined}). These scores quantify representational performance within the two synthetic experiments, in the sense specified in Sec.~\ref{sec:method}, and both inverse-quadratic fits carry a prior-boundary qualification, with $\epsilon$ close to its upper limit for each target.

The fitted generalized axion-like representation of each target requires an additional physical qualification. In both cases, the inferred zero-phase exponent lies below $1/2$, implying that $V_{,\phi}$ diverges at the included periodic-minimum endpoint $\widetilde\phi=0$. Their evidence scores therefore assess the representation of potential values over the sampled field intervals, but the fitted members do not provide differentiable Klein--Gordon completions on the corresponding closed intervals. The regular $n=1$ offset-axion solutions used in the forward dynamical examples are distinct smooth members of the broader template family.

The forward integrations form a separate, non-observational layer. The displayed sigmoid--Gaussian curves use three accepted $H_0=73\,{\rm km\,s^{-1}\,Mpc^{-1}}$ shooting rows; in the offset-axion continuation scan, only the $H_0=75\,{\rm km\,s^{-1}\,Mpc^{-1}}$ row satisfies the strict cost threshold, while the $H_0=73\,{\rm km\,s^{-1}\,Mpc^{-1}}$ and $H_0=71\,{\rm km\,s^{-1}\,Mpc^{-1}}$ curves are explicitly retained as diagnostic continuations. In every case $H_0$ is prescribed, no observational likelihood is evaluated, and the trajectories are neither posterior-summary potentials nor dynamical reconstructions of the ECDM or SSCDM targets. The regular examples nevertheless exhibit complete sign changes in the scalar energy density. Because $\widetilde\Omega_K\leq0$ on the phantom branch, the potential zero precedes the density zero, $z_{\rm t}>z_\dagger$, for the displayed time orientation; the separation is trajectory dependent and is not claimed to be universal. The poles in $w_\phi$ at $z_\dagger$ are ratio singularities, while the underlying background variables remain finite.

The closure test then connects the analytic reconstruction, potential-space comparison, and forward dynamics. From reconstructed ECDM data, the top-ranked template follows the target only approximately, with $\max|\widetilde\Omega_\phi-\widetilde\Omega_{\rm de}|=0.16$, $\Delta z_\dagger\simeq0.11$, and a $4.3\%$ mismatch in $E(0)$. For compact SSCDM, the failure is structural: members with nonvanishing endpoint force leave the plateau immediately, whereas a regular $n=1$ potential launched from the exact extremum remains frozen by uniqueness. Its displayed seeded trajectory evolves immediately, with the seed controlling the crossing timescale. Thus, potential-value rankings and dynamical adequacy need not coincide, and the analytic hierarchy---regular finite-redshift ECDM, non-Lipschitz compact SSCDM, and the distributional Ladder---does not select a unique global scalar theory.

The practical lesson is that observational viability and field-theoretic realizability are complementary questions. Increasingly precise expansion measurements, including the DESI DR2 Ly$\alpha$ Alcock--Paczy\'nski anchor at $z_{\rm eff}=2.33$ \cite{DESI:2026lnd}, make the redshift regime relevant to these histories progressively testable, but a successful fit to distance data does not guarantee membership in the configuration space of a conventional scalar theory. A useful screening sequence is instead to work first with regular stress-tensor variables, impose the fixed-sign condition, test field-map invertibility, and then classify the endpoints of $V(\phi)$---their regularity, and their stability through the $a^{-6}$ arrival criterion of Eq.~\eqref{eq:endpoint_sign}. This prevents an EoS ratio pole from being mistaken for a physical singularity, or a nonunique compact embedding or distributional weak limit from being mistaken for ordinary scalar dynamics. The screen is complementary to perturbation and likelihood studies and is especially useful because background observations do not uniquely determine scalar microphysics \cite{Garcia-Garcia:2026nzy}.

Finally, the phantom representation remains an effective homogeneous proxy. For $\xi=-1$, the two-derivative action has a negative kinetic eigenvalue and is a genuine quantum ghost; neither its unit rest-frame sound speed nor regular potential curvature removes the unbounded Hamiltonian. The potential curves and homogeneous trajectories, therefore, do not constitute a perturbative-stability analysis or a fundamental completion.

Future work should test the robustness of both potential-space rankings under alternative independently specified physical priors, multiple synthetic realizations, field-coordinate conventions, and full-endpoint fits for all five SSCDM families; extend the closure test of Sec.~\ref{subsec:closure} to full posterior samples and, if warranted, to an evidence calculation under a precisely specified bounded-force, $C^1$-potential, or locally-Lipschitz-force prior; and confront the resulting dynamical histories with perturbations and joint cosmological likelihoods. A fundamental interpretation would additionally require a controlled ghost-free completion.

%%-----------------------------------------------------------------------------%%
\section*{Data and code availability}
%%-----------------------------------------------------------------------------%%

Numerical data, configuration files, selected validation scripts, and derived products supporting this work are publicly available at \url{https://github.com/shan1525/Scalar-field_sign-switch}. No proprietary observational data were used.

%%-----------------------------------------------------------------------------%%
\begin{acknowledgments}
M. B.-L. is supported by the Basque Foundation for Science, Ikerbasque. M. B.-L. and B. I.-U. are supported by Spanish grant PID2023-149016NB-I00, funded by MCIN/AEI/10.13039/501100011033 and by the ERDF program ``A way of making Europe.'' They are also supported by the Basque Government grant No.~IT1977-26 (Spain). S.A.A. acknowledges the support of the DGAPA postdoctoral fellowship program at ICF-UNAM, Mexico, and the High Performance Computing facility Pegasus at IUCAA, Pune, India. J.A.V. acknowledges support from UNAM-DGAPA-PAPIIT IN109126, IN110325, HTC project LANCAD-UNAM-DGTIC-477 and C\'{a}tedra de Investigaci\'{o}n Marcos Moshinsky. \"{O}.A. acknowledges the support from the Turkish Academy of Sciences in the scheme of the Outstanding Young Scientist Award  (T\"{U}BA-GEB\.{I}P). The authors acknowledge the contribution of COST Action CA21136, ``Addressing observational tensions in cosmology with systematics and fundamental physics'' (CosmoVerse). 
\end{acknowledgments}

%%-----------------------------------------------------------------------------%%
\appendix
\section{Diagnostics of the forward shooting calculations}
\label{app:diag}

This appendix collects the operational diagnostics of the shooting calculations of Sec.~\ref{sec:dynamics}: endpoint residuals and cost diagnostics for every retained row and, for the sigmoid--Gaussian scans, optimizer status and tolerance-refinement changes. They document numerical convergence to the compressed targets under the stated prescriptions and are not observational statistics. The numerical environment used Python~3.9.6, NumPy~1.26.4~\cite{Harris:2020xlr},
and SciPy~1.13.1~\cite{Virtanen:2019joe}; the tables and figures were prepared
with pandas~2.3.1~\cite{McKinney:2010nts} and
Matplotlib~3.9.4~\cite{Hunter:2007ouj}.

\onecolumngrid
\begin{center}
\centering
\scriptsize
\renewcommand{\arraystretch}{1.04}
\setlength{\tabcolsep}{1.2pt}
\begin{minipage}[t]{0.49\textwidth}
\centering
\textbf{(a) Endpoint residuals}\\[2pt]
\begin{tabular}{@{}ccrrrr@{}}
\toprule
Block & $H_0$ & $\widetilde\Lambda$ & $E_0-1$ & $\Delta D_{\rm M}\,[{\rm Mpc}]$ & $\mathcal O_{\rm F}$\\
\midrule
 \multirow{5}{*}{A} &75&0.748102&$0$&$-7.28{\times}10^{-12}$&$0$\\
 &73&0.733052&$-1.89{\times}10^{-15}$&$-3.46{\times}10^{-9}$&$-1.11{\times}10^{-16}$\\
 &71&0.716892&$0$&$-1.82{\times}10^{-11}$&$0$\\
 &69&0.699639&$0$&$ 1.82{\times}10^{-12}$&$0$\\
 &68&0.691738&$0$&$ 2.36{\times}10^{-11}$&$0$\\
\midrule
\multirow{5}{*}{B} 
 &75&0.750096&$-2.00{\times}10^{-15}$&$ 1.22{\times}10^{-10}$&$0$\\
 &73&0.734200&$-2.22{\times}10^{-16}$&$ 1.27{\times}10^{-11}$&$0$\\
 &71&0.717439&$0$&$-1.31{\times}10^{-10}$&$-1.11{\times}10^{-16}$\\
 &69&0.699840&$-1.11{\times}10^{-16}$&$ 3.46{\times}10^{-11}$&$0$\\
 &68&0.691868&$ 2.51{\times}10^{-8}$&$ 5.82{\times}10^{-1}$&$ 2.22{\times}10^{-16}$\\
\midrule
\multirow{2}{*}{C}
 &75&0.747294&$0$&$0$&$0$\\
 &73&0.732195&$0$&$-3.64{\times}10^{-12}$&$0$\\
\bottomrule
\end{tabular}
\end{minipage}\hfill
\begin{minipage}[t]{0.49\textwidth}
\centering
\textbf{(b) Optimizer and convergence diagnostics}\\[2pt]
\begin{tabular}{@{}ccrrrc@{}}
\toprule
Block & $H_0$ & $\mathcal C_{\rm LS}^{({\rm sig})}$ & status & $N_{\rm fev}$ & $\delta D_{\rm M}^{({\rm conv})}\,[{\rm Mpc}]$\\
\midrule
\multirow{5}{*}{A} 
 &75&$4.24{\times}10^{-26}$&1&8 &$-2.31{\times}10^{-7}$\\
 &73&$9.77{\times}10^{-21}$&3&8 &$ 9.24{\times}10^{-7}$\\
 &71&$2.65{\times}10^{-25}$&1&9 &$-7.54{\times}10^{-8}$\\
 &69&$2.65{\times}10^{-27}$&1&9 &$ 1.57{\times}10^{-6}$\\
 &68&$4.47{\times}10^{-25}$&1&21&$-1.01{\times}10^{-6}$\\
\midrule
\multirow{5}{*}{B}
 &75&$2.12{\times}10^{-22}$&3&15&$ 3.32{\times}10^{-7}$\\
 &73&$2.59{\times}10^{-24}$&3&11&$-3.65{\times}10^{-7}$\\
 &71&$1.37{\times}10^{-23}$&1&55&$-6.72{\times}10^{-7}$\\
 &69&$1.57{\times}10^{-24}$&3&10&$ 1.14{\times}10^{-6}$\\
 &68&$2.71{\times}10^{-4}$ &2&24&$ 4.99{\times}10^{-7}$\\
\midrule
\multirow{2}{*}{C} 
 &75&$0$&1&8 &$-8.04{\times}10^{-8}$\\
 &73&$1.06{\times}10^{-26}$&1&8 &$ 5.70{\times}10^{-7}$\\
\bottomrule
\end{tabular}
\end{minipage}
\captionof{table}{Numerical diagnostics for the retained rows of Table~\ref{tab:tanh_gaussian_simple}, with $H_0$ in $\mathrm{km\,s^{-1}\,Mpc^{-1}}$. Here $\Delta D_{\rm M}=D_{\rm M}-D_{\rm M}^{({\rm cal})}$ and $\delta D_{\rm M}^{({\rm conv})}$ is the change produced by the tighter integration settings stated in Sec.~\ref{sec:dynamics}. Status 1 denotes termination by the gradient criterion, status 2 termination by the cost-change criterion, and status 3 termination by the step-size criterion. For every retained row, the optimizer terminated successfully with $\mathcal C_{\rm LS}^{({\rm sig})}<10^{-3}$, the integration was finite with positive $E^2$, both required crossings were finite, and neither adjusted parameter was pinned to a bound. The corresponding convergence changes in $E_0$, $z_{\rm t}$, and $z_\dagger$ are below the displayed precision. The tolerance-refinement column tests integration stability, whereas $\mathcal O_{\rm F}$ is an internal constraint-consistency diagnostic rather than an independent accuracy test. The residuals and costs are numerical shooting diagnostics, not likelihood residuals or observational goodness-of-fit statistics.}
\label{tab:tanh_gaussian_diagnostics}

\vspace{4pt}
\centering
\scriptsize
\renewcommand{\arraystretch}{1.10}
\setlength{\tabcolsep}{2.2pt}
\begin{tabular}{@{}ccccc@{}}
\toprule
$H_0$ & $E_0-1$ & $\Delta D_{\rm M}\,[{\rm Mpc}]$ & $\mathcal O_{\rm F}$ & $\mathcal C_{\rm LS}$\\
\midrule
$\mathbf{75}$ & $4.441{\times}10^{-16}$ & $-1.273{\times}10^{-11}$ & $0$ & $1.002{\times}10^{-23}$\\
$73^\dagger$ & $4.102{\times}10^{-7}$ & $11.5252$ & $0$ & $1.133{\times}10^{-1}$\\
$71^\dagger$ & $2.502{\times}10^{-6}$ & $70.0538$ & $-2.220{\times}10^{-16}$ & $4.146$\\
$69^\dagger$ & $4.697{\times}10^{-6}$ & $131.2137$ & $0$ & $14.422$\\
$68^\dagger$ & $5.832{\times}10^{-6}$ & $162.7646$ & $0$ & $22.107$\\
$67^\dagger$ & $7.004{\times}10^{-6}$ & $194.9583$ & $-2.220{\times}10^{-16}$ & $31.604$\\
\bottomrule
\end{tabular}
\captionof{table}{Endpoint and optimization diagnostics for the offset-axion continuation scan of Table~\ref{tab:axion_V0_scan}, with $H_0$ in $\mathrm{km\,s^{-1}\,Mpc^{-1}}$. Here $\Delta D_{\rm M}\equiv D_{\rm M}-D_{\rm M}^{({\rm cal})}$, and the three-term cost $\mathcal C_{\rm LS}^{({\rm ax})}$ of Eq.~\eqref{eq:numerical_merit} contains the normalized present-day expansion, distance, and $z_\dagger$ residuals. Because the residual denominators are numerical weights rather than observational uncertainties, the cost is a shooting diagnostic and not a statistical goodness-of-fit measure.}
\label{tab:axion_diagnostics}
\end{center}
\clearpage
\twocolumngrid

\bibliography{bibliography}

\end{document}